\documentclass[onecolumn,showpacs,preprintnumbers]{revtex4}
\usepackage{mathrsfs}
\usepackage{graphicx}
\usepackage{dcolumn}
\usepackage{bm}
\usepackage{epsfig}
\usepackage{amsmath,amssymb}
\usepackage{array}
\usepackage{tabularx}

\begin{document}
\title{\textbf{Study of $\Lambda_c$$\Lambda_c$ dibaryon and $\Lambda_c$$\bar{\Lambda}_c$ baryonium states via QCD sum rules}}
\author{Xiu-Wu Wang$^{1,2}$}
\email{wangxiuwu2020@163.com}
\author{Zhi-Gang Wang$^{1}$}
\email{zgwang@aliyun.com}
\author{Guo-liang Yu$^{1}$}

\affiliation{$^1$ School of Mathematics and Physics, North China
Electric power university, Baoding, 071003, People's Republic of
China\\$^2$ School of Nuclear Science and Engineering, North China Electric power University, Beijing, 102206, People's Republic of China}

\begin{abstract}
In this article, we construct the six-quark currents with the $J^P=0^+$, $0^-$, $1^+$ and $1^-$ to study the $\Lambda_c$$\Lambda_c$ dibaryon and $\Lambda_c$$\bar{\Lambda}_c$ baryonium states via QCD sum rules. We consider the vacuum condensates up to dimension 16 and truncation of the order $\mathcal{O}(\alpha_s^k )$ with $k\leq3$. The predicted masses are $5.11_{-0.12}^{+0.15}$GeV, $4.66_{-0.06}^{+0.10}$GeV, $4.99_{-0.09}^{+0.10}$GeV $4.68^{+0.08}_{-0.08}$GeV for the $J^P=0^+$, $0^-$, $1^+$ and $1^-$ states, respectively, which can be confronted to the experimental data in the future considering
the high integrated luminosity at the center-of-mass energy about $4.8\,\rm{GeV}$ at the BESIII. We find the terms with $\frac{3}{2}< k \leq 3$ do play a tiny role, and we can ignore these terms safely in the QCD sum rules.
\newline
\newline
\noindent{Keywords: }dibaryon states; QCD sum rules
\pacs{12.39.Mk, 12.38.Lg}
\end{abstract}

\maketitle

\begin{large}
\noindent\textbf{1 Introduction}
\end{large}

In 2003, $X(3872)$ was discovered by the Belle collaboration \cite{Choi}, later on, this state was confirmed by the BABAR collaboration \cite{Aubert}. After that, many charmonium-like and bottomonium-like states were observed. Due to the extraordinary properties  of these exotic $X$, $Y$ and $Z$ states, it is hard to embed them into the conventional charmonium spectrum. The charged charmonium-like states are good candidates for the multiquark states, for example, tetraquark states and molecular states \cite{Guo}. Among these exotic states, the $\Lambda_c$$\bar{\Lambda}_c$  baryonium state was introduced to interpret the decays of the $Y(4260)$ \cite{Qiao1,Qiao2}. In 2006, the $X(1835)$ was tentatively interpreted  as a proton-antiproton baryonium state based on the QCD sum rules \cite{Wang1}. Interestingly, the BESIII collaboration  searched for the strange cousin of the $Z_c(3885)$ in the process $e^++e^-\rightarrow K^+(D_s^-D^{*0}+D_s^{*-}D^0)$ \cite{Ablikim}, they observed the strongest signal  at the center-of-mass energy $\sqrt{s}=4.681$GeV,  but they did not see any evidence of exotic states in the $D_s^{*-}D_{s2}^*(2573)^+$ mass spectrum.

The dibaryon  state is composed by two baryons. The well-known dibaryon state is the deuteron, the bound state of a proton and neutron with binding energy $E_B=2.225$ MeV. In 1994, the $H$-dibaryon state was studied with the QCD sum rules \cite{Kodama}, thereafter,  applying the QCD sum rules to study the dibaryon and baryonium states has triggered much attention. In Ref.\cite{Chen}, the masses of the $d^*(2380)$ et al were calculated using the $\Delta\Delta$-like six-quark local interpolating  currents. In Ref.\cite{Wang2}, the scalar and axialvector $\Xi_{cc}\Sigma_c$ dibaryon states were studied using the color-singlet-color-singlet type interpolating currents. And very recently, the hidden-bottom and hidden-charm hexaquark molecular states with the quantum numbers $J^{PC}=0^{++},0^{-+},1^{++}$, and $1^{--}$ have  been studied in the framework of the QCD sum rules \cite{Wan}.  It is proposed via the Bethe-Salpeter equation that the $\Lambda_c$$\bar{\Lambda}_c$ baryonium states should contain at least two states: the $\Lambda_c$$\bar{\Lambda}_c$ molecular state and another one with mass around $4.65$GeV for the thresholds up to $4.7$GeV \cite{Dong}. But the masses of the $\Lambda_c$$\bar{\Lambda}_c$ baryonium states in Ref.\cite{Wan} are all above $4.7$GeV.

The QCD sum rules is a successful method in studying the hadronic properties\cite{xzAgaev,xzChen1,xzChen2,xzOzdem,xzLee1,xzZhang1,xzZhang2,xzMatheus,xzMihara,xzAlbuquerque,xzDi,xzLee2,xzShi0}. It could be used to calculate the masses of hadrons, decay constants, hadronic form-factors, coupling constants, etc \cite{Shifman,Reinders}. This method has been applied extensively to study  the hidden-charm (bottom) tetraquark (molecular) states \cite{xzAgaev,xzChen1,xzChen2,xzOzdem,xzLee1,xzZhang1,xzZhang2,xzMatheus,xzMihara,xzAlbuquerque,xzDi,Wang3,Wang4,Wang5} and pentaquark (molecular) states \cite{Wang6,Wang7}. In this paper, we construct four six-quark local interpolating currents to study both the $\Lambda_c$$\Lambda_c$ dibaryon and $\Lambda_c$$\bar{\Lambda}_c$ baryonium states with the $J^P=0^+,0^-,1^+$ and $1^-$,  respectively.

The article is arranged as follows: we derive the QCD sum rules for the masses  and pole residues of the dibaryon and baryonium states in Sect.2; we present the numerical results and discussions in Sect.3; Sect.4 is reserved for our conclusion.

\begin{large}
\noindent\textbf{2 QCD sum rules for the dibaryon and baryonium states}
\end{large}

We begin  our calculation from the following two-point correlation functions,
\begin{eqnarray}
\notag &&\Pi_{1}(p)=i\int d^4x e^{ip\cdot x}\langle 0 | J_1 (x) J_1^\dag(0) | 0\rangle \, ,\\
\notag &&\Pi_{2}(p)=i\int d^4x e^{ip\cdot x}\langle 0 | J_2 (x) J_2^\dag(0) | 0\rangle \, ,\\
\notag &&\Pi_{3;\mu\nu}(p)=i\int d^4x e^{ip\cdot x}\langle 0 | J_{3,\mu} (x) J_{3,\nu}^\dag(0) | 0\rangle \, ,\\
&&\Pi_{4;\mu\nu}(p)=i\int d^4x e^{ip\cdot x}\langle 0 | J_{4,\mu}(x)J_{4,\nu}^\dag(0) | 0\rangle \, ,
\end{eqnarray}
where
\begin{eqnarray}
\notag && J_1(x)=J_{\Lambda}^{T}(x)C\gamma_5 J_{\Lambda}(x)\, ,\\
\notag && J_2(x)=\overline{J}_{\Lambda}(x)i\gamma_5 J_{\Lambda}(x)\, ,\\
\notag && J_{3,\mu}(x)=\overline{J}_{\Lambda}(x)\gamma_5\gamma_\mu J_{\Lambda}(x)\, ,\\
 && J_{4,\mu}(x)=\overline{J}_{\Lambda}(x)\gamma_\mu J_{\Lambda}(x)\, ,
\end{eqnarray}
 $J_\Lambda(x)=\varepsilon^{ijk}u^{iT}(x)C\gamma_5d^j(x)c^k(x)$, the superscripts $i, j, k$ are color indices and $C$ represents the charge conjugation matrix. Under parity transformation $\widehat{P}$ , the four currents satisfy the following equations,
\begin{eqnarray}
\notag &&\widehat{P}J_1(x)\widehat{P}^{-1}=J_1(\widetilde{x}) \, ,\\
\notag &&\widehat{P}J_2(x)\widehat{P}^{-1}=-J_2(\widetilde{x})\, , \\
\notag &&\widehat{P}J_{3,\mu}(x)\widehat{P}^{-1}=-J_3^\mu(\widetilde{x})\, ,\\
&&\widehat{P}J_{4,\mu}(x)\widehat{P}^{-1}=J_4^\mu(\widetilde{x}) \, ,
\end{eqnarray}
\noindent where  $x^\mu=(x^0,x^1,x^2,x^3)$ and $\widetilde{x}^\mu=(x^0,-x^1,-x^2,-x^3)$. As the phenomenological expressions of the correlation functions are concerned, we  insert  a complete set of intermediate hadronic states with the same quantum numbers as the current operators $J_{1,2,3,4}(x)$ into the correlation functions and  get,
\begin{eqnarray}
\notag && \Pi_{1,2}(p)=\frac{\lambda_{1,2}^2}{M_{1,2}^2-p^2}+...\, , \\
&&\Pi_{3,4;\mu\nu}(p)=A_{3,4}(p)\left(-g_{\mu\nu}+\frac{p_\mu p_\nu}{p^2}\right)+... \, ,
\end{eqnarray}
where
\begin{eqnarray}
\notag && A_{3,4}(p)=\frac{\lambda_{3,4}^2}{M_{3,4}^2-p^2}+...\, , \\
\notag && \langle0 | J_{1,2}(0)|Z_{1,2}(p) \rangle =\lambda_{1,2}\, ,\\
&&\langle0 | J_{3,4;\mu}(0)|Z_{3,4}(p) \rangle =\lambda_{3,4}\epsilon_\mu \, ,
\end{eqnarray}
\noindent $|Z_{1,2,3,4}\rangle$ denote the ground states and $\lambda_{1,2,3,4}$ are the pole residues, which show the couplings  of the currents $J_{1,2,3,4}$ to the molecular states, the $\epsilon_\mu$ are the polarization vectors.

We apply the Wick theorem to contract the quark fields and acquire  the expressions of the correlation functions in terms of full quark propagators,
\begin{eqnarray}
\notag &&\Pi_1(p)=2\varepsilon^{ijk}\varepsilon^{lmn}\varepsilon^{i'j'k'}\varepsilon^{l'm'n'}i\int d^4x e^{ip\cdot x}  \\
\notag && \Big\{-Tr\Big\{\gamma_5B^{kk'}(x)\gamma_5CB^{nn'T}(x)C\Big\}Tr\Big\{\gamma_5Q^{jj'}(x)\gamma_5CQ^{ii'T}(x)C\Big\}Tr\Big\{\gamma_5Q^{mm'}(x)\gamma_5CQ^{ll'T}(x)C\Big\} \\
\notag && -Tr\Big\{\gamma_5B^{kn'}(x)\gamma_5CB^{nk'T}(x)C\Big\}Tr\Big\{\gamma_5Q^{jj'}(x)\gamma_5CQ^{ii'T}(x)C\Big\}Tr\Big\{\gamma_5Q^{mm'}(x)\gamma_5CQ^{ll'T}(x)C\Big\} \\
\notag && +Tr\Big\{\gamma_5B^{kk'}(x)\gamma_5CB^{nn'T}(x)C\Big\}Tr\Big\{\gamma_5Q^{li'}(x)\gamma_5CQ^{jj'T}(x)C\gamma_5Q^{il'}(x)\gamma_5CQ^{mm'T}(x)C\Big\} \\
\notag && +Tr\Big\{\gamma_5B^{kk'}(x)\gamma_5CB^{nn'T}(x)C\Big\}Tr\Big\{\gamma_5Q^{ii'}(x)\gamma_5CQ^{mj'T}(x)C\gamma_5Q^{ll'}(x)\gamma_5CQ^{jm'T}(x)C\Big\}\Big\}\, ,
\end{eqnarray}
\begin{eqnarray}
\notag &&\Pi_2(p)=\varepsilon^{ijk}\varepsilon^{lmn}\varepsilon^{i'j'k'}\varepsilon^{l'm'n'}i\int d^4x e^{ip\cdot x}\\
\notag && Tr\Big\{\gamma_5B^{nn'}(x)\gamma_5B^{k'k}(-x)\Big\}Tr\Big\{\gamma_5Q^{mm'}(x)\gamma_5CQ^{ll'T}(x)C\Big\}Tr\Big\{\gamma_5Q^{j'j}(-x)\gamma_5CQ^{i'iT}(-x)C\Big\}\, ,
\end{eqnarray}
\begin{eqnarray}
\notag &&\Pi_{3;\mu\nu}(p)=-\varepsilon^{ijk}\varepsilon^{lmn}\varepsilon^{i'j'k'}\varepsilon^{l'm'n'}i\int d^4x e^{ip\cdot x}  \\
\notag && Tr\Big\{\gamma_5\gamma_{\mu}B^{kk'}(x)\gamma_5\gamma_{\nu}B^{n'n}(-x)\Big\}Tr\Big\{\gamma_5Q^{jj'}(x)\gamma_5CQ^{ii'T}(x)C\Big\}Tr\Big\{\gamma_5Q^{m'm}(-x)\gamma_5CQ^{l'lT}(-x)C\Big\}\, ,
\end{eqnarray}
\begin{eqnarray}
\notag &&\Pi_{4;\mu\nu}(p)=-\varepsilon^{ijk}\varepsilon^{lmn}\varepsilon^{i'j'k'}\varepsilon^{l'm'n'}i\int d^4x e^{ip\cdot x}  \\
&& Tr\Big\{\gamma_{\mu}B^{kk'}(x)\gamma_{\nu}B^{n'n}(-x)\Big\}Tr\Big\{\gamma_5Q^{jj'}(x)\gamma_5CQ^{ii'T}(x)C\Big\}Tr\Big\{\gamma_5Q^{m'm}(-x)\gamma_5CQ^{l'lT}(-x)C\Big\}\, ,
\end{eqnarray}
where $Q^{ab}(x)$ and $B^{ab}(x)$ are the full light and heavy quark propagators respectively, and they can be  written as,
\begin{eqnarray}
\notag\ Q^{ab}(x)=&& \frac{ix\!\!\!/\delta^{ab}}{2\pi^{2}x^{4}}-\frac{\delta^{ab}}{12}\langle\overline{q}q\rangle-\frac{\delta^{ab}x^2}{192}\langle\overline{q}g_s\sigma G q\rangle-\frac{i\delta^{ab}x^2x\!\!\!/g_s^2\langle\overline{q}q\rangle^2}{7776}-t^n_{ab}(x\!\!\!/\sigma^{\alpha\beta}+\sigma^{\alpha\beta}x\!\!\!/)\frac{i}{32\pi^2x^2}g_s G_{\alpha\beta}^n  \\
\notag\ &&-\frac{\delta^{ab}x^4\langle\overline{q}q\rangle\langle GG \rangle}{27648}-\frac{1}{8}\langle\overline{q}^b\sigma^{\alpha\beta}q^a\rangle\sigma_{\alpha\beta}-\frac{1}{4}\langle\overline{q}^b\gamma_\mu q^a\rangle\gamma^\mu+\cdots\, ,
\end{eqnarray}

\begin{eqnarray}
\notag\
B_{ab}(x)&&=\frac{i}{(2\pi)^{4}}\int d^{4}ke^{-ik\cdot x}\bigg\{\frac{\delta_{ab}}{k\!\!\!/-m_{c}}-\frac{g_{s}G_{\alpha\beta }^{h}t_{ab}^{h}}{4}\frac{\sigma^{\alpha\beta}(k\!\!\!/+m_{c})+(k\!\!\!/+m_{c})\sigma ^{\alpha
\beta }}{(k^{2}-m_{c}^{2})^{2}}\\
\notag\
&&+\frac{g_{s}D_{\alpha}G_{\beta\lambda}^{h}t_{ab}^{h}(f^{\lambda\beta\alpha}+f^{\lambda\alpha\beta})}{3(k^{2}-m_{c}^{2})^{4}}
-\frac{g_{s}^{2}(t^{h}t^{r})_{ab}G_{\alpha \beta }^{h}G_{\mu \nu
}^{r}(f^{\alpha \beta \mu \nu }+f^{\alpha \mu \beta \nu }+f^{\alpha
\mu \nu \beta })}{4(k^{2}-m_{c}^{2})^{5}}\\
\notag\
&&\frac{\langle GGG\rangle}{48(k^2-m_c^2)^6}(k\!\!\!/+m_c)[k\!\!\!/(k^2-3m_c^2)+2m_c(2k^2-m_c^2)](k\!\!\!/+m_c)+\cdot\cdot\cdot\bigg \}\, ,
\end{eqnarray}

\begin{eqnarray}
\notag\ &&f^{\lambda \alpha \beta }=(k\!\!\!/+m_{c})\gamma ^{\lambda
}(k\!\!\!/+m_{c})\gamma ^{\alpha }(k\!\!\!/+m_{c})\gamma ^{\beta
}(k\!\!\!/+m_{c}) \, ,\\
&&f^{\alpha \beta \mu \nu }=(k\!\!\!/+m_{c})\gamma ^{\alpha
}(k\!\!\!/+m_{c})\gamma ^{\beta }(k\!\!\!/+m_{c})\gamma ^{\mu
}(k\!\!\!/+m_{c})\gamma ^{\nu }(k\!\!\!/+m_{c}) \, ,
\end{eqnarray}

\noindent where $t^n=\frac{\lambda^n}{2}$, $\lambda^n$ is the Gell-Mann matrix and $D_\alpha=\partial_\alpha-ig_sG_\alpha^ht^h$ \cite{Reinders,Wang3,Pascual}. Here, we consider the term $\langle\overline{q}^b\sigma^{\alpha\beta}q^a\rangle$ ($\langle\overline{q}^b\gamma^{\mu}q^a\rangle$) which is derived from Fierz rearrangement of the quark-antiquark pair $\langle q^a\overline{q}^b\rangle$, it is used to absorb the gluons emitted from other quark lines to form the mixed condensates $\langle\overline{q}g_s\sigma Gq\rangle$, $\langle\overline{q}g_s\sigma Gq\rangle^2$ and $\langle\overline{q}g_s\sigma Gq\rangle^3$, respectively (the four-quark condensate $g_s^2\langle\bar{q}q\rangle^2$) \cite{Wang3}. For the light quark propagator, we calculate the integrals in the coordinate space, as for the heavy quark counterpart, we consider the momentum space.

\begin{figure}[ht]
\centering
\includegraphics[height=5cm,width=6.5cm]{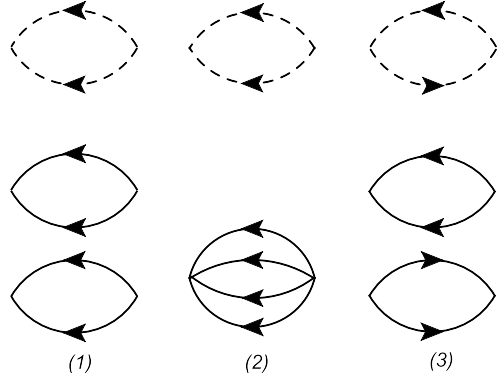}
\centering\caption{The Feynman diagrams (1) and (2) (diagram (3)) are the lowest order contributions for the  current $J_1$ (currents $J_{2,3,4}$), where the solid and dashed lines represent the light and heavy quark propagators respectively.}
\label{fig:label}
\end{figure}

\begin{figure}[ht]
\centering
\includegraphics[height=5.5cm,width=8.5cm]{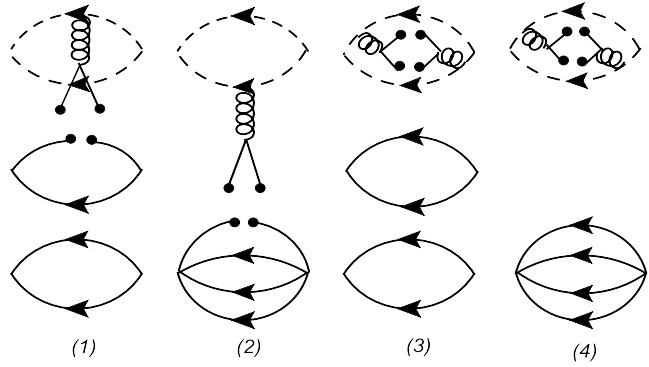}
\centering\caption{The Feynman diagrams contribute to the four-quark condensates, where (1)(2) for $\langle\overline{q}q\rangle^2$ and (3)(4) for $\langle\overline{\psi}\psi\rangle^2$, respectively.}
\label{fig:label}
\end{figure}

\begin{figure}[ht]
\centering
\includegraphics[height=5.5cm,width=9.9cm]{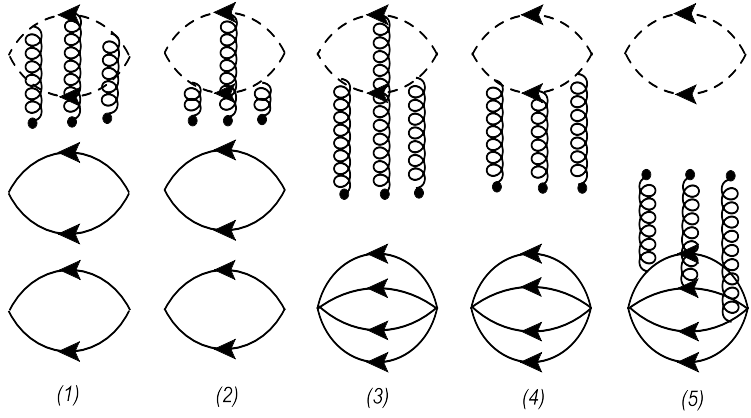}
\centering\caption{Typical Feynman graphs for three-gluon condensates, truncation $k=\frac{3}{2}$.}
\label{fig:label}
\end{figure}

\begin{figure}[ht]
\centering
\includegraphics[height=5.5cm,width=8.8cm]{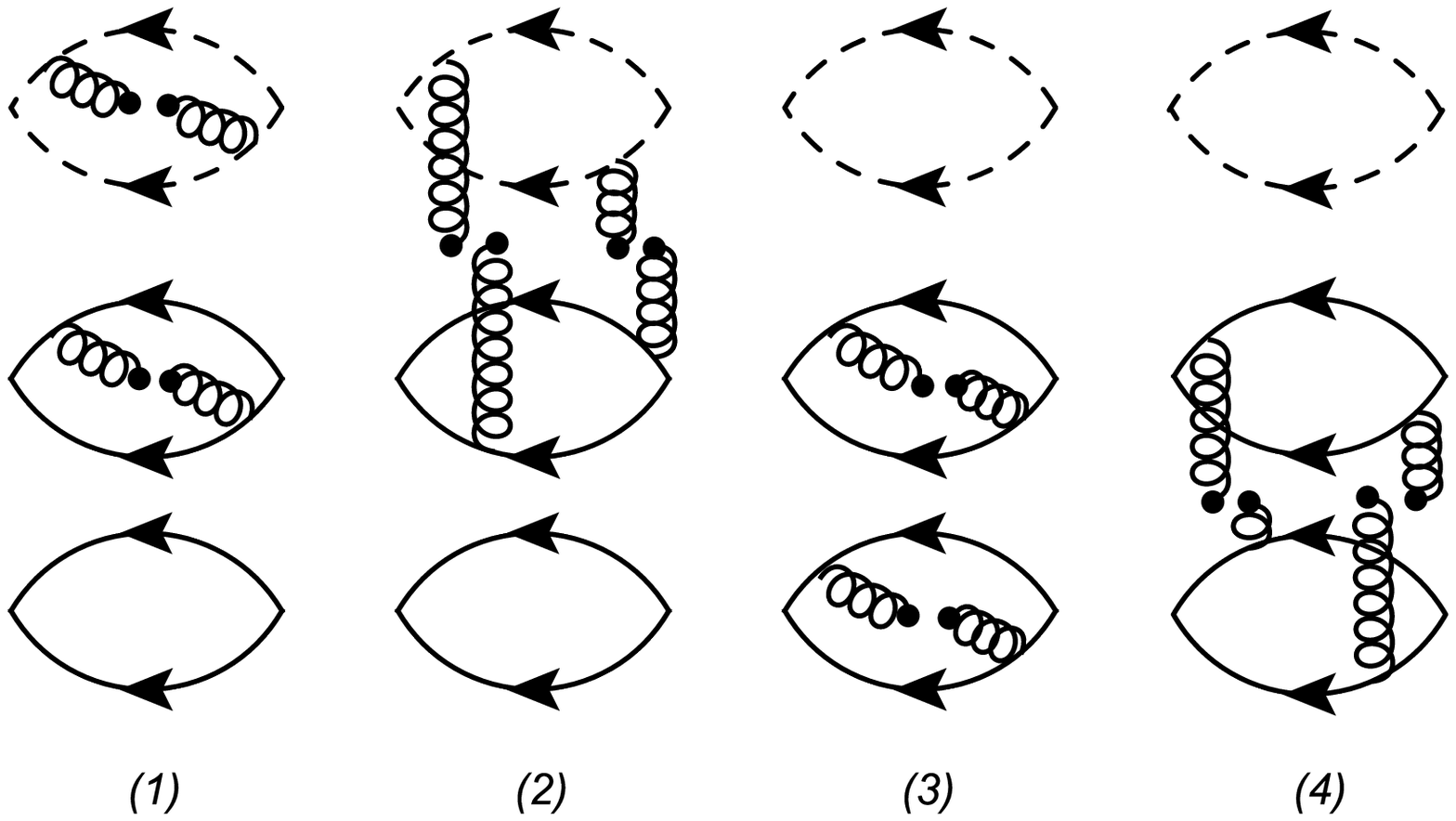}
\centering\caption{Typical Feynman graphs for the $\langle\frac{\alpha_s}{\pi}GG\rangle^2$ condensates, truncation  $k=2$.}
\label{fig:label}
\end{figure}

\begin{figure}[ht]
\centering
\includegraphics[height=5.5cm,width=6.6cm]{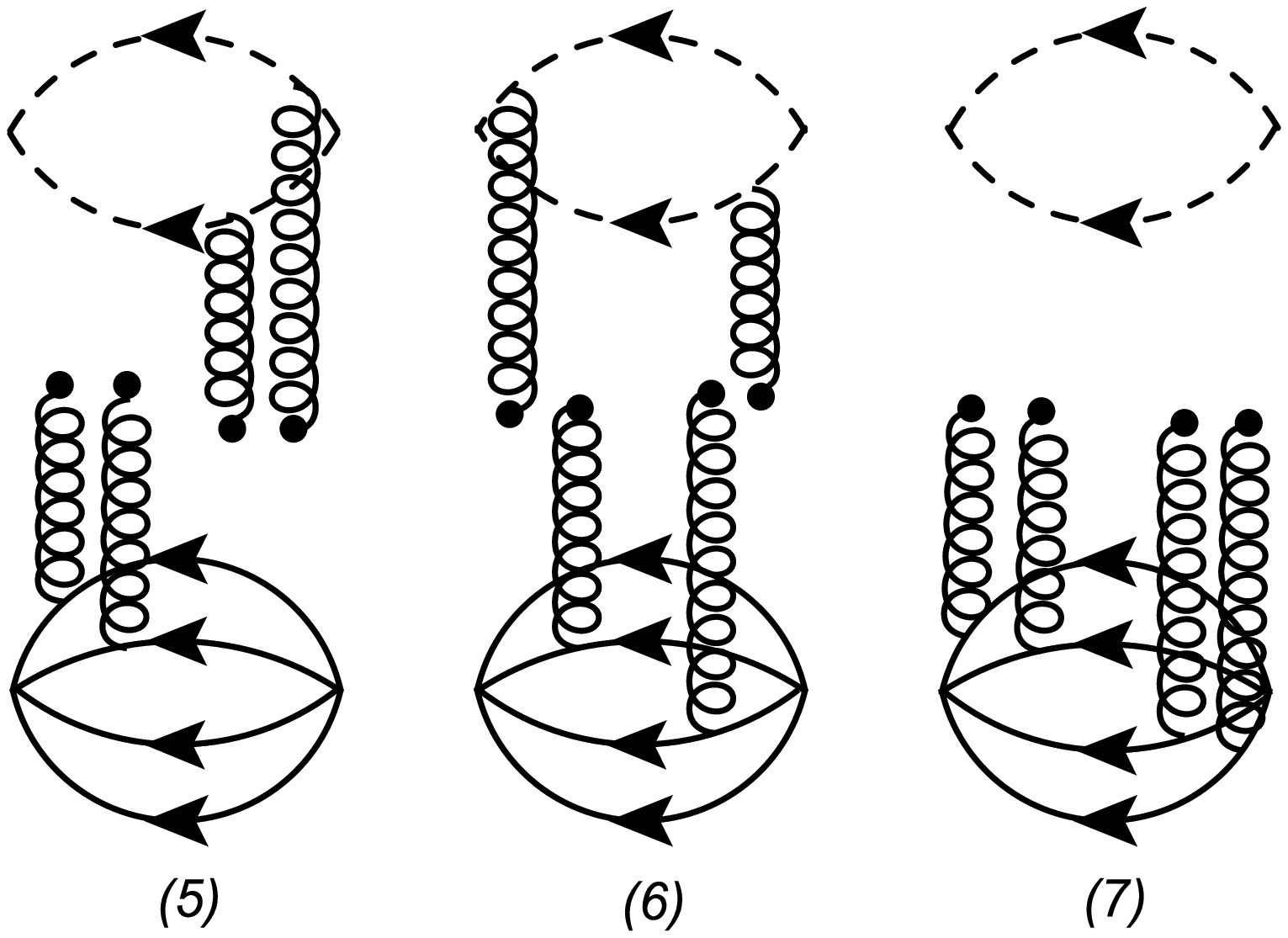}
\centering\caption{Typical Feynman graphs for the $\langle\frac{\alpha_s}{\pi}GG\rangle^2$, truncation  $k=2$.}
\label{fig:label}
\end{figure}

\begin{figure}[ht]
\centering
\includegraphics[height=5.5cm,width=6.6cm]{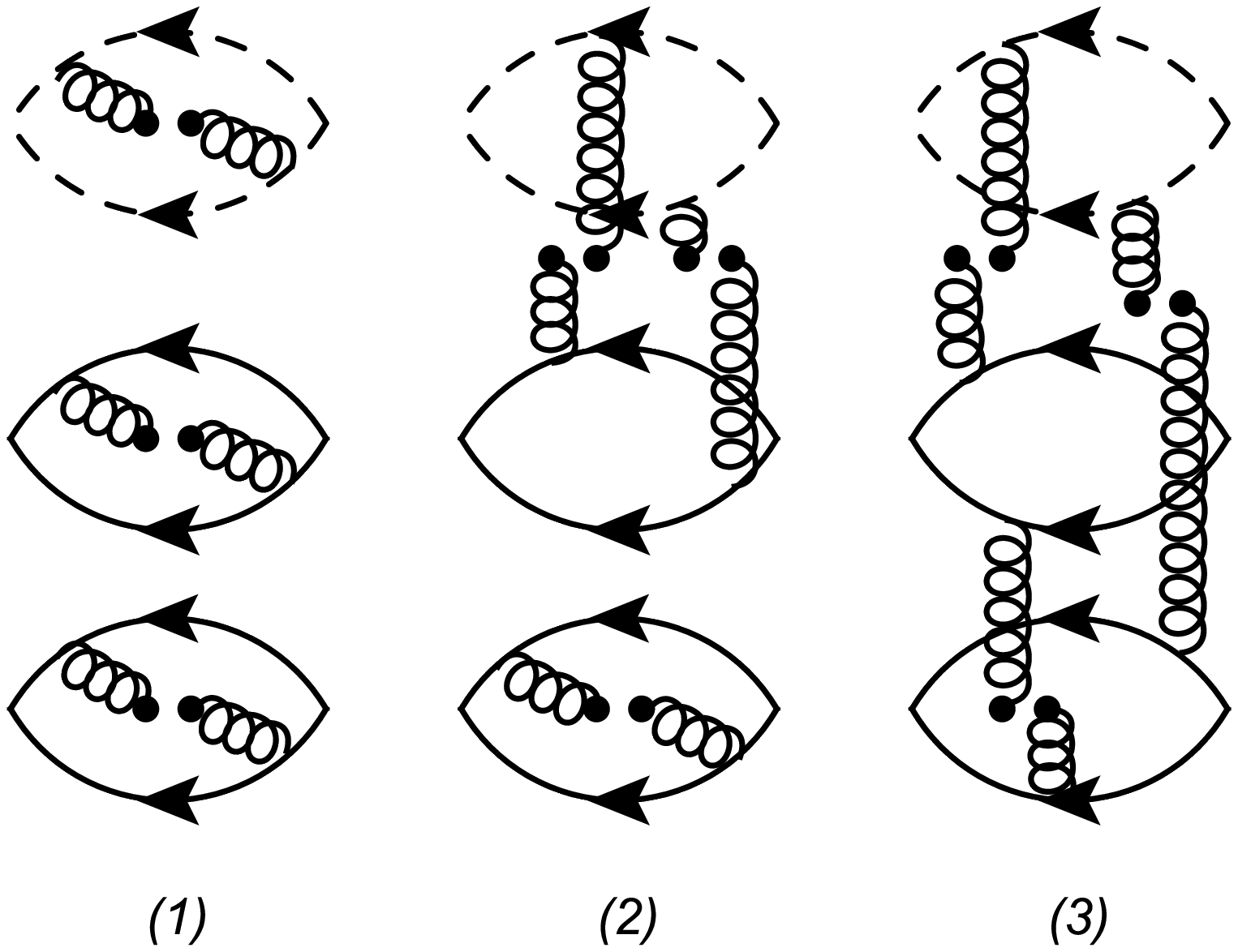}
\centering\caption{Parts of typical Feynman graphs for the $\langle\frac{\alpha_s}{\pi}GG\rangle^3$ condensates with truncation  $k=3$.}
\label{fig:label}
\end{figure}

Now, let us draw the Feynman diagrams for the leading order contribution of  these correlation functions, see Fig.1, there are four light quark lines and two heavy quark lines. Considering each heavy quark line emits a gluon and each light quark line contributes quark-antiquark pair, we obtain the quark-gluon operator $g_sG_{\alpha\beta}g_sG_{\eta\tau}\overline{q}q\overline{q}q\overline{q}q\overline{q}q$ with dimension 16. This operator leads to the vacuum condensates $\langle\frac{\alpha_s}{\pi}GG\rangle\langle\overline{q}q\rangle^4$ and $\langle\overline{q}g_s\sigma Gq\rangle^2\langle\overline{q}q\rangle^2$.  In order to find the exact contributions  of different kinds of the vacuum condensates, we  choose the $J_1(x)$ for detailed calculations and consider all kinds of the vacuum condensates up to  dimension 16 and take the truncation of the order $\mathcal{O}(\alpha_s^k)$ with $k\leq3$. The considered terms include $\langle\frac{\alpha_s}{\pi}GG\rangle$, $\langle\overline{q}q\rangle^2$, $\langle\overline{\psi}\psi\rangle^2=\sum\limits_{u,s,d}\langle\overline{q}q\rangle^2$, $\langle g_s^3fGGG\rangle$, $\langle\overline{q}g_s\sigma Gq\rangle\langle\overline{q}q\rangle$, $\langle\frac{\alpha_s}{\pi}GG\rangle^2$, $\langle\overline{q}g_s\sigma Gq\rangle^2$, $\langle\frac{\alpha_s}{\pi}GG\rangle\langle\overline{q}q\rangle^2$, $\langle\overline{q}q\rangle^4$, $\langle\overline{q}g_s\sigma Gq\rangle\langle\frac{\alpha_s}{\pi}GG\rangle\langle\overline{q}q\rangle$, $\langle g_s^3fGGG\rangle^2$, $\langle\frac{\alpha_s}{\pi}GG\rangle^3$, $\langle\overline{q}q\rangle^2\langle\overline{\psi}\psi\rangle^2$, $\langle\overline{q}g_s\sigma Gq\rangle\langle\overline{q}q\rangle^3$, $\langle\overline{q}q\rangle\langle\overline{q}g_s\sigma Gq\rangle\langle\overline{\psi}\psi\rangle^2$, $\langle\frac{\alpha_s}{\pi}GG\rangle^2\langle\overline{q}q\rangle^2$, $\langle\overline{q}g_s\sigma Gq\rangle^2\langle\frac{\alpha_s}{\pi}GG\rangle$, $\langle\overline{q}g_s\sigma Gq\rangle^2\langle\overline{q}q\rangle^2$, $\langle\frac{\alpha_s}{\pi}GG\rangle\langle\overline{q}q\rangle^4$,  $\langle\overline{q}g_s\sigma Gq\rangle\langle\frac{\alpha_s}{\pi}GG\rangle^2\langle\overline{q}q\rangle$ and  $\langle\overline{q}g_s\sigma Gq\rangle^2\langle\overline{\psi}\psi\rangle^2$. The vacuum condensates $\langle\overline{q}q\rangle$, $\langle\overline{q}g_s\sigma Gq\rangle$ and $\langle\overline{q}g_s\sigma Gq\rangle^3$ have no contribution for all the four currents. From the numerical results, we find that those terms with  $k\geq\frac{3}{2}$ do play a tiny role. So, for the currents $J_{2,3,4}$, it is safe for us to take account of only the terms $\langle\frac{\alpha_s}{\pi}GG\rangle$, $\langle\overline{q}q\rangle^2$, $\langle\overline{q}g_s\sigma Gq\rangle\langle\overline{q}q\rangle$, $\langle\overline{q}g_s\sigma Gq\rangle^2$, $\langle\frac{\alpha_s}{\pi}GG\rangle\langle\overline{q}q\rangle^2$, $\langle\overline{q}q\rangle^4$, $\langle\overline{q}g_s\sigma Gq\rangle\langle\overline{q}q\rangle^3$, $\langle\overline{q}g_s\sigma Gq\rangle^2\langle\overline{q}q\rangle^2$ and $\langle\frac{\alpha_s}{\pi}GG\rangle\langle\overline{q}q\rangle^4$.

For the current $J_1$, we take into account of the four-quark condensates, whose typical Feynman diagrams  are shown in Fig.2; the three-gluon condensates, whose typical Feynman diagrams  are shown in Fig.3. We also take into consideration of the $\langle\frac{\alpha_s}{\pi}GG\rangle^2$  of the order $\mathcal{O}(\alpha_s^2)$, the corresponding Feynman diagrams are shown in  Fig.4 and Fig.5. The typical terms with the highest truncation $k=3$ are the vacuum  condensates $\langle\frac{\alpha_s}{\pi}GG\rangle^3$ and $\langle g_s^3fGGG\rangle^2$. Parts of the Feynman diagrams are shown in Fig.6, one can draw  other diagrams in a similar way as displayed.

After accomplishing the operator product expansion, we perform the Borel transform in regard to $P^2=-p^2$,
\begin{eqnarray}
\widehat{B}{(P^2)}\Pi(p)=\int_{\Delta^2}^{\infty}ds\rho_{QCD}(s)\exp\left(-\frac{s}{T^2}\right)\, ,
\end{eqnarray}
\noindent where the $\rho_{QCD}(s)$ are the QCD spectral densities, the $T^2$ is the Borel parameter and $\widehat{B}{(P^2)}$ is the Borel operator which is defined as,

\begin{eqnarray}
\widehat{B}_{T^2}(P^2)=\lim_{-p^2,n\rightarrow\infty\atop -p^2/n\rightarrow T^2}\frac{(-p^2)^{(n+1)}}{n!}\left(\frac{d}{dp^2}\right)^n\, .
\end{eqnarray}

Matching  Eq.(8) with the spectral densities on the hadron side, we get the QCD sum rules,
\begin{eqnarray}
\lambda_{1,2,3,4}^2\exp\left(-\frac{M_{1,2,3,4}^2}{T^2}\right)=\int_{\Delta^2}^{s_0}ds \rho_{1,2,3,4}(s)\exp\left(-\frac{s}{T^2}\right)\, ,
\end{eqnarray}
\noindent where the $\rho_{1,2,3,4}(s)$ are the QCD spectral densities for the currents $J_{1,2,3,4}$ and $\Delta^2=4m_c^2$.
An appendix is attached at the end of this paper, one can check the exact expressions of the spectral densities of the four currents there.
As for the continuum threshold parameters $s_0$, we take  experiential values,
\begin{eqnarray}
\sqrt{s_0}=M_{Z}+(0.5\sim0.7){\rm GeV}\, ,
\end{eqnarray}
where $M_{Z}$ are the masses of the ground states.

We differentiate the Eq.(10) with respect to $\tau=-\frac{1}{T^2}$ and eliminate the pole residues, the extracted masses can be written as,
\begin{eqnarray}
M_{1,2,3,4}^2=\frac{-\frac{d}{d\tau}\int_{\Delta^2}^{s_0}ds \rho_{1,2,3,4}(s)\exp\left(-s\tau\right)}{\int_{\Delta^2}^{s_0}ds \rho_{1,2,3,4}(s)\exp\left(-s\tau\right)}\, .
\end{eqnarray}

\begin{large}
\noindent\textbf{3 Numerical result and discussion}
\end{large}

We apply the standard values of the vacuum condensates $\langle
\bar{q}q \rangle=-(0.24\pm 0.01\, \rm{GeV})^3$,   $\langle
\bar{q}g_s\sigma G q \rangle=m_0^2\langle \bar{q}q \rangle$,
$m_0^2=(0.8 \pm 0.1)\,\rm{GeV}^2$, $\langle \frac{\alpha_s
GG}{\pi}\rangle=(0.33\,\rm{GeV})^4 $  at the energy scale $\mu=1{\rm GeV}$ \cite{Reinders,PDG}, and choose the $
\overline{MS}$ mass $m_c(m_c)=1.275\pm0.025{\rm GeV}$ from the Particle Data Group \cite{PDG}. Consider the energy-scale dependence for the above condensates \cite{Narison},
\begin{eqnarray}
  \langle\bar{q}q \rangle(\mu)&=&\langle\bar{q}q \rangle({\rm 1 GeV})\left[\frac{\alpha_{s}({\rm 1 GeV})}{\alpha_{s}(\mu)}\right]^{\frac{12}{33-2n_f}}\, , \nonumber\\
   \langle\bar{q}g_s \sigma Gq \rangle(\mu)&=&\langle\bar{q}g_s \sigma Gq \rangle({\rm 1 GeV})\left[\frac{\alpha_{s}({\rm 1 GeV})}{\alpha_{s}(\mu)}\right]^{\frac{2}{33-2n_f}}\, ,\nonumber\\
m_c(\mu)&=&m_c(m_c)\left[\frac{\alpha_{s}(\mu)}{\alpha_{s}(m_c)}\right]^{\frac{12}{33-2n_f}} \, ,\nonumber\\
\alpha_s(\mu)&=&\frac{1}{b_0t}\left[1-\frac{b_1}{b_0^2}\frac{\log t}{t} +\frac{b_1^2(\log^2{t}-\log{t}-1)+b_0b_2}{b_0^4t^2}\right]\, ,
\end{eqnarray}
where the related parameters are written as,
where $t=\log \frac{\mu^2}{\Lambda^2}$, $b_0=\frac{33-2n_f}{12\pi}$, $b_1=\frac{153-19n_f}{24\pi^2}$, $b_2=\frac{2857-\frac{5033}{9}n_f+\frac{325}{27}n_f^2}{128\pi^3}$,  $\Lambda=213\,\rm{MeV}$, $296\,\rm{MeV}$  and  $339\,\rm{MeV}$ for the quark flavors  $n_f=5$, $4$ and $3$, respectively  \cite{PDG}. In this paper, for the four constructed currents, we choose the flavor number $n_f=4$. We set $\alpha_s\langle\overline{\psi}\psi\rangle^2=-1.8\times10^{-4}$GeV$^6$, $\langle g_{s}^3fGGG\rangle=0.045$GeV$^6$ \cite{Shifman,Reinders,Colangelo1}. Compared with the heavy quarks, the masses of the light quarks $u$ and $d$ are too small to make much difference. So, we ignore the masses of the light quarks. The masses of baryon and anti-baryon are $M_{\Lambda_c}=M_{\overline\Lambda_c}=2.29$GeV from the Particle Data Group \cite{PDG}.
Note that, Eq.(12) depends on the energy scale $\mu$ \cite{Wang4,Wang5},   we apply the energy scale formula \cite{Wang3},
\begin{eqnarray}
\mu=\sqrt{M_Z^2-4\mathbb{M}_c^2}\, ,
\end{eqnarray}
to determine the best energy scales of the QCD spectral densities, where the $\mathbb{M}_c$ is the effective charmed quark mass, its updated value is $1.85$GeV \cite{wangzg}.
Combining above equations, we now find that the $M_{Z}$ depend    on the threshold parameter $s_0$ and Borel parameter $T^2$. Of course, at first, we do not know the suitable energy scale, so we adjust the value of the $\mu$ via trial and error. From the energy scale formula Eq.(14), one can determine that the mass $M_Z$ increases with the parameter $\mu$, however, numerical results show that mass derived from Eq.(12) decreases with the increment of the parameter $\mu$, we can always find a best energy scale and determine the mass which satisfies the energy scale formula. As for the values of the $s_0$ and $T^2$, they are determined by the following two principles,
\newline
firstly, the pole dominance criterion, here we define the pole contributions (PC) of the QCD sum rules as,
\begin{eqnarray}
{\rm PC}=\frac{\int_{4m_c^2}^{s_0}ds\rho_{QCD}(s)\exp\left(-\frac{s}{T^2}\right)}{\int_{4m_c^2}^{\infty}ds\rho_{QCD}(s)\exp\left(-\frac{s}{T^2}\right)}\, ,
\end{eqnarray}
secondly, the convergence of the  operator product expansion, for which, we define the contributions of the  vacuum condensates of dimension $n$,
\begin{eqnarray}
D(n)=\frac{\int_{4m_c^2}^{s_0}ds\rho_{QCD;n}(s)\exp\left(-\frac{s}{T^2}\right)}{\int_{4m_c^2}^{s_0}ds\rho_{QCD}(s)\exp\left(-\frac{s}{T^2}\right)}\, .
\end{eqnarray}

\begin{figure}[h]
\begin{minipage}[h]{0.45\linewidth}
\centering
\includegraphics[height=5cm,width=7cm]{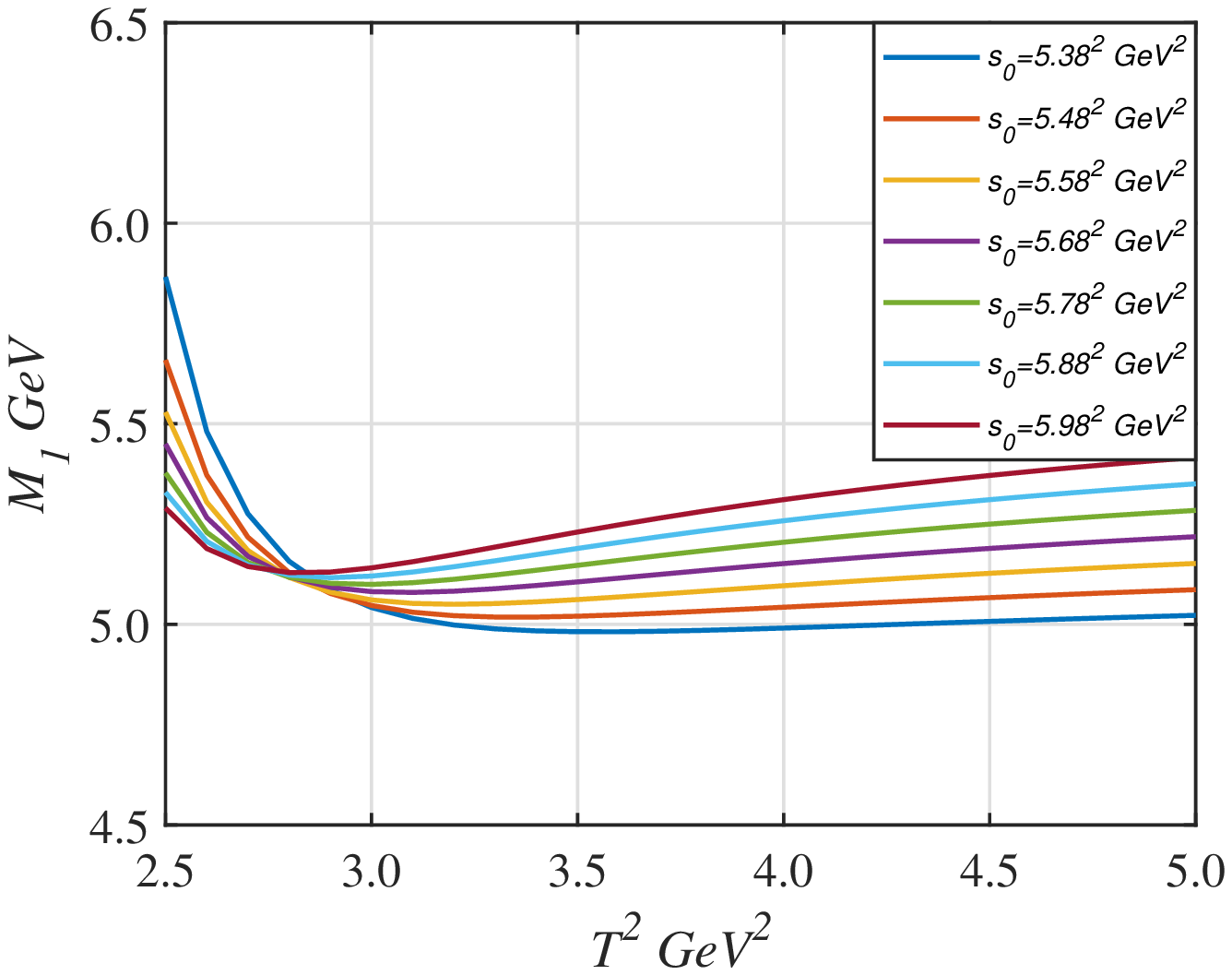}
\caption{The $M_1-T^2$ curves of the $J_1$ under different values of $\sqrt{s_0}$. \label{your label}}
\end{minipage}
\hfill
\begin{minipage}[h]{0.45\linewidth}
\centering
\includegraphics[height=5cm,width=7cm]{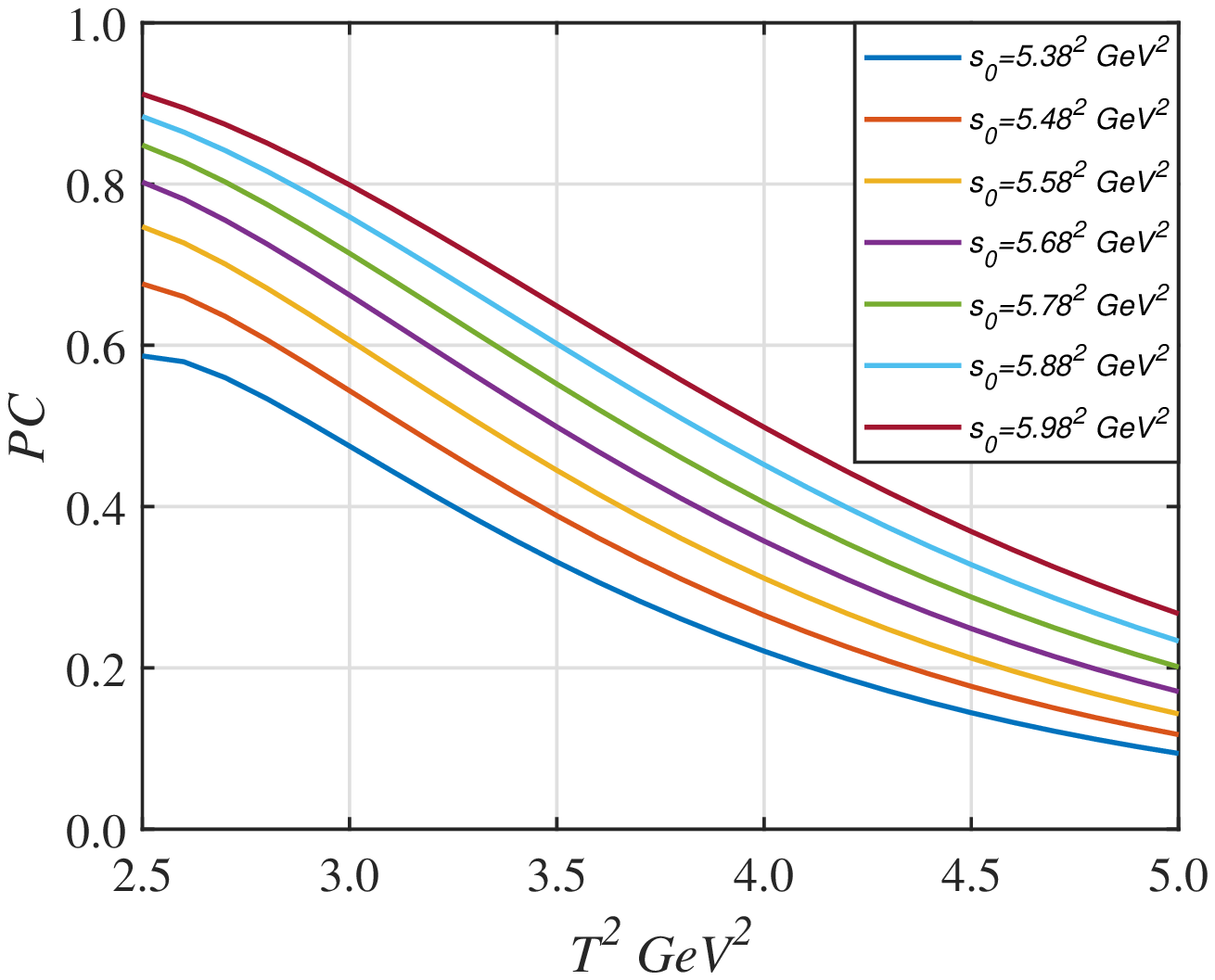}
\caption{The ${\rm PC}-T^2$ curves of the $J_1$ under different values of $\sqrt{s_0}$. \label{your label}}
\end{minipage}
\end{figure}

Since the extracted mass from  the QCD sum rules relies on the energy scale $\mu$ which differs with the threshold $s_0$, one should choose different suitable energy scale $\mu$ for each $s_0$, thus, we show the $M-T^2$ curves rather than $M-s_0$.
We take the current $J_1$ as an example to show the detailed  steps to determine the threshold $s_0$.

In Fig.7 and Fig.8, the numerical results of the $M_1-T^2$ and ${\rm PC}-T^2$ curves under different thresholds $\sqrt{s_0}$ are displayed. For $\sqrt{s_0}=5.78 \rm GeV$, according to the energy scale formula Eq.(14), the Borel window ranges from $3.9-4.3 \rm GeV^2$, the corresponding pole contribution is $43-35\%$, which does not satisfy the pole dominance criterion. In the case of $\sqrt{s_0}=5.88\rm GeV$ or $\sqrt{s_0}=5.98\rm GeV$, the pole contribution has the smaller value compared with that of $\sqrt{s_0}=5.78 \rm GeV$. As for $\sqrt{s_0}=5.58 \rm GeV$, the mass should be $5.01\rm GeV$ according to the energy scale formula Eq.(14),  however,  the minimum value of the $M-T^2$ curve is $5.05\rm GeV$. The difference between the minimum value of the $M-T^2$ curve and mass satisfying Eq.(14) is even larger for either $\sqrt{s_0}=5.48 \rm GeV$ or $\sqrt{s_0}=5.38 \rm GeV$. Thus we have solid reason to set the threshold being $\sqrt{s_0}=5.68 \rm GeV$.

The $M_1-T^2$ graph of the current $J_1$ is shown in Fig.9, where the error bounds come from the uncertainties of the input parameters. A stable Borel platform is obtained with the parameter $T^2$ ranging from $3.4{\rm GeV}^2$ to $3.8{\rm GeV}^2$. The central value of the mass equals to $5.11{\rm GeV}$ which satisfies the energy scale formula very well. As is discussed in Ref.\cite{Wang2}, the contribution of the scattering  states  plays a tiny role. For a color-singlet-color-singlet type hexaquark bound state, naively we expect that its mass should be lower than the sum of the two baryon constituents, here, we find that the mass of the state is about $530$MeV above that of the two $\Lambda_c$ baryons. All these facts show that the $\Lambda_c\Lambda_c$-type resonance with the $J^P=0^+$ is a dibaryon state. The detailed numerical results are shown in Table \uppercase\expandafter{\romannumeral1}.

\begin{table*}[t]
\begin{ruledtabular}\caption{The related numerical results extracted from the Borel windows of the four currents. }
\begin{tabular}{c c c c c c c c}
 & \ $J^P$ &\ $\mu $(GeV)    &\ $T^2$ (GeV$^2$) &\ $\sqrt{s_0}$(GeV) &\ pole &\ $M$(GeV) &\  $\lambda(10^{-4}$GeV$^8$) \\
\hline
$J_1$ & \ $0^+$ & \ 3.52  &  \   $3.4-3.8$  & \ $5.68\pm0.1$  & \  $(41-53)\% $ & \ $5.11^{+0.15}_{-0.12}$  &  \   $17.55_{-2.68}^{+3.16}$ \\
$J_2$ & \ $0^-$ & \ 2.83  &  \   $3.2-3.6$  & \ $5.28\pm0.1$  & \  $(42-55)\%$ & \ $4.66^{+0.10}_{-0.06}$  &  \   $6.60^{+0.17}_{-0.74}$ \\
$J_3$ & \ $1^+$ & \ 3.35  &  \   $3.4-3.8$  & \ $5.58\pm0.1$  & \  $(41-53)\%$ & \ $4.99^{+0.10}_{-0.09}$  &  \   $6.16^{+1.06}_{-0.97}$ \\
$J_4$& \ $1^-$ & \ 2.87  &  \   $3.3-3.7$  & \ $5.38\pm0.1$  & \  $(42-56)\%$ & \ $4.68^{+0.08}_{-0.08}$  &  \   $6.75_{-1.06}^{+1.12}$ \\
\end{tabular}
\end{ruledtabular}
\end{table*}

Especially, in this paper, we consider the contributions of the truncation of the order $\mathcal{O}(\alpha_s^k)$ for $\frac{3}{2}\leq k\leq 3$ in detail. The vacuum condensates of the order  $\mathcal{O}(\alpha_s^{\frac{3}{2}})$ are $\langle\frac{\alpha_s}{\pi}GG\rangle\langle\overline{q}g_s\sigma Gq\rangle$, $\langle g_s^3fGGG\rangle$ and $\langle\overline{q}g_s\sigma Gq\rangle\langle\frac{\alpha_s}{\pi}GG\rangle\langle\overline{q}q\rangle$. For $k=2$ , the related vacuum condensates  are $\langle\frac{\alpha_s}{\pi}GG\rangle^2$, $\langle\frac{\alpha_s}{\pi}GG\rangle^2\langle\overline{q}q\rangle^2$ and $\langle\frac{\alpha_s}{\pi}GG\rangle\langle\overline{q}g_s\sigma Gq\rangle^2$. For $k=\frac{5}{2}$, the corresponding vacuum condensate is $\langle\frac{\alpha_s}{\pi}GG\rangle^2\langle\overline{q}g_s\sigma Gq\rangle\langle\overline{q}q\rangle$. For $k=3$, the vacuum condensates in consideration are $\langle\frac{\alpha_s}{\pi}GG\rangle^3$ and $\langle g_s^3fGGG\rangle^2$. The contributions of the orders $\mathcal{O}(\alpha_s^k)$ are determined by the following equation,
\begin{eqnarray}
D(k)=\frac{\int_{4m_c^2}^{s_0}ds\rho_{QCD;k}(s)\exp\left(-\frac{s}{T^2}\right)}{\int_{4m_c^2}^{s_0}ds\rho_{QCD}(s)\exp\left(-\frac{s}{T^2}\right)}\, ,
\end{eqnarray}
where $k=\frac{3}{2},2,\frac{5}{2},3$ respectively. In Table \uppercase\expandafter{\romannumeral2}, all the condensate terms with $\frac{3}{2}< k\leq 3$ do play tiny roles, thus, it is reasonable for the QCD sum rules to choose the terms with $k\leq\frac{3}{2}$, they play the decisive roles. Furthermore,  we calculate the contribution of the condensates related to the $\langle\overline{\psi}\psi\rangle^2$, similar to the $D(k)$, we define the $D(\langle\overline{\psi}\psi\rangle)$ as,
\begin{eqnarray}
D(\langle\overline{\psi}\psi\rangle)=\frac{\int_{4m_c^2}^{s_0}ds\rho_{QCD;\langle\overline{\psi}\psi\rangle}(s)\exp\left(-\frac{s}{T^2}\right)}
{\int_{4m_c^2}^{s_0}ds\rho_{QCD}(s)\exp\left(-\frac{s}{T^2}\right)}\, ,
\end{eqnarray}
where the $\rho_{QCD;\langle\overline{\psi}\psi\rangle}$ refers to the spectral density contains the vacuum condensate $\langle\overline{\psi}\psi\rangle^2$. Its numerical value at the Borel window is $D(\langle\overline{\psi}\psi\rangle)\approx8.7\times10^{-5}$, it shows that the vacuum condensate $\langle\overline{\psi}\psi\rangle^2$   does play a tiny role.
\begin{figure}[h]
\begin{minipage}[h]{0.45\linewidth}
\centering
\includegraphics[height=5cm,width=7cm]{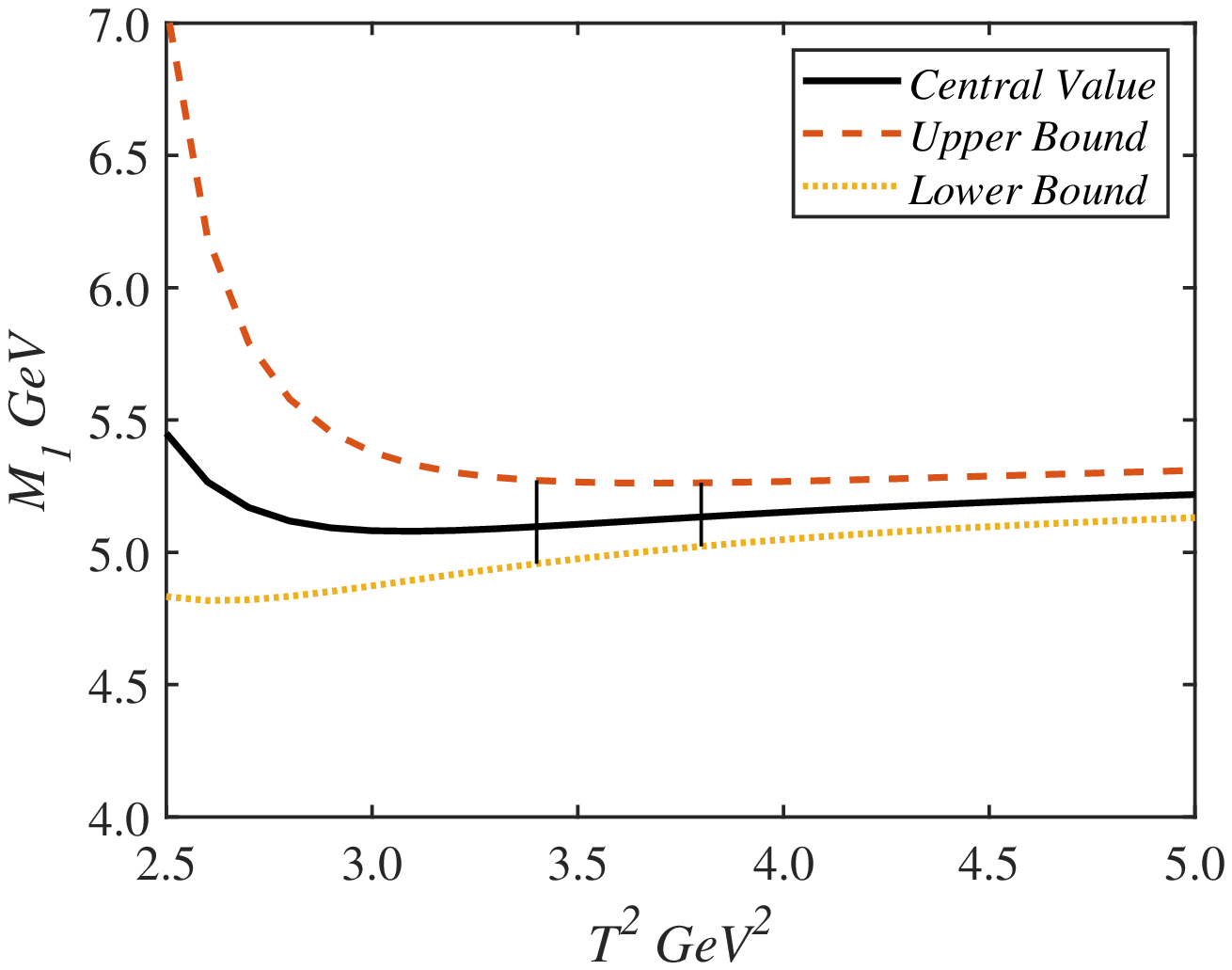}
\caption{The $M_1-T^2$ graph, dashed curve is drawn from the upper value of the input parameters and dotted curve is from the lower value.\label{your label}}
\end{minipage}
\hfill
\begin{minipage}[h]{0.45\linewidth}
\centering
\includegraphics[height=5cm,width=7cm]{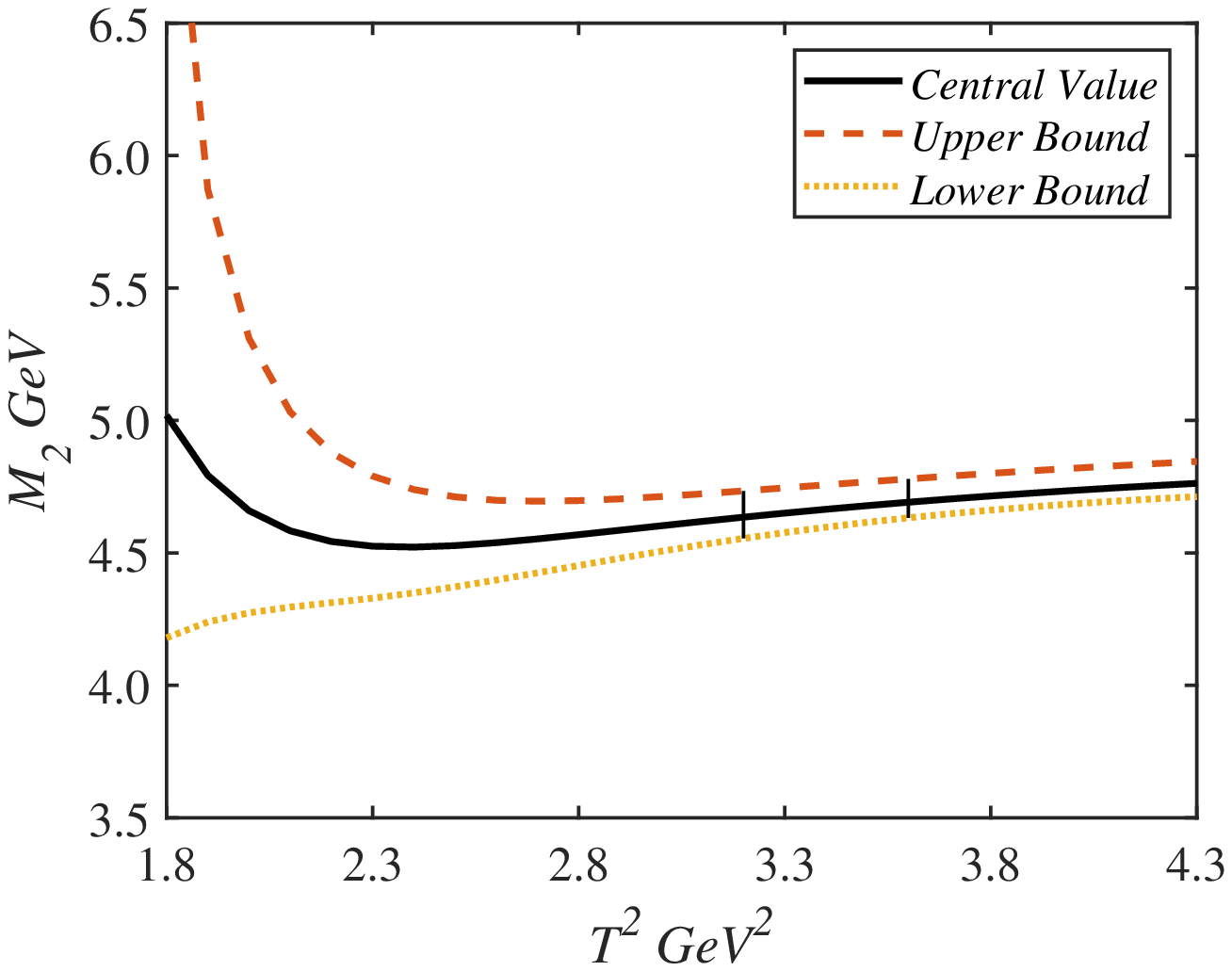}
\caption{The $M_2-T^2$ graph, dashed curve is drawn from the upper value of the input parameters and dotted curve is from the lower value.\label{your label}}
\end{minipage}
\end{figure}

\begin{figure}[h]
\begin{minipage}[h]{0.45\linewidth}
\centering
\includegraphics[height=5cm,width=7cm]{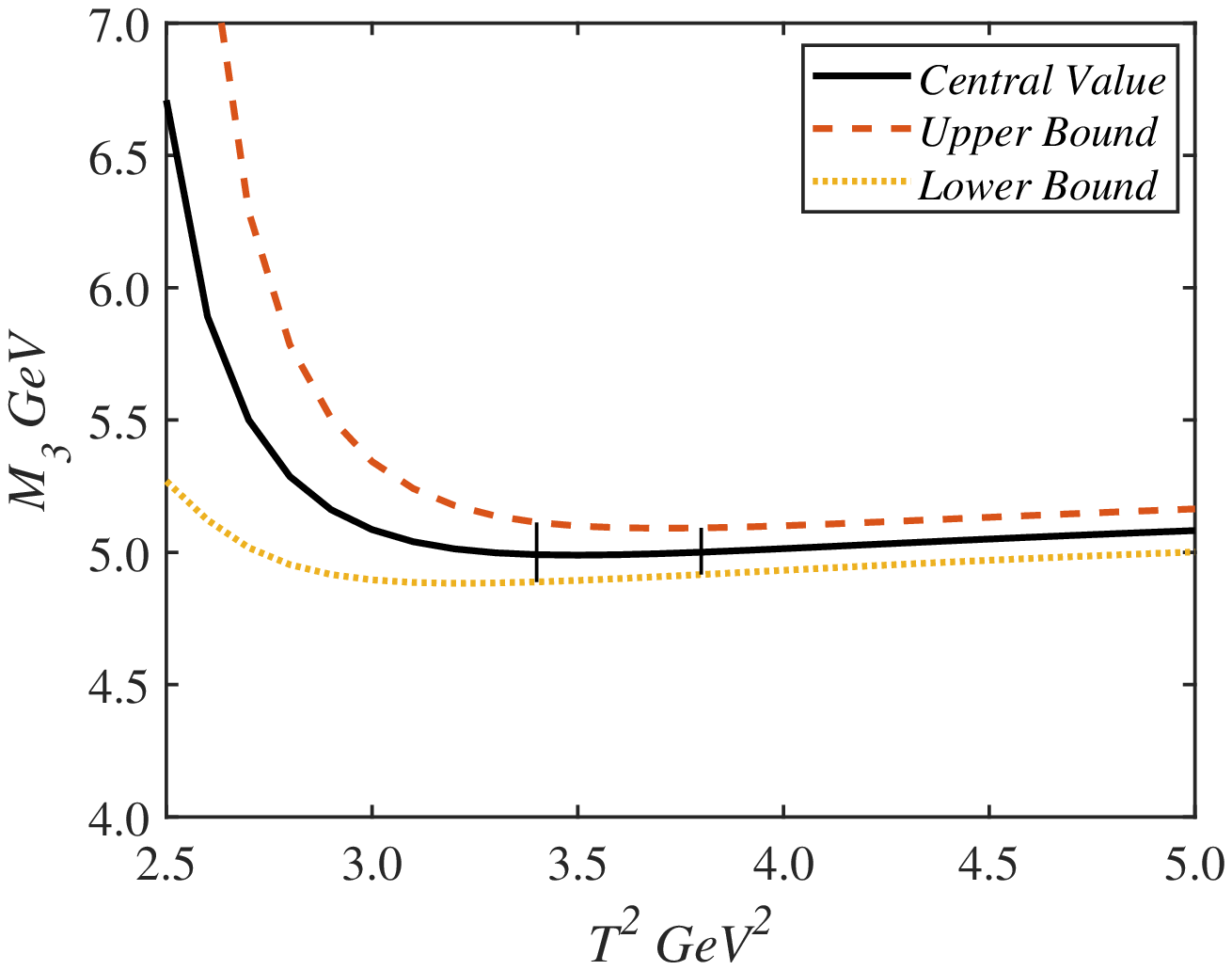}
\caption{The $M_3-T^2$ graph, dashed curve is drawn from the upper value of the input parameters and dotted curve is from the lower value.\label{your label}}
\end{minipage}
\hfill
\begin{minipage}[h]{0.45\linewidth}
\centering
\includegraphics[height=5cm,width=7cm]{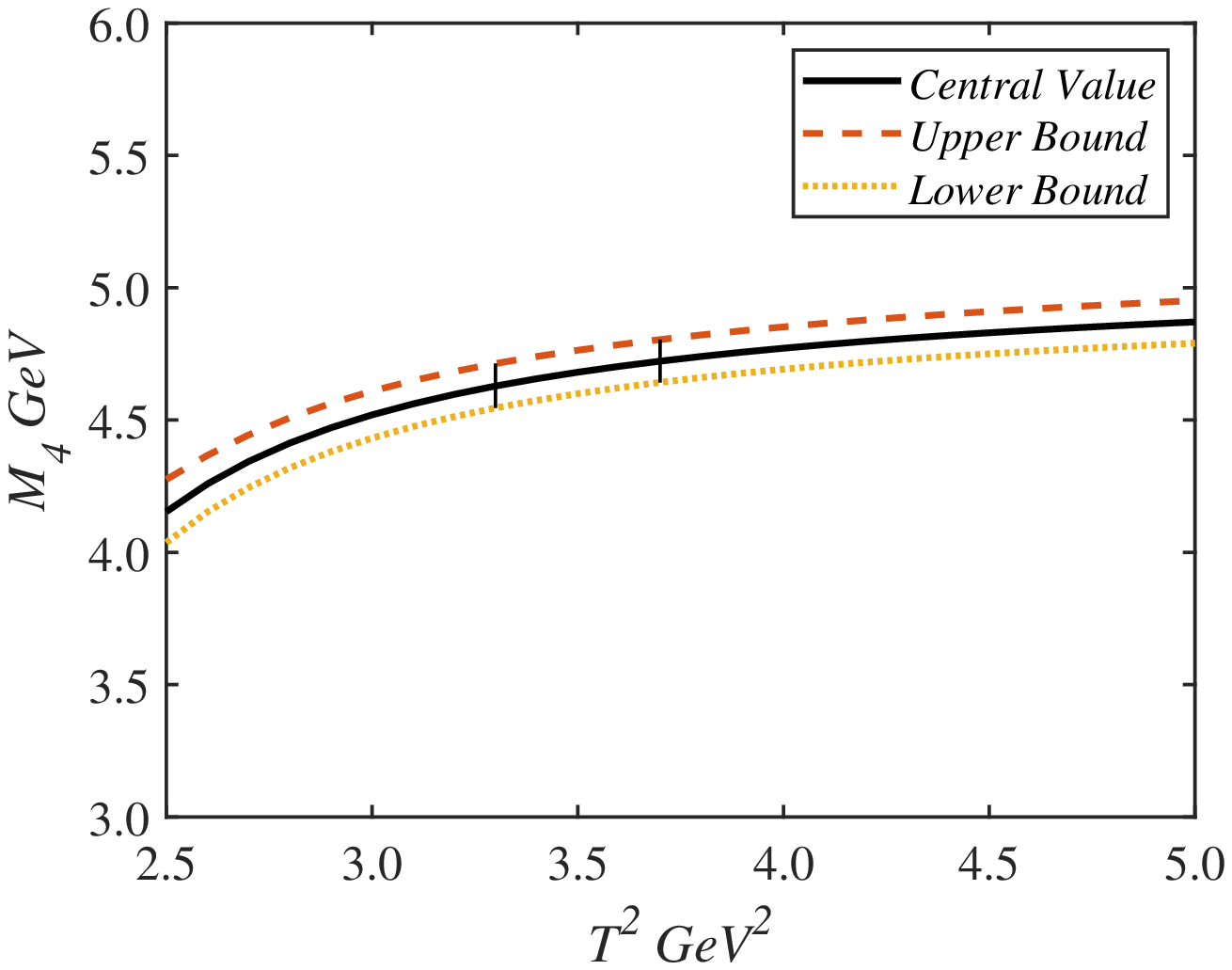}
\caption{The $M_4-T^2$ graph, dashed curve is drawn from the upper value of the input parameters and dotted curve is from the lower value.\label{your label}}
\end{minipage}
\end{figure}

\begin{figure}[h]
\begin{minipage}[h]{0.45\linewidth}
\centering
\includegraphics[height=5cm,width=7cm]{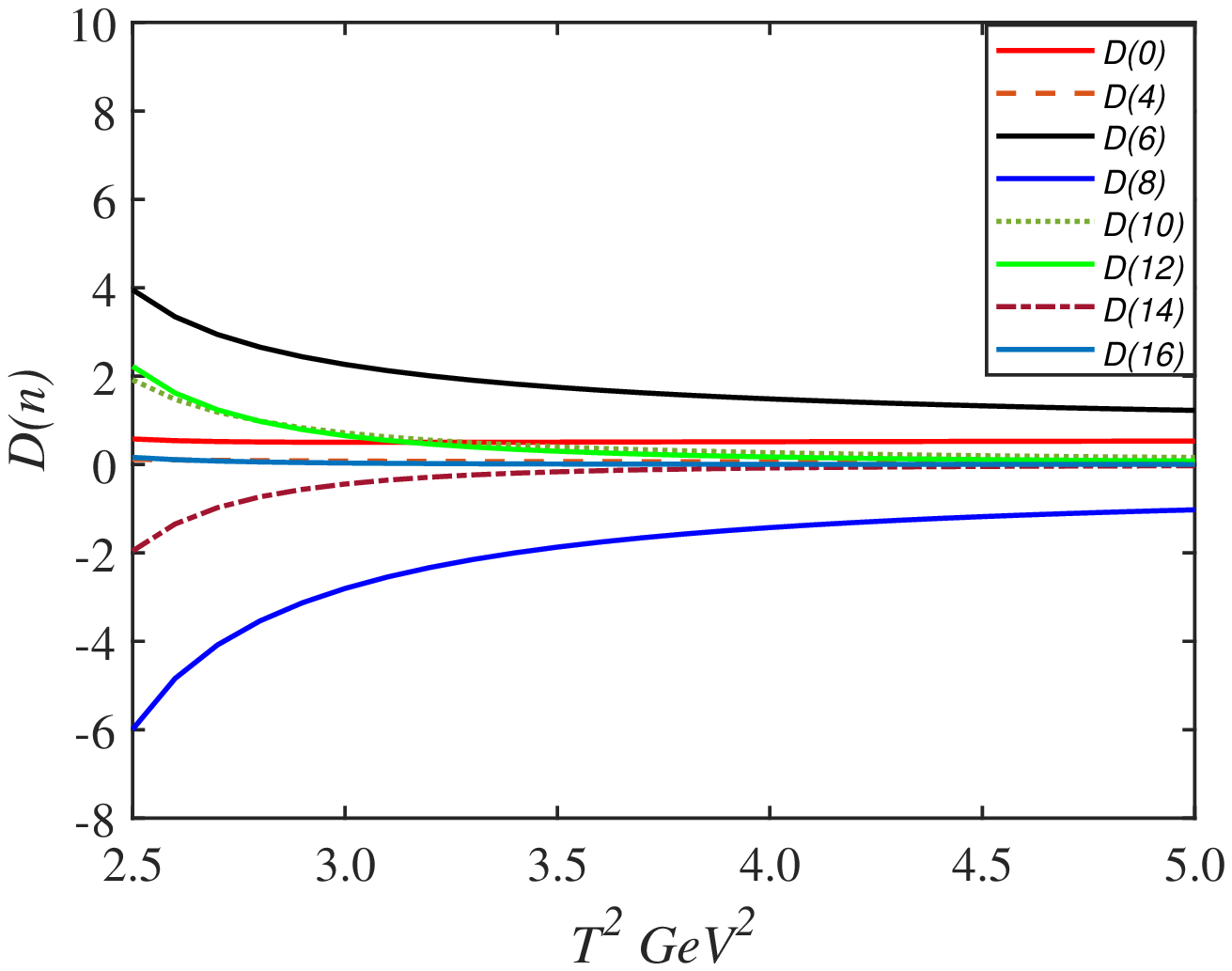}
\caption{The $D(n)-T^2$ graph  for the $J_1$.\label{your label}}
\end{minipage}
\hfill
\begin{minipage}[h]{0.45\linewidth}
\centering
\includegraphics[height=5cm,width=7cm]{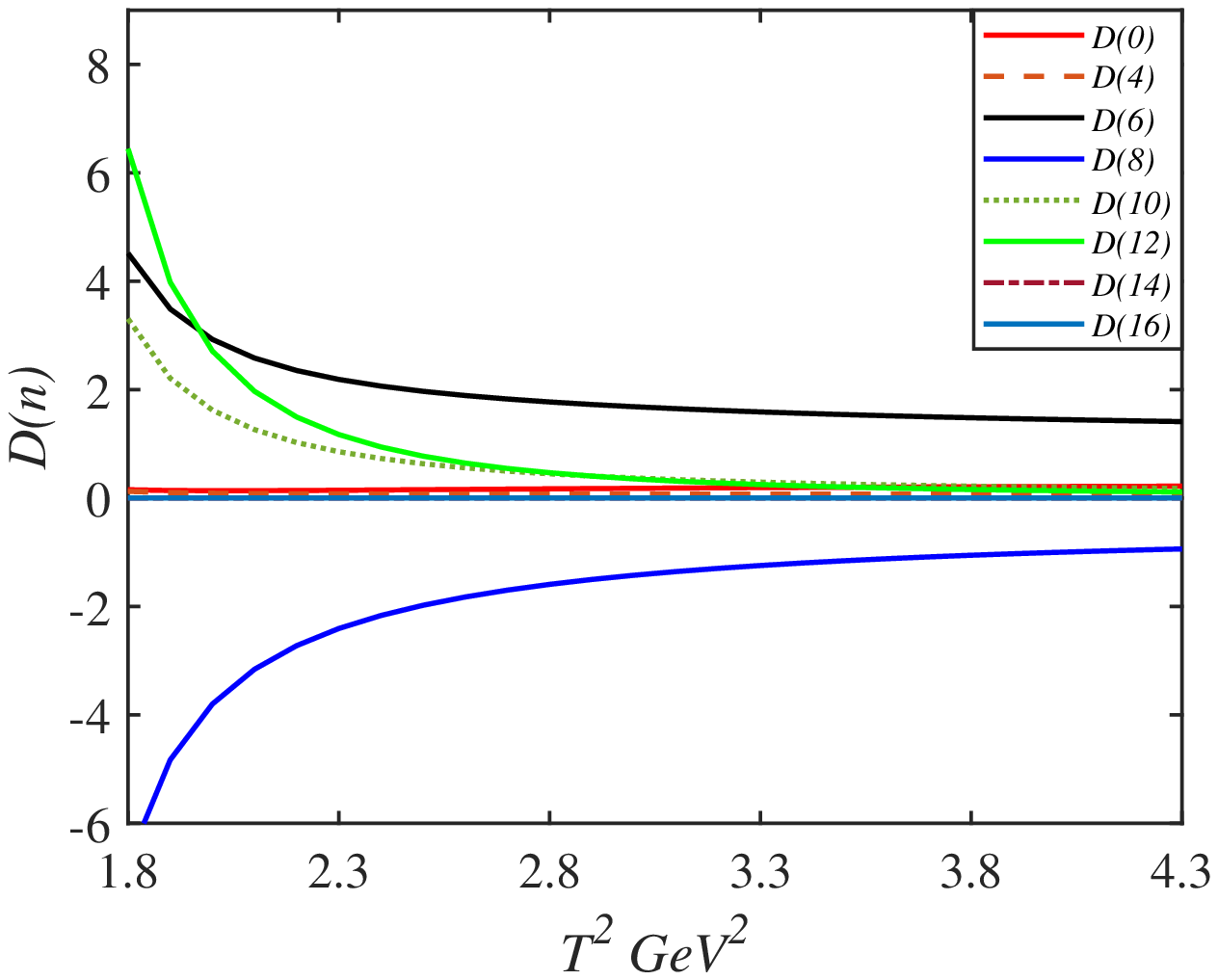}
\caption{The $D(n)-T^2$ graph  for the $J_2$.\label{your label}}
\end{minipage}
\end{figure}

\begin{figure}[h]
\begin{minipage}[h]{0.45\linewidth}
\centering
\includegraphics[height=5cm,width=7cm]{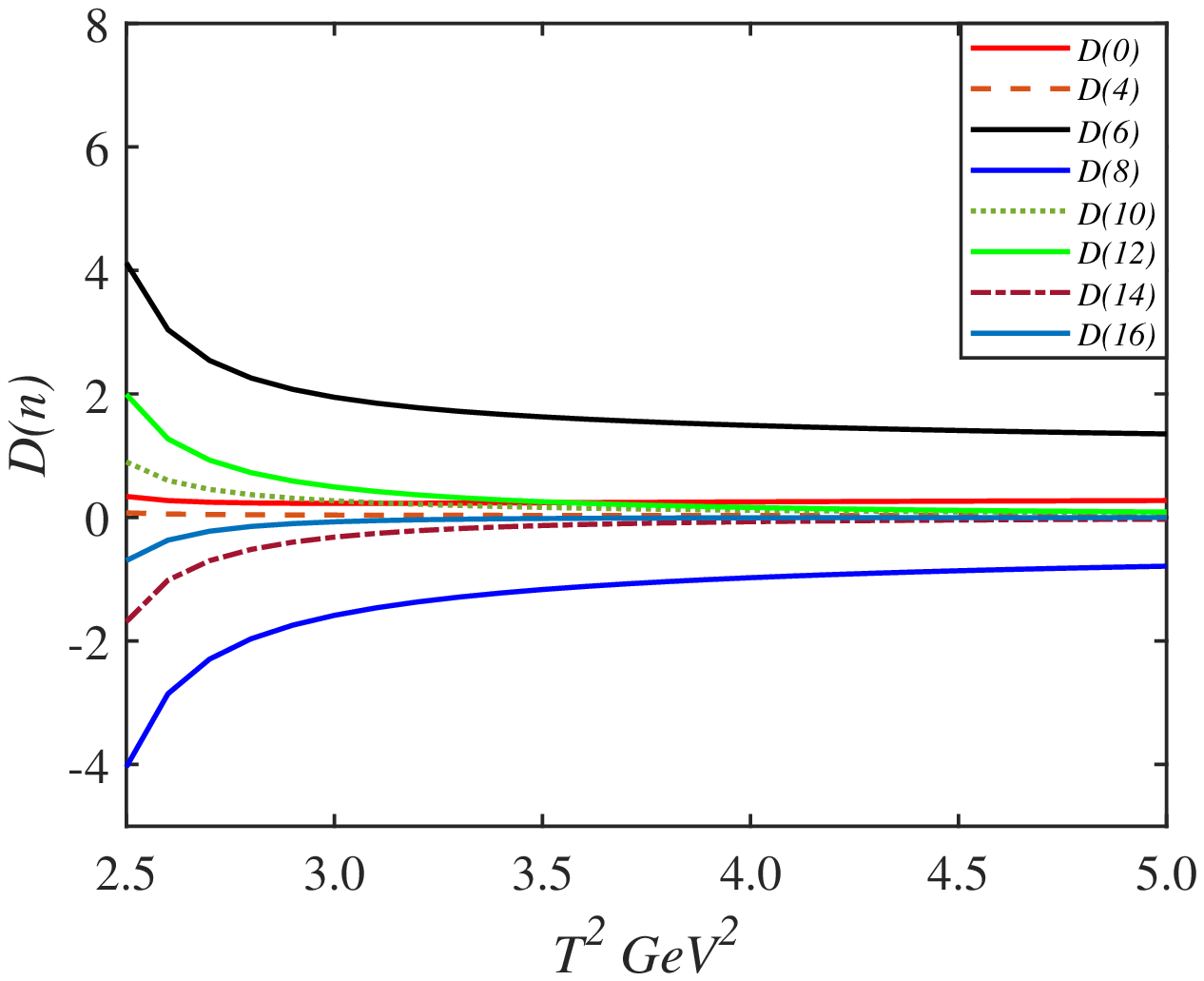}
\caption{The $D(n)-T^2$ graph  for the $J_3$.\label{your label}}
\end{minipage}
\hfill
\begin{minipage}[h]{0.45\linewidth}
\centering
\includegraphics[height=5cm,width=7cm]{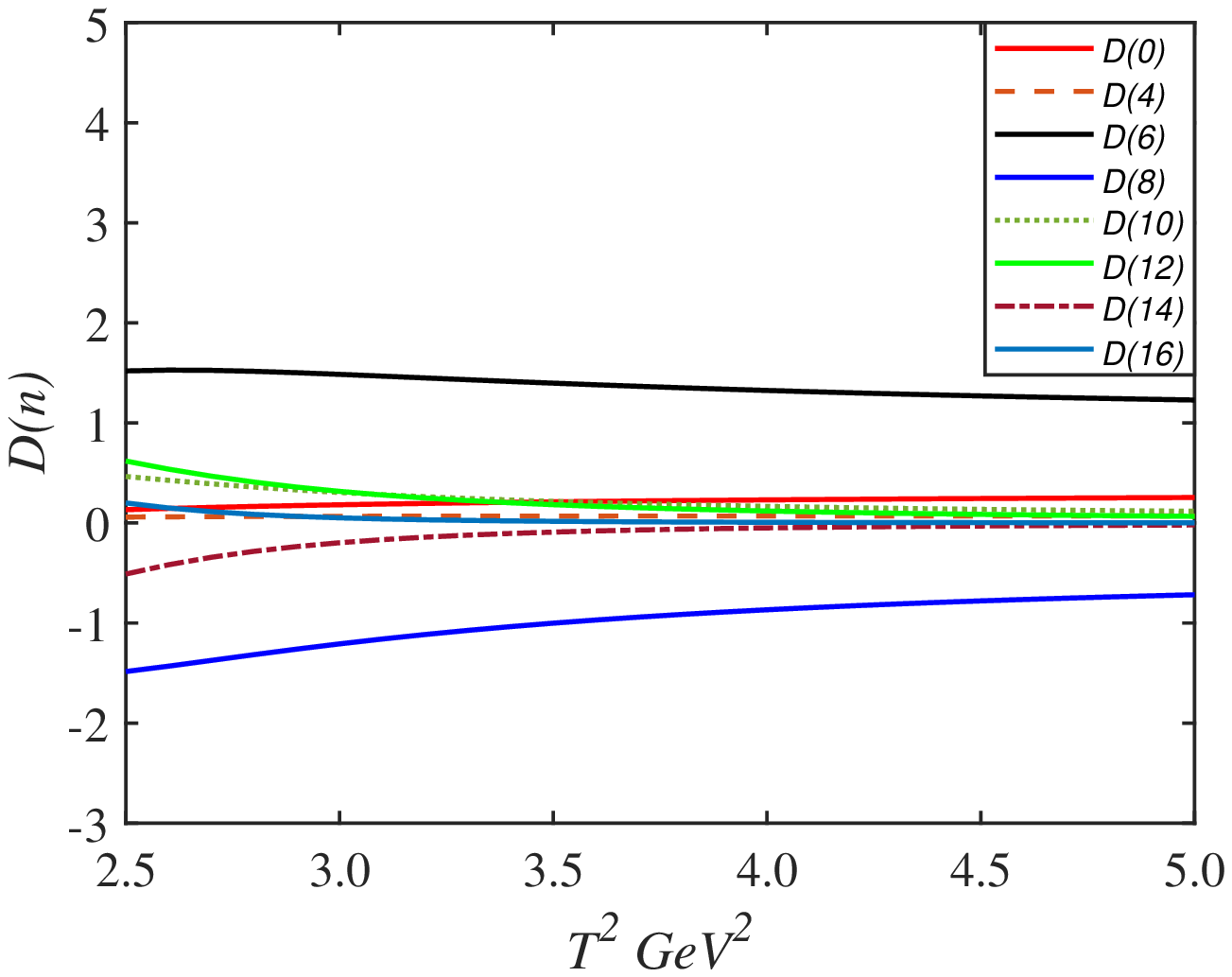}
\caption{The $D(n)-T^2$ graph  for the $J_4$.\label{your label}}
\end{minipage}
\end{figure}

\begin{table*}[t]
\begin{ruledtabular}\caption{Contributions of $D(k)$ for $\frac{3}{2}\leq k\leq3$ in the Borel window of the $J_1$.}
\begin{tabular}{c c c c c c c}
   &\ $k=\frac{3}{2}$    &\ $k=2$  &\ $k=\frac{5}{2}$ &\  $k=3$ \\
\hline
$D(k)$ & \ $-3.07\times10^{-2}$  &  \   $1.04\times10^{-2}$ & \ $-3.06\times10^{-5}$  &  \   $-1.85\times10^{-5}$ \\
\end{tabular}
\end{ruledtabular}
\end{table*}

The  numerical results of the masses for the currents  $J_2,J_3$ and $J_4$ are shown in Fig.8$\sim$Fig.10 respectively. We also find Borel platforms for the $J_2,J_3$ and $J_4$. For the pseudoscalar current $J_2$, the central value of the mass in the platform is $4.66{\rm GeV}$, just about $80$MeV above the mass of the two baryon constituents. For the axialvector current $J_3$, the mass extracted from the Borel window is $4.99{\rm GeV}$. As for the vector current $J_4$, the extracted  mass is $4.68{\rm GeV}$, which coincides very well with the center-of mass energy of the electron-positron annihilation with  high  integrated luminosity \cite{Ablikim}.
  We are interested to work out whether or not there is a $Y$ state at around $4.68{\rm GeV}$, thus it may be possible for us to find the relation between the $Y$ state and $\Lambda_c\overline{\Lambda}_c$ baryonium state with the $J^P=1^-$. The  contributions $D(n)$ in the operator product expansion  of each currents are displayed in  Fig.13$\sim$Fig.16, as can be seen clearly on these graphs, for higher dimensions ($n\geq10$) in the regions of the Borel windows, they play the less important roles compared with the condensates with the low ones. The convergency of the operator product expansion holds well for the QCD sum rules. It is worth mentioning that it does not mean the consideration of the higher  dimensions are meaningless. They maybe  play important roles in stabilizing the Borel windows from which to extract the masses and pole residues and so on. As is shown in Fig.13$\sim$Fig.16, the vacuum condensates of dimension 6 (mainly $\langle\overline{q}q\rangle^2$) and dimension 8
  ($\langle\overline{q}g_s\sigma Gq\rangle\langle\overline{q}q\rangle$) play the most important roles for these four states.

\begin{large}
\noindent\textbf{4 Conclusion}
\end{large}

In this paper, we construct four currents $J_{1,2,3,4}$ with the $J^P=0^+,0^-,1^+,1^-$ respectively to study the $\Lambda_c\Lambda_c$ dibaryon and $\Lambda_c\overline{\Lambda}_c$ baryonium states. We acquire the Borel windows for these currents and extract the corresponding masses  $5.11{\rm GeV},4.66{\rm GeV},4.99{\rm GeV}$ and $4.68{\rm GeV}$ respectively. They are all above the thresholds of the two baryon constituents,  which is qualitatively in agreement with the results obtained in Ref.\cite{Wan}. In contrary to the conclusion obtained  in Ref.\cite{Wan}, we find the currents with the quantum numbers $J^P=0^-$ and $1^+$ also exist Borel platforms. The vacuum condensates are calculated  up to dimension 16, especially, for the current $J_1$, we consider all the vacuum condensates up to dimension 16 and choose the truncation of the operator product expansion of the order $\mathcal{O}(\alpha_s^k)$ with $k\leq3$. The numerical results show that those vacuum condensates with $\frac{3}{2}< k\leq3$ do play tiny roles and it is reasonable to discard these  terms in the QCD sum rules. For the vector current $J_4$, the predicted mass  $4.68{\rm GeV}$ coincides very well
with the center-of mass energy of the electron-positron annihilation with high integrated luminosity \cite{Ablikim}, we can confronted the prediction to the experimental data in the BESIII in the future.


\clearpage
\begin{large}
\noindent\textbf{Appendix}
\end{large}

The right side of Eq.(10) can be  expressed in the form \textbf{S},
\begin{eqnarray}
\notag \textbf{S}=\sum\limits_{A,B,C,D,n,i}\textbf{S}_{A,B,C,D}(n)_i
\end{eqnarray}
where $A$, $B$, $C$ and $D$ refer to four types of integrals, $n$ are the dimension of the vacuum condensates. For the following expressions, $\tilde{m}^2=\frac{m_c^2}{(1-y)y}$ and $\bar{m}^2=\frac{(y+z)m_c^2}{y z}$, $y$ and $z$ are parameters introduced during the calculations of the integrals in the momentum space of the two heavy quarks, For type $A$ and $B$, $y_i=\frac{1}{2}(1-\sqrt{1-4m_c^2/s})$, $y_f=\frac{1}{2}(1+\sqrt{1-4m_c^2/s})$ and $z_i=\frac{y m_c^2}{y s-m_c^2}$. For type $C$ and $D$, $y_i=0$, $y_f=1$ and $z_i=0$.

\noindent========================================================
Type A for $J_1$
\\

\begin{eqnarray}
\notag \textbf{S}_A(12)_1&&=\langle \bar{q}q\rangle^4\int_{4m_c^2}^{s_0}ds\int_{y_i}^{y_f}dye^{-\frac{s}{T^2}}\{y(y-1)\}\{\frac{-7(s-\tilde{m}^2)}{216\pi^2 }-\frac{7s}{432\pi^2}\}  \\
\notag && +m_c^2\langle \bar{q}q\rangle^4\int_{4m_c^2}^{s_0}ds\int_{y_i}^{y_f}dye^{-\frac{s}{T^2}}\{\frac{7}{432\pi^2} \} \\
\notag && +m_c^2\langle \bar{q}q\rangle^4\int_{4m_c^2}^{s_0}ds\int_{y_i}^{y_f}dye^{-\frac{s}{T^2}}\{\frac{7}{839808\pi^6 }\} \\
\notag && +\langle \bar{q}q\rangle^4\int_{4m_c^2}^{s_0}ds\int_{y_i}^{y_f}dye^{-\frac{s}{T^2}}\{(y-1)y\}\{\frac{7(\tilde{m}^2-s)}{419904\pi^6 }-\frac{7s}{839808\pi^6}\}  \\
\notag && +m_c^2\langle \bar{q}q\rangle^4\int_{4m_c^2}^{s_0}ds\int_{y_i}^{y_f}dye^{-\frac{s}{T^2}}\{\frac{5}{11664\pi^4}\}  \\
\notag && +\langle \bar{q}q\rangle^4\int_{4m_c^2}^{s_0}ds\int_{y_i}^{y_f}dye^{-\frac{s}{T^2}}\{(y-1)y\}\{\frac{5(\tilde{m}^2-s)}{5832\pi^4 }-\frac{5s}{11664\pi^4}\} \, , \\
\notag
\end{eqnarray}

\begin{eqnarray}
\notag \textbf{S}_A(12)_2&& =\langle\bar{q}q\rangle \langle g_s^2GG\rangle m_c^2 \langle\bar{q}g_s\sigma Gq\rangle \int_{4m_c^2}^{s_0}ds\int_{y_i}^{y_f}dye^{-\frac{s}{T^2}} \{-\frac{5}{110592 \pi^6}\}  \\
\notag && +\langle\bar{q}q\rangle \langle g_s^2GG\rangle \langle\bar{q}g_s\sigma Gq\rangle \int_{4m_c^2}^{s_0}ds\int_{y_i}^{y_f}dye^{-\frac{s}{T^2}} \{(y-1) y\} \{\frac{5 s}{110592 \pi^6}-\frac{5 (\tilde{m}^2-s)}{55296 \pi^6}\}  \\
\notag && +\langle g_s^2GG\rangle m_c^2 \langle\bar{q}q\rangle \langle\bar{q}g_s\sigma Gq\rangle \int_{4m_c^2}^{s_0}ds\int_{y_i}^{y_f}dye^{-\frac{s}{T^2}} \{-\frac{19}{147456 \pi^6}\}  \\
\notag && +\langle g_s^2GG\rangle \langle\bar{q}q\rangle \langle\bar{q}g_s\sigma Gq\rangle \int_{4m_c^2}^{s_0}ds\int_{y_i}^{y_f}dye^{-\frac{s}{T^2}} \{(y-1) y\} \{\frac{19 s}{49152 \pi^6}-\frac{19 \tilde{m}^2}{73728 \pi^6}\}  \, ,\\
\notag
\end{eqnarray}

\begin{eqnarray}
\notag \textbf{S}_A(14)&&=\langle\bar{q}g_s\sigma Gq\rangle \langle\bar{q}q\rangle^3 \int_{4m_c^2}^{s_0}ds\int_{y_i}^{y_f}dye^{-\frac{s}{T^2}} \{(y-1) y\}\{ \frac{7}{72 \pi^2} \}\\
\notag && +\langle g_s^2GG\rangle \langle\bar{q}g_s\sigma Gq\rangle^2 \int_{4m_c^2}^{s_0}ds\int_{y_i}^{y_f}dye^{-\frac{s}{T^2}} \{(y-1) y\} \{-\frac{19}{196608 \pi^6}\}   \\
\notag && +\langle\bar{q}q\rangle^2 \langle g_s^2GG\rangle^2 \int_{4m_c^2}^{s_0}ds\int_{y_i}^{y_f}dye^{-\frac{s}{T^2}} \{(y-1) y\} \{-\frac{5}{1327104 \pi^6}\}  \\
\notag && +\langle\bar{q}q\rangle^3 \langle\bar{q}g_s\sigma Gq\rangle \int_{4m_c^2}^{s_0}ds\int_{y_i}^{y_f}dye^{-\frac{s}{T^2}} \{(y-1) y\} \{\frac{5}{3888 \pi^4}\}  \, .\\
\notag
\end{eqnarray}

Type B for $J_1$

\begin{eqnarray}
\notag \textbf{S}_B(0)=&& \int_{4m_c^2}^{s_0}ds\int_{y_i}^{y_f}dy\int_{z_i}^{1-y}dze^{-\frac{s}{T^2}} \{-y z (y+z-1)^5\} \{\frac{(s-\bar{m}^2)^7}{58982400 \pi^{10}}+\frac{7 s (s-\bar{m}^2)^6}{117964800 \pi^{10}}\}  \\
\notag &&+m_c^2 \int_{4m_c^2}^{s_0}ds\int_{y_i}^{y_f}dy\int_{z_i}^{1-y}dze^{-\frac{s}{T^2}} \{-(y+z-1)^5\} \{\frac{7 (s-\bar{m}^2)^6}{117964800 \pi^{10}}\} \, , \\
\notag
\end{eqnarray}

\begin{eqnarray}
\notag \textbf{S}_B(4)&&=2 m_c^2 \langle g_s^2GG\rangle \int_{4m_c^2}^{s_0}ds\int_{y_i}^{y_f}dy\int_{z_i}^{1-y}dze^{-\frac{s}{T^2}} \{\frac{(y+z-1)^5}{y^2}\} \{\frac{7 y (s-\bar{m}^2)^4}{141557760 \pi^{10}}+\frac{7 s y (s-\bar{m}^2)^3}{70778880 \pi^{10}}\}\\
\notag &&+2 m_c^2 \langle g_s^2GG\rangle \int_{4m_c^2}^{s_0}ds\int_{y_i}^{y_f}dy\int_{z_i}^{1-y}dze^{-\frac{s}{T^2}} \{\frac{(y+z-1)^5}{y^2}\}\{-\frac{7 (s-\bar{m}^2)^4}{141557760 \pi^{10}}\}  \\
\notag
\notag && +2 m_c^2 \langle g_s^2GG\rangle \int_{4m_c^2}^{s_0}ds\int_{y_i}^{y_f}dy\int_{z_i}^{1-y}dze^{-\frac{s}{T^2}}\{\frac{-z (y+z-1)^5}{y^2}\}\{-\frac{7 (s-\bar{m}^2)^4}{141557760 \pi^{10}}-\frac{7 s (s-\bar{m}^2)^3}{70778880 \pi^{10}}\}  \\
\notag
\notag && +m_c^2\langle g_s^2GG\rangle  \int_{4m_c^2}^{s_0}ds\int_{y_i}^{y_f}dy\int_{z_i}^{1-y}dze^{-\frac{s}{T^2}}\{-(y+z-1)^3\} \{\frac{19 (s-\bar{m}^2)^4}{18874368 \pi^{10}}\}  \\
\notag
\notag &&  +\langle g_s^2GG\rangle \int_{4m_c^2}^{s_0}ds\int_{y_i}^{y_f}dy\int_{z_i}^{1-y}dze^{-\frac{s}{T^2}}\{-y z (y+z-1)^3\} \{\frac{19 (s-\bar{m}^2)^5}{47185920 \pi^{10}}+\frac{19 s (s-\bar{m}^2)^4}{18874368 \pi^{10}}\}  \\
\notag
\notag &&+m_c^2 \langle g_s^2GG\rangle \int_{4m_c^2}^{s_0}ds\int_{y_i}^{y_f}dy\int_{z_i}^{1-y}dze^{-\frac{s}{T^2}}\{\frac{-(y+z-1)^5}{y z}\} \{-\frac{7 (s-\bar{m}^2)^4}{188743680 \pi^{10}}\}   \\
\notag
\notag &&  +\langle g_s^2GG\rangle \int_{4m_c^2}^{s_0}ds\int_{y_i}^{y_f}dy\int_{z_i}^{1-y}dze^{-\frac{s}{T^2}}\{(y+z-1)^5\} \{\frac{7 (s-\bar{m}^2)^5}{943718400 \pi^{10}}+\frac{7 s (s-\bar{m}^2)^4}{377487360 \pi^{10}}\} \, , \\
\notag
\end{eqnarray}

\begin{eqnarray}
\notag \textbf{S}_B(6)_1&& =\langle\bar{q}q\rangle^2 \int_{4m_c^2}^{s_0}ds\int_{y_i}^{y_f}dy\int_{z_i}^{1-y}dze^{-\frac{s}{T^2}}\{y z (y+z-1)^2\} \{\frac{5 (s-\bar{m}^2)^4}{9216 \pi^6}+\frac{5 s (s-\bar{m}^2)^3}{4608 \pi^6}\}  \\
\notag
\notag && +m_c^2 \langle\bar{q}q\rangle^2 \int_{4m_c^2}^{s_0}ds\int_{y_i}^{y_f}dy\int_{z_i}^{1-y}dze^{-\frac{s}{T^2}} \{(y+z-1)^2\} \{\frac{5 (s-\bar{m}^2)^3}{4608 \pi^6}\}  \\
\notag
\notag && +m_c^2 2 \langle\bar{q}q\rangle^2 \int_{4m_c^2}^{s_0}ds\int_{y_i}^{y_f}dy\int_{z_i}^{1-y}dze^{-\frac{s}{T^2}}\{\frac{(y+z-1)^4}{y}\}\{-\frac{35 (s-\bar{m}^2)^3}{7962624 \pi^8}-\frac{35 s (s-\bar{m}^2)^2}{5308416 \pi^8}\}    \\
\notag
\notag && +2 m_c^2\langle\bar{q}q\rangle^2  \int_{4m_c^2}^{s_0}ds\int_{y_i}^{y_f}dy\int_{z_i}^{1-y}dze^{-\frac{s}{T^2}}\{\frac{z (y+z-1)^4}{y^2}\}\{-\frac{35 (s-\bar{m}^2)^3}{7962624 \pi^8}-\frac{35 s (s-\bar{m}^2)^2}{5308416 \pi^8 }\}    \\
\notag
\notag && +2 \langle\bar{q}q\rangle^2 \int_{4m_c^2}^{s_0}ds\int_{y_i}^{y_f}dy\int_{z_i}^{1-y}dze^{-\frac{s}{T^2}}\{\frac{z (y+z-1)^4}{y}\} \{\frac{35 s^2 y (s-\bar{m}^2)^2}{7962624 \pi^8}+\frac{35 y (s-\bar{m}^2)^4}{15925248 \pi^8 }\}\\
\notag && +2 \langle\bar{q}q\rangle^2 \int_{4m_c^2}^{s_0}ds\int_{y_i}^{y_f}dy\int_{z_i}^{1-y}dze^{-\frac{s}{T^2}}\{\frac{z (y+z-1)^4}{y}\}\{\frac{35 s y (s-\bar{m}^2)^3}{3981312 \pi^8}-\frac{35 (s-\bar{m}^2)^4}{23887872 \pi^8}\}\\
\notag && +2 \langle\bar{q}q\rangle^2 \int_{4m_c^2}^{s_0}ds\int_{y_i}^{y_f}dy\int_{z_i}^{1-y}dze^{-\frac{s}{T^2}}\{\frac{z (y+z-1)^4}{y}\}\{-\frac{35 s (s-\bar{m}^2)^3}{11943936 \pi^8}\}    \\
\notag
\notag && +2 \langle\bar{q}q\rangle^2 \int_{4m_c^2}^{s_0}ds\int_{y_i}^{y_f}dy\int_{z_i}^{1-y}dze^{-\frac{s}{T^2}}\{\frac{z (y+z-1)^4}{y}\} \{-\frac{35 s^2 y (s-\bar{m}^2)^2}{7962624 \pi^8}-\frac{35 y (s-\bar{m}^2)^4}{15925248 \pi^8}\}\\
\notag && +2 \langle\bar{q}q\rangle^2 \int_{4m_c^2}^{s_0}ds\int_{y_i}^{y_f}dy\int_{z_i}^{1-y}dze^{-\frac{s}{T^2}}\{\frac{z (y+z-1)^4}{y}\}\{-\frac{35 s y (s-\bar{m}^2)^3}{3981312 \pi^8}+\frac{35 (s-\bar{m}^2)^4}{95551488 \pi^8}\}\\
\notag && +2 \langle\bar{q}q\rangle^2 \int_{4m_c^2}^{s_0}ds\int_{y_i}^{y_f}dy\int_{z_i}^{1-y}dze^{-\frac{s}{T^2}}\{\frac{z (y+z-1)^4}{y}\}\{\frac{35 s (s-\bar{m}^2)^3}{47775744 \pi^8}\}  \\
\notag
\notag &&  +m_c^2 \langle\bar{q}q\rangle^2 \int_{4m_c^2}^{s_0}ds\int_{y_i}^{y_f}dy\int_{z_i}^{1-y}dze^{-\frac{s}{T^2}}\{(y+z-1)^2\} \{\frac{7 (s-\bar{m}^2)^3}{248832 \pi^8}\}    \\
\notag
\notag &&  +\langle\bar{q}q\rangle^2 \int_{4m_c^2}^{s_0}ds\int_{y_i}^{y_f}dy\int_{z_i}^{1-y}dze^{-\frac{s}{T^2}}\{y z (y+z-1)^2\} \{\frac{7 (s-\bar{m}^2)^4}{497664 \pi^8}+\frac{7 s (s-\bar{m}^2)^3}{248832 \pi^8}\}   \, , \\
\notag
\end{eqnarray}

\begin{eqnarray}
\notag \textbf{S}_B(6)_2&& =m_c^4\langle \bar{\psi}\psi\rangle^2  \int_{4m_c^2}^{s_0}ds\int_{y_i}^{y_f}dy\int_{z_i}^{1-y}dze^{-\frac{s}{T^2}}\{-\frac{(y+z-1)^5}{y^2 z^2}\} \{-\frac{7 (s-\bar{m}^2)^2}{159252480 \pi^{10}}\}    \\
\notag
\notag && +m_c^2 \langle \bar{\psi}\psi\rangle^2 \int_{4m_c^2}^{s_0}ds\int_{y_i}^{y_f}dy\int_{z_i}^{1-y}dze^{-\frac{s}{T^2}}\{\frac{(y+z-1)^5}{y z^2}\} \{-\frac{7 (y-1) (s-\bar{m}^2)^3}{238878720 \pi^{10}}-\frac{7 s y (s-\bar{m}^2)^2}{159252480 \pi^{10}}\}    \\
\notag
\notag && +\langle \bar{\psi}\psi\rangle^2 \int_{4m_c^2}^{s_0}ds\int_{y_i}^{y_f}dy\int_{z_i}^{1-y}dze^{-\frac{s}{T^2}}\{\frac{(y+z-1)^6}{(y-1) y z}\} \{\frac{7 s y (s-\bar{m}^2)^3}{716636160 \pi^{10}}-\frac{7 (1-y) (s-\bar{m}^2)^4}{1433272320 \pi^{10}}\}    \\
\notag
\notag && +\langle \bar{\psi}\psi\rangle^2 \int_{4m_c^2}^{s_0}ds\int_{y_i}^{y_f}dy\int_{z_i}^{1-y}dze^{-\frac{s}{T^2}}\{\frac{(y+z-1)^5}{(y-1) y}\} \{-\frac{7 s (s-\bar{m}^2)^3}{716636160 \pi^{10}}\}    \\
\notag
\notag && +\langle \bar{\psi}\psi\rangle^2 \int_{4m_c^2}^{s_0}ds\int_{y_i}^{y_f}dy\int_{z_i}^{1-y}dze^{-\frac{s}{T^2}}\{(y+z-1)^5\} \{-\frac{7 s^2 (s-\bar{m}^2)^2}{477757440 \pi^{10}}-\frac{7 (s-\bar{m}^2)^4}{955514880 \pi^{10}}\}\\
\notag && +\langle \bar{\psi}\psi\rangle^2 \int_{4m_c^2}^{s_0}ds\int_{y_i}^{y_f}dy\int_{z_i}^{1-y}dze^{-\frac{s}{T^2}}\{(y+z-1)^5\} \{-\frac{7 s (s-\bar{m}^2)^3}{238878720 \pi^{10}}\}    \\
\notag
\notag && +m_c^2\langle \bar{\psi}\psi\rangle^2  \int_{4m_c^2}^{s_0}ds\int_{y_i}^{y_f}dy\int_{z_i}^{1-y}dze^{-\frac{s}{T^2}}\{-\frac{(y+z-1)^5}{y z}\} \{-\frac{7 (s-\bar{m}^2)^3}{59719680 \pi^{10}}-\frac{7 s (s-\bar{m}^2)^2}{39813120 \pi^{10}}\}    \\
\notag
\notag && +\langle \bar{\psi}\psi\rangle^2 \int_{4m_c^2}^{s_0}ds\int_{y_i}^{y_f}dy\int_{z_i}^{1-y}dze^{-\frac{s}{T^2}}\{\frac{(y+z-1)^6}{(y-1) y z}\}\{\frac{49 (1-y) (s-\bar{m}^2)^4}{5733089280 \pi^{10}}-\frac{49 s y (s-\bar{m}^2)^3}{2866544640 \pi^{10}}\}    \\
\notag
\notag && +\langle \bar{\psi}\psi\rangle^2 \int_{4m_c^2}^{s_0}ds\int_{y_i}^{y_f}dy\int_{z_i}^{1-y}dze^{-\frac{s}{T^2}}\{\frac{(y+z-1)^5}{(y-1) y}\} \{\frac{49 s (s-\bar{m}^2)^3}{2866544640 \pi^{10}}\}    \\
\notag
\notag &&  +\langle \bar{\psi}\psi\rangle^2 \int_{4m_c^2}^{s_0}ds\int_{y_i}^{y_f}dy\int_{z_i}^{1-y}dze^{-\frac{s}{T^2}}\{(y+z-1)^5\} \{\frac{49 s^2 (s-\bar{m}^2)^2}{477757440 \pi^{10}}+\frac{49 (s-\bar{m}^2)^4}{955514880 \pi^{10}}\}\\
\notag && +\langle \bar{\psi}\psi\rangle^2 \int_{4m_c^2}^{s_0}ds\int_{y_i}^{y_f}dy\int_{z_i}^{1-y}dze^{-\frac{s}{T^2}}\{(y+z-1)^5\} \{\frac{49 s (s-\bar{m}^2)^3}{238878720 \pi^{10}}\}   \, ,\\
\notag
\end{eqnarray}

\begin{eqnarray}
\notag \textbf{S}_B(6)_3&&   =m_c^2 \langle g_s^3fGGG\rangle \int_{4m_c^2}^{s_0}ds\int_{y_i}^{y_f}dy\int_{z_i}^{1-y}dze^{-\frac{s}{T^2}}\{(y+z-1)^2\} \{-\frac{(s-\bar{m}^2)^3}{524288 \pi^{10}}\}   \\
\notag
\notag &&  +\langle g_s^3fGGG\rangle \int_{4m_c^2}^{s_0}ds\int_{y_i}^{y_f}dy\int_{z_i}^{1-y}dze^{-\frac{s}{T^2}}\{y z (y+z-1)^2\} \{-\frac{(s-\bar{m}^2)^4}{1048576 \pi^{10}}-\frac{s (s-\bar{m}^2)^3}{524288 \pi^{10}}\}    \\
\notag
\notag &&    +m_c^4\langle g_s^3fGGG\rangle  \int_{4m_c^2}^{s_0}ds\int_{y_i}^{y_f}dy\int_{z_i}^{1-y}dze^{-\frac{s}{T^2}}\{-\frac{(y+z-1)^5}{y^4}\} \{\frac{7 (s-\bar{m}^2)^2}{94371840 \pi^{10}}\}   \\
\notag
\notag &&   +m_c^2\langle g_s^3fGGG\rangle  \int_{4m_c^2}^{s_0}ds\int_{y_i}^{y_f}dy\int_{z_i}^{1-y}dze^{-\frac{s}{T^2}}\{\frac{(y+z-1)^5}{y^3}\} \{\frac{7 (1-y) (s-\bar{m}^2)^3}{47185920 \pi^{10}}-\frac{7 s y (s-\bar{m}^2)^2}{31457280 \pi^{10}}\}   \\
\notag
\notag &&   +m_c^2\langle g_s^3fGGG\rangle  \int_{4m_c^2}^{s_0}ds\int_{y_i}^{y_f}dy\int_{z_i}^{1-y}dze^{-\frac{s}{T^2}}\{-\frac{z (y+z-1)^5}{y^3}\} \{\frac{49 (s-\bar{m}^2)^3}{283115520 \pi^{10}}+\frac{49 s (s-\bar{m}^2)^2}{188743680 \pi^{10}}\}   \\
\notag
\notag &&   +\langle g_s^3fGGG\rangle \int_{4m_c^2}^{s_0}ds\int_{y_i}^{y_f}dy\int_{z_i}^{1-y}dze^{-\frac{s}{T^2}}\{-\frac{z (y+z-1)^5}{y^2}\}\{-\frac{7 (s-\bar{m}^2)^4}{283115520 \pi^{10}}-\frac{7 s (s-\bar{m}^2)^3}{141557760 \pi^{10}} \}  \\
\notag
\notag &&   +m_c^2\langle g_s^3fGGG\rangle  \int_{4m_c^2}^{s_0}ds\int_{y_i}^{y_f}dy\int_{z_i}^{1-y}dze^{-\frac{s}{T^2}}\{-\frac{z (y+z-1)^5}{y^3}\} \{\frac{7 (s-\bar{m}^2)^3}{283115520 \pi^{10}}+\frac{7 s (s-\bar{m}^2)^2}{188743680 \pi^{10}}\}   \\
\notag
\notag &&  +m_c^4\langle g_s^3fGGG\rangle  \int_{4m_c^2}^{s_0}ds\int_{y_i}^{y_f}dy\int_{z_i}^{1-y}dze^{-\frac{s}{T^2}}\{-\frac{(y+z-1)^5}{y^3 z}\} \{-\frac{7 (s-\bar{m}^2)^2}{94371840 \pi^{10}}\}    \\
\notag
\notag &&  +m_c^2 \langle g_s^3fGGG\rangle \int_{4m_c^2}^{s_0}ds\int_{y_i}^{y_f}dy\int_{z_i}^{1-y}dze^{-\frac{s}{T^2}}\{-\frac{(y+z-1)^5}{y^2 z}\}\{\frac{7 s y (s-\bar{m}^2)^2}{47185920 \pi^{10}}-\frac{7 (1-y) (s-\bar{m}^2)^3}{70778880 \pi^{10}}\}    \\
\notag
\notag &&   +m_c^2\langle g_s^3fGGG\rangle  \int_{4m_c^2}^{s_0}ds\int_{y_i}^{y_f}dy\int_{z_i}^{1-y}dze^{-\frac{s}{T^2}}\{\frac{(y+z-1)^5}{y^2}\} \{-\frac{7 (s-\bar{m}^2)^3}{283115520 \pi^{10}}-\frac{7 s (s-\bar{m}^2)^2}{188743680 \pi^{10}}\}   \\
\notag
\notag &&  +\langle g_s^3fGGG\rangle \int_{4m_c^2}^{s_0}ds\int_{y_i}^{y_f}dy\int_{z_i}^{1-y}dze^{-\frac{s}{T^2}}\{\frac{(y+z-1)^5}{y}\} \{\frac{7 (s-\bar{m}^2)^4}{377487360 \pi^{10}}+\frac{7 s (s-\bar{m}^2)^3}{188743680 \pi^{10}}\}    \\
\notag
\notag &&  +m_c^2 \langle g_s^3fGGG\rangle \int_{4m_c^2}^{s_0}ds\int_{y_i}^{y_f}dy\int_{z_i}^{1-y}dze^{-\frac{s}{T^2}}\{\frac{(y+z-1)^5}{y^2}\} \{-\frac{7 (s-\bar{m}^2)^3}{283115520 \pi^{10}}-\frac{7 s (s-\bar{m}^2)^2}{188743680 \pi^{10}}\}   \, , \\
\notag
\end{eqnarray}

\begin{eqnarray}
\notag \textbf{S}_B(8)_1&& =\langle\bar{q}g_s\sigma Gq\rangle \langle\bar{q}q\rangle \int_{4m_c^2}^{s_0}ds\int_{y_i}^{y_f}dy\int_{z_i}^{1-y}dze^{-\frac{s}{T^2}} \{-y z (y+z-1)\} \{-\frac{5 (s-\bar{m}^2)^3}{2304 \pi^6}-\frac{5 s (s-\bar{m}^2)^2}{1536 \pi^6}\}  \\
\notag
\notag && +m_c^2 \langle\bar{q}g_s\sigma Gq\rangle \langle\bar{q}q\rangle \int_{4m_c^2}^{s_0}ds\int_{y_i}^{y_f}dy\int_{z_i}^{1-y}dze^{-\frac{s}{T^2}} \{-y-z+1\} \{-\frac{5 (s-\bar{m}^2)^2}{1536 \pi^6}\}  \\
\notag
\notag && +\langle\bar{q}g_s\sigma Gq\rangle m_c^2 \langle\bar{q}q\rangle \int_{4m_c^2}^{s_0}ds\int_{y_i}^{y_f}dy\int_{z_i}^{1-y}dze^{-\frac{s}{T^2}}\{-y-z+1\} \{-\frac{13 (s-\bar{m}^2)^2}{6144 \pi^6}\}  \\
\notag
\notag && +\langle\bar{q}g_s\sigma Gq\rangle \langle\bar{q}q\rangle \int_{4m_c^2}^{s_0}ds\int_{y_i}^{y_f}dy\int_{z_i}^{1-y}dze^{-\frac{s}{T^2}}\{-y z (y+z-1)\} \{-\frac{13 (s-\bar{m}^2)^3}{9216 \pi^6}-\frac{13 s (s-\bar{m}^2)^2}{6144 \pi^6}\}   \, , \\
\notag
\end{eqnarray}

\begin{eqnarray}
\notag \textbf{S}_B(8)_2&& =m_c^4 \langle g_s^2GG\rangle^2 \int_{4m_c^2}^{s_0}ds\int_{y_i}^{y_f}dy\int_{z_i}^{1-y}dze^{-\frac{s}{T^2}}\{\frac{(y+z-1)^5}{y^2 z^2}\} \{\frac{7 (\bar{m}^2-s)}{424673280 \pi^{10}}-\frac{7 s}{849346560 \pi^{10}}\}    \\
\notag
\notag &&  +m_c^2\langle g_s^2GG\rangle^2  \int_{4m_c^2}^{s_0}ds\int_{y_i}^{y_f}dy\int_{z_i}^{1-y}dze^{-\frac{s}{T^2}}\{-\frac{(y+z-1)^6}{(y-1) y^2 z^2}\}\{\frac{7 s (1-y) (s-\bar{m}^2)^2}{424673280 \pi^{10}}-\frac{7 s y (s-\bar{m}^2)}{424673280 \pi^{10}}\}   \\
\notag
\notag &&  + m_c^2 \langle g_s^2GG\rangle^2 \int_{4m_c^2}^{s_0}ds\int_{y_i}^{y_f}dy\int_{z_i}^{1-y}dze^{-\frac{s}{T^2}}\{-\frac{(y+z-1)^5}{(y-1) y^2 z}\} \{\frac{7 s (s-\bar{m}^2)}{424673280 \pi^{10}}\}   \\
\notag
\notag && +m_c^2\langle g_s^2GG\rangle^2  \int_{4m_c^2}^{s_0}ds\int_{y_i}^{y_f}dy\int_{z_i}^{1-y}dze^{-\frac{s}{T^2}}\{-\frac{(y+z-1)^5}{y z}\} \{\frac{7 s (s-\bar{m}^2)}{141557760 \pi^{10}}+\frac{7 (s-\bar{m}^2)^2}{283115520 \pi^{10}}\}\\
\notag && +m_c^2\langle g_s^2GG\rangle^2  \int_{4m_c^2}^{s_0}ds\int_{y_i}^{y_f}dy\int_{z_i}^{1-y}dze^{-\frac{s}{T^2}}\{-\frac{(y+z-1)^5}{y z}\}\{\frac{7 s^2}{849346560 \pi^{10}}\}    \\
\notag
\notag && + m_c^2 \langle g_s^2GG\rangle^2 \int_{4m_c^2}^{s_0}ds\int_{y_i}^{y_f}dy\int_{z_i}^{1-y}dze^{-\frac{s}{T^2}}\{-\frac{(y+z-1)^3}{y z}\} \{-\frac{145 (s-\bar{m}^2)^2}{905969664 \pi^{10}}\}    \\
\notag
\notag &&  +\langle g_s^2GG\rangle^2 \int_{4m_c^2}^{s_0}ds\int_{y_i}^{y_f}dy\int_{z_i}^{1-y}dze^{-\frac{s}{T^2}}\{(y+z-1)^3\} \{\frac{259 (s-\bar{m}^2)^3}{4076863488 \pi^{10}}+\frac{259 s (s-\bar{m}^2)^2}{2717908992 \pi^{10}}\}   \\
\notag
\notag && +\langle g_s^2GG\rangle^2 \int_{4m_c^2}^{s_0}ds\int_{y_i}^{y_f}dy\int_{z_i}^{1-y}dze^{-\frac{s}{T^2}}\{-(y+z-1)^4\}\{-\frac{83 s^2 (s-\bar{m}^2)}{2717908992 \pi^{10}}-\frac{83 (s-\bar{m}^2)^3}{2717908992 \pi^{10}}\} \\
\notag && +\langle g_s^2GG\rangle^2 \int_{4m_c^2}^{s_0}ds\int_{y_i}^{y_f}dy\int_{z_i}^{1-y}dze^{-\frac{s}{T^2}}\{-(y+z-1)^4\}\{-\frac{83 s (s-\bar{m}^2)^2}{905969664 \pi^{10}}\}    \\
\notag
\notag && +m_c^2 \langle g_s^2GG\rangle^2 \int_{4m_c^2}^{s_0}ds\int_{y_i}^{y_f}dy\int_{z_i}^{1-y}dze^{-\frac{s}{T^2}}\{-y-z+1\} \{\frac{199 (s-\bar{m}^2)^2}{150994944 \pi^{10}}\}    \\
\notag
\notag && +\langle g_s^2GG\rangle^2 \int_{4m_c^2}^{s_0}ds\int_{y_i}^{y_f}dy\int_{z_i}^{1-y}dze^{-\frac{s}{T^2}}\{-y z (y+z-1)\} \{\frac{199 (s-\bar{m}^2)^3}{226492416 \pi^{10}}+\frac{199 s (s-\bar{m}^2)^2}{150994944 \pi^{10}}\}   \, , \\
\notag
\end{eqnarray}

\begin{eqnarray}
\notag \textbf{S}_B(10)_1&& =\langle\bar{q}g_s\sigma Gq\rangle^2 \int_{4m_c^2}^{s_0}ds\int_{y_i}^{y_f}dy\int_{z_i}^{1-y}dze^{-\frac{s}{T^2}}\{y z\} \{\frac{5 (s-\bar{m}^2)^2}{6144 \pi^6}+\frac{5 s (s-\bar{m}^2)}{6144 \pi^6 }\}  \\
\notag
\notag && +m_c^2 \langle\bar{q}g_s\sigma Gq\rangle^2 \int_{4m_c^2}^{s_0}ds\int_{y_i}^{y_f}dy\int_{z_i}^{1-y}dze^{-\frac{s}{T^2}} \{-\frac{5 (\bar{m}^2-s)}{6144 \pi^6}\}  \\
\notag
\notag && +m_c^2\langle\bar{q}g_s\sigma Gq\rangle^2  \int_{4m_c^2}^{s_0}ds\int_{y_i}^{y_f}dy\int_{z_i}^{1-y}dze^{-\frac{s}{T^2}}\{\frac{-13 (\bar{m}^2-s)}{12288 \pi^6 }\}    \\
\notag
\notag && +\langle\bar{q}g_s\sigma Gq\rangle^2 \int_{4m_c^2}^{s_0}ds\int_{y_i}^{y_f}dy\int_{z_i}^{1-y}dze^{-\frac{s}{T^2}}\{y z\} \{\frac{13 (s-\bar{m}^2)^2}{12288 \pi^6}+\frac{13 s (s-\bar{m}^2)}{12288 \pi^6}\}    \\
\notag
\notag && +m_c^2\langle\bar{q}g_s\sigma Gq\rangle^2  \int_{4m_c^2}^{s_0}ds\int_{y_i}^{y_f}dy\int_{z_i}^{1-y}dze^{-\frac{s}{T^2}}\{\frac{59 (\bar{m}^2-s)}{1179648 \pi^6}\}    \\
\notag
\notag && +\langle\bar{q}g_s\sigma Gq\rangle^2 \int_{4m_c^2}^{s_0}ds\int_{y_i}^{y_f}dy\int_{z_i}^{1-y}dze^{-\frac{s}{T^2}}\{y z\} \{-\frac{59 (s-\bar{m}^2)^2}{1179648 \pi^6}-\frac{59 s (s-\bar{m}^2)}{1179648 \pi^6}\}    \\
\notag
\notag && +m_c^2\langle\bar{q}g_s\sigma Gq\rangle^2  \int_{4m_c^2}^{s_0}ds\int_{y_i}^{y_f}dy\int_{z_i}^{1-y}dze^{-\frac{s}{T^2}}\{\frac{(y+z-1)^2}{y z}\} \{\frac{-(\bar{m}^2-s)}{24576 \pi^6}\}    \\
\notag
\notag && +\langle\bar{q}g_s\sigma Gq\rangle^2 \int_{4m_c^2}^{s_0}ds\int_{y_i}^{y_f}dy\int_{z_i}^{1-y}dze^{-\frac{s}{T^2}}\{-(y+z-1)^2\} \{-\frac{109 (s-\bar{m}^2)}{7077888 \pi^6}-\frac{109 s (s-\bar{m}^2)}{7077888 \pi^6}\}    \\
\notag
\notag && +\langle\bar{q}g_s\sigma Gq\rangle^2 \int_{4m_c^2}^{s_0}ds\int_{y_i}^{y_f}dy\int_{z_i}^{1-y}dze^{-\frac{s}{T^2}}\{(y+z-1)^3\}\{-\frac{35 s (s-\bar{m}^2)}{884736 \pi^6}-\frac{35 (s-\bar{m}^2)^2}{1769472 \pi^6}\}\\
\notag && +\langle\bar{q}g_s\sigma Gq\rangle^2 \int_{4m_c^2}^{s_0}ds\int_{y_i}^{y_f}dy\int_{z_i}^{1-y}dze^{-\frac{s}{T^2}}\{(y+z-1)^3\}\{-\frac{35 s^2}{5308416 \pi^6}\}    \, ,\\
\notag
\end{eqnarray}

\begin{eqnarray}
\notag \textbf{S}_B(10)_2&& =2 m_c^2 \langle\bar{q}q\rangle^2 \langle g_s^2GG\rangle \int_{4m_c^2}^{s_0}ds\int_{y_i}^{y_f}dy\int_{z_i}^{1-y}dze^{-\frac{s}{T^2}} \{-\frac{(y+z-1)^2}{y^2}\} \{-\frac{5 y (\bar{m}^2-s)}{27648 \pi^6 }+\frac{5 (\bar{m}^2-s)}{27648 \pi^6}\}\\
\notag && +2 m_c^2 \langle\bar{q}q\rangle^2 \langle g_s^2GG\rangle \int_{4m_c^2}^{s_0}ds\int_{y_i}^{y_f}dy\int_{z_i}^{1-y}dze^{-\frac{s}{T^2}} \{-\frac{(y+z-1)^2}{y^2}\}\{\frac{5 s y}{55296 \pi^6}\}  \\
\notag
\notag && +2 m_c^2 \langle\bar{q}q\rangle^2 \langle g_s^2GG\rangle \int_{4m_c^2}^{s_0}ds\int_{y_i}^{y_f}dy\int_{z_i}^{1-y}dze^{-\frac{s}{T^2}} \{\frac{z (y+z-1)^2}{y^2}\} \{\frac{5 (\bar{m}^2-s)}{27648 \pi^6 }-\frac{5 s}{55296 \pi^6}\}  \\
\notag
\notag && +m_c^2 \langle g_s^2GG\rangle \langle\bar{q}q\rangle^2 \int_{4m_c^2}^{s_0}ds\int_{y_i}^{y_f}dy\int_{z_i}^{1-y}dze^{-\frac{s}{T^2}}\{-\frac{19 (\bar{m}^2-s)}{73728 \pi^6}\}  \\
\notag
\notag && +\langle g_s^2GG\rangle \langle\bar{q}q\rangle^2 \int_{4m_c^2}^{s_0}ds\int_{y_i}^{y_f}dy\int_{z_i}^{1-y}dze^{-\frac{s}{T^2}}\{y z\} \{\frac{19 (s-\bar{m}^2)^2}{73728 \pi^6}+\frac{19 s (s-\bar{m}^2)}{73728 \pi^6}\}  \\
\notag
\notag &&  +m_c^2\langle g_s^2GG\rangle  \langle\bar{q}q\rangle^2 \int_{4m_c^2}^{s_0}ds\int_{y_i}^{y_f}dy\int_{z_i}^{1-y}dze^{-\frac{s}{T^2}}\{\frac{(y+z-1)^2}{y z}\}\{\frac{5 (\bar{m}^2-s)}{36864 \pi^6}\}  \\
\notag
\notag && +\langle g_s^2GG\rangle \langle\bar{q}q\rangle^2 \int_{4m_c^2}^{s_0}ds\int_{y_i}^{y_f}dy\int_{z_i}^{1-y}dze^{-\frac{s}{T^2}}\{-(y+z-1)^2\} \{\frac{5 (s-\bar{m}^2)^2}{73728 \pi^6}+\frac{5 s (s-\bar{m}^2)}{73728 \pi^6}\}  \\
\notag
\notag &&+m_c^2\langle\bar{q}q\rangle^2 \langle g_s^2GG\rangle  \int_{4m_c^2}^{s_0}ds\int_{y_i}^{y_f}dy\int_{z_i}^{1-y}dze^{-\frac{s}{T^2}}\{-\frac{5 (\bar{m}^2-s)}{27648 \pi^6}\}   \\
\notag
\notag && +\langle\bar{q}q\rangle^2 \langle g_s^2GG\rangle \int_{4m_c^2}^{s_0}ds\int_{y_i}^{y_f}dy\int_{z_i}^{1-y}dze^{-\frac{s}{T^2}}\{y z\} \{\frac{5 (s-\bar{m}^2)^2}{27648 \pi^6 }+\frac{5 s (s-\bar{m}^2)}{27648 \pi^6}\}  \\
\notag
\notag && \langle g_s^2GG\rangle \langle\bar{q}q\rangle^2 \int_{4m_c^2}^{s_0}ds\int_{y_i}^{y_f}dy\int_{z_i}^{1-y}dze^{-\frac{s}{T^2}}\{\frac{z (y+z-1)^2}{y^2}\} \{\frac{\bar{m}^2-s}{186624 \pi^8}-\frac{s}{373248 \pi^8}\}   \\
\notag
\notag && \langle g_s^2GG\rangle \langle\bar{q}q\rangle^2 \int_{4m_c^2}^{s_0}ds\int_{y_i}^{y_f}dy\int_{z_i}^{1-y}dze^{-\frac{s}{T^2}}\{\frac{-(y+z-1)^2}{y^2}\} \{\frac{(1-y) (\bar{m}^2-s)}{186624 \pi^8}+\frac{s y}{373248 \pi^8}\} \, ,   \\
\notag
\end{eqnarray}

\begin{eqnarray}
\notag \textbf{S}_B(12)_1&&  =2 m_c^2 \langle\bar{q}g_s\sigma Gq\rangle \langle\bar{q}q\rangle \langle g_s^2GG\rangle \int_{4m_c^2}^{s_0}ds\int_{y_i}^{y_f}dy\int_{z_i}^{1-y}dze^{-\frac{s}{T^2}} \{\frac{y+z-1}{y^2}\} \{-\frac{5 y}{27648 \pi^6}+\frac{5}{27648 \pi^6}\}  \\
\notag
\notag &&  +2 m_c^2 \langle\bar{q}g_s\sigma Gq\rangle \langle\bar{q}q\rangle \langle g_s^2GG\rangle \int_{4m_c^2}^{s_0}ds\int_{y_i}^{y_f}dy\int_{z_i}^{1-y}dze^{-\frac{s}{T^2}} \{-\frac{z (y+z-1)}{y^2}\} \{\frac{5}{27648 \pi^6}\}  \\
\notag
\notag &&   +\langle g_s^2GG\rangle \langle\bar{q}g_s\sigma Gq\rangle \langle\bar{q}q\rangle \int_{4m_c^2}^{s_0}ds\int_{y_i}^{y_f}dy\int_{z_i}^{1-y}dze^{-\frac{s}{T^2}}\{y+z-1\} \{\frac{5 \bar{m}^2}{36864 \pi^6 }-\frac{5 s}{24576 \pi^6}\}    \\
\notag
\notag &&  +m_c^2 \langle g_s^2GG\rangle \langle\bar{q}g_s\sigma Gq\rangle \langle\bar{q}q\rangle \int_{4m_c^2}^{s_0}ds\int_{y_i}^{y_f}dy\int_{z_i}^{1-y}dze^{-\frac{s}{T^2}}\{-\frac{y+z-1}{y z}\}\{\frac{5}{36864 \pi^6}\}   \, , \\
\notag
\end{eqnarray}

\begin{eqnarray}
\notag \textbf{S}_B(12)_2&&=2  m_c^2 \langle\bar{q}q\rangle^4 \int_{4m_c^2}^{s_0}ds\int_{y_i}^{y_f}dy\int_{z_i}^{1-y}dze^{-\frac{s}{T^2}}\{\frac{y+z-1}{y}\} \{\frac{25}{31104 \pi^4}\}     \\
\notag
\notag && +2 m_c^2 \langle\bar{q}q\rangle^4 \int_{4m_c^2}^{s_0}ds\int_{y_i}^{y_f}dy\int_{z_i}^{1-y}dze^{-\frac{s}{T^2}}\{-\frac{z (z+y-1)}{y^2}\} \{-\frac{25}{31104 \pi^4}\}    \\
\notag
\notag &&+2 \langle\bar{q}q\rangle^4 \int_{4m_c^2}^{s_0}ds\int_{y_i}^{y_f}dy\int_{z_i}^{1-y}dze^{-\frac{s}{T^2}}\{-\frac{z (y+z-1)}{y}\} \{-\frac{25 y (\bar{m}^2-s)}{15552 \pi^4}+\frac{25 (\bar{m}^2-s)}{23328 \pi^4}\}\\
\notag && +2 \langle\bar{q}q\rangle^4 \int_{4m_c^2}^{s_0}ds\int_{y_i}^{y_f}dy\int_{z_i}^{1-y}dze^{-\frac{s}{T^2}}\{-\frac{z (y+z-1)}{y}\}\{\frac{25 s y}{15552 \pi^4}-\frac{25 s}{46656 \pi^4}\}     \\
\notag
\notag && +2 \langle\bar{q}q\rangle^4 \int_{4m_c^2}^{s_0}ds\int_{y_i}^{y_f}dy\int_{z_i}^{1-y}dze^{-\frac{s}{T^2}}\{-\frac{z (y+z-1)}{y}\} \{\frac{25 y (\bar{m}^2-s)}{15552 \pi^4}-\frac{25 (\bar{m}^2-s)}{93312 \pi^4}\}\\
\notag && +2 \langle\bar{q}q\rangle^4 \int_{4m_c^2}^{s_0}ds\int_{y_i}^{y_f}dy\int_{z_i}^{1-y}dze^{-\frac{s}{T^2}}\{-\frac{z (y+z-1)}{y}\}\{-\frac{25 s y}{15552 \pi^4}+\frac{25 s}{186624 \pi^4}\}   \, , \\
\notag
\end{eqnarray}

\begin{eqnarray}
\notag \textbf{S}_B(12)_3&&  =\langle \bar{\psi}\psi\rangle^2 \langle\bar{q}q\rangle^2 \int_{4m_c^2}^{s_0}ds\int_{y_i}^{y_f}dy\int_{z_i}^{1-y}dze^{-\frac{s}{T^2}}\{-\frac{(y+z-1)^3}{(y-1) y z})\} \{\frac{35 (y-1) (\bar{m}^2-s)}{1119744 \pi^6}-\frac{35 s y}{2239488 \pi^6}\}   \\
\notag
\notag && +\langle \bar{\psi}\psi\rangle^2 \langle\bar{q}q\rangle^2 \int_{4m_c^2}^{s_0}ds\int_{y_i}^{y_f}dy\int_{z_i}^{1-y}dze^{-\frac{s}{T^2}}\{-\frac{(y+z-1)^2}{(y-1) y}\} \{\frac{35 s}{2239488 \pi^6}\}    \\
\notag
\notag && +\langle \bar{\psi}\psi\rangle^2 \langle\bar{q}q\rangle^2 \int_{4m_c^2}^{s_0}ds\int_{y_i}^{y_f}dy\int_{z_i}^{1-y}dze^{-\frac{s}{T^2}}\{-(y+z-1)^2\} \{-\frac{35 (\bar{m}^2-s)}{186624 \pi^6}+\frac{35 s}{186624 \pi^6}\}    \\
\notag
\notag && + m_c^2 \langle \bar{\psi}\psi\rangle^2 \langle\bar{q}q\rangle^2 \int_{4m_c^2}^{s_0}ds\int_{y_i}^{y_f}dy\int_{z_i}^{1-y}dze^{-\frac{s}{T^2}}\{-\frac{(y+z-1)^2}{y z^2}\} \{\frac{5 (1-y)}{186624 \pi^6}\}    \\
\notag
\notag && +\langle \bar{\psi}\psi\rangle^2 \langle\bar{q}q\rangle^2 \int_{4m_c^2}^{s_0}ds\int_{y_i}^{y_f}dy\int_{z_i}^{1-y}dze^{-\frac{s}{T^2}}\{-\frac{(y+z-1)^3}{(y-1) y z}\} \{\frac{5 (1-y) (\bar{m}^2-s)}{279936 \pi^6}+\frac{5 s y}{559872 \pi^6}\}    \\
\notag
\notag && +\langle \bar{\psi}\psi\rangle^2 \langle\bar{q}q\rangle^2 \int_{4m_c^2}^{s_0}ds\int_{y_i}^{y_f}dy\int_{z_i}^{1-y}dze^{-\frac{s}{T^2}}\{-\frac{(y+z-1)^2}{(y-1) y}\} \{-\frac{5 s}{559872 \pi^6}\}    \\
\notag
\notag && +\langle \bar{\psi}\psi\rangle^2 \langle\bar{q}q\rangle^2 \int_{4m_c^2}^{s_0}ds\int_{y_i}^{y_f}dy\int_{z_i}^{1-y}dze^{-\frac{s}{T^2}}\{-(y+z-1)^2\} \{\frac{5 (\bar{m}^2-s)}{186624 \pi^6}-\frac{5 s}{186624 \pi^6 }\}    \\
\notag
\notag && +m_c^2\langle \bar{\psi}\psi\rangle^2 \langle\bar{q}q\rangle^2  \int_{4m_c^2}^{s_0}ds\int_{y_i}^{y_f}dy\int_{z_i}^{1-y}dze^{-\frac{s}{T^2}}\{\frac{(y+z-1)^2}{y z}\} \{-\frac{5}{46656 \pi^6}\}   \, , \\
\notag
\end{eqnarray}

\begin{eqnarray}
\notag \textbf{S}_B(12)_4&&  = m_c^2\langle g_s^2GG\rangle^3 \int_{4m_c^2}^{s_0}ds\int_{y_i}^{y_f}dy\int_{z_i}^{1-y}dze^{-\frac{s}{T^2}}\{-\frac{y+z-1}{y z}\} \{-\frac{3143}{65229815808 \pi^{10}}\}   \\
\notag
\notag &&  +\langle g_s^2GG\rangle^3 \int_{4m_c^2}^{s_0}ds\int_{y_i}^{y_f}dy\int_{z_i}^{1-y}dze^{-\frac{s}{T^2}}\{y+z-1\} \{\frac{2383 s}{97844723712 \pi^{10}}-\frac{2383 (\bar{m}^2-s)}{48922361856 \pi^{10}}\}    \\
\notag
\notag && +\langle g_s^2GG\rangle^3 \int_{4m_c^2}^{s_0}ds\int_{y_i}^{y_f}dy\int_{z_i}^{1-y}dze^{-\frac{s}{T^2}}\{-(y+z-1)^2\} \{\frac{103 (\bar{m}^2-s)}{32614907904 \pi^{10}}-\frac{103 s}{32614907904 \pi^{10}}\}  \, , \\
\notag
\end{eqnarray}

\begin{eqnarray}
\notag \textbf{S}_B(12)_5&&   =m_c^2\langle g_s^3fGGG\rangle^2  \int_{4m_c^2}^{s_0}ds\int_{y_i}^{y_f}dy\int_{z_i}^{1-y}dze^{-\frac{s}{T^2}}\{-\frac{y+z-1}{y z}\} \{-\frac{1}{50331648 \pi^{10}}\}    \\
\notag
\notag &&   +\langle g_s^3fGGG\rangle^2 \int_{4m_c^2}^{s_0}ds\int_{y_i}^{y_f}dy\int_{z_i}^{1-y}dze^{-\frac{s}{T^2}}\{y+z-1\} \{\frac{s}{150994944 \pi^{10}}-\frac{\bar{m}^2-s}{75497472 \pi^{10}}\}    \\
\notag
\notag &&   +\langle g_s^3fGGG\rangle^2 \int_{4m_c^2}^{s_0}ds\int_{y_i}^{y_f}dy\int_{z_i}^{1-y}dze^{-\frac{s}{T^2}}\{-(y+z-1)^2\} \{-\frac{\bar{m}^2-s}{25165824 \pi^{10}}+\frac{s}{25165824 \pi^{10}}\}  \, ,  \\
\notag
\end{eqnarray}

\begin{eqnarray}
\notag \textbf{S}_B(14)_1&&  =\langle g_s^2GG\rangle^2 \langle\bar{q}q\rangle^2 \int_{4m_c^2}^{s_0}ds\int_{y_i}^{y_f}dy\int_{z_i}^{1-y}dze^{-\frac{s}{T^2}}\{\frac{259}{63700992 \pi^6}\}   \\
\notag
\notag && +\langle g_s^2GG\rangle^2 \langle\bar{q}q\rangle^2 \int_{4m_c^2}^{s_0}ds\int_{y_i}^{y_f}dy\int_{z_i}^{1-y}dze^{-\frac{s}{T^2}}\{y+z-1\} \{-\frac{83}{10616832 \pi^6}\}   \, ,\\
\notag
\end{eqnarray}

\begin{eqnarray}
\notag &&   \textbf{S}_B(14)_2=\langle g_s^2GG\rangle \langle\bar{q}g_s\sigma Gq\rangle^2 \int_{4m_c^2}^{s_0}ds\int_{y_i}^{y_f}dy\int_{z_i}^{1-y}dze^{-\frac{s}{T^2}}\{-\frac{5}{294912 \pi^6 }\} \, ,  \\
\notag
\end{eqnarray}

\begin{eqnarray}
\notag \textbf{S}_B(14)_3&&   =2 \langle\bar{q}g_s\sigma Gq\rangle \langle\bar{q}q\rangle^3 \int_{4m_c^2}^{s_0}ds\int_{y_i}^{y_f}dy\int_{z_i}^{1-y}dze^{-\frac{s}{T^2}}\{\frac{z}{y}\}\{-\frac{25 y}{31104 \pi^4}+\frac{25}{46656 \pi^4}\}   \\
\notag
\notag &&    +2 \langle\bar{q}g_s\sigma Gq\rangle \langle\bar{q}q\rangle^3 \int_{4m_c^2}^{s_0}ds\int_{y_i}^{y_f}dy\int_{z_i}^{1-y}dze^{-\frac{s}{T^2}}\{\frac{z}{y}\} \{\frac{25 y}{31104 \pi^4}-\frac{25}{186624 \pi^4}\} \, , \\
\notag
\end{eqnarray}

\begin{eqnarray}
\notag \textbf{S}_B(14)_4&&    =\langle\bar{q}q\rangle \langle\bar{q}g_s\sigma Gq\rangle \langle \bar{\psi}\psi\rangle^2 \int_{4m_c^2}^{s_0}ds\int_{y_i}^{y_f}dy\int_{z_i}^{1-y}dze^{-\frac{s}{T^2}}\{\frac{(y+z-1)^2}{(y-1) y z}\} \{-\frac{5 y}{279936 \pi^6}+\frac{5}{279936 \pi^6}\}   \\
\notag
\notag &&   +\langle\bar{q}q\rangle \langle\bar{q}g_s\sigma Gq\rangle \langle \bar{\psi}\psi\rangle^2 \int_{4m_c^2}^{s_0}ds\int_{y_i}^{y_f}dy\int_{z_i}^{1-y}dze^{-\frac{s}{T^2}}\{y+z-1\} \{\frac{5}{186624 \pi^6}\}    \\
\notag
\notag &&   +\langle\bar{q}q\rangle \langle\bar{q}g_s\sigma Gq\rangle \langle \bar{\psi}\psi\rangle^2 \int_{4m_c^2}^{s_0}ds\int_{y_i}^{y_f}dy\int_{z_i}^{1-y}dze^{-\frac{s}{T^2}}\{\frac{(y+z-1)^2}{(y-1) y z}\} \{\frac{35 y}{1119744 \pi^6}-\frac{35}{1119744 \pi^6}\}   \\
\notag
\notag &&    +\langle\bar{q}q\rangle \langle\bar{q}g_s\sigma Gq\rangle \langle \bar{\psi}\psi\rangle^2 \int_{4m_c^2}^{s_0}ds\int_{y_i}^{y_f}dy\int_{z_i}^{1-y}dze^{-\frac{s}{T^2}}\{y+z-1\}\{-\frac{35}{186624 \pi^6}\} \, .  \\
\notag
\end{eqnarray}

Type C for $J_1$

\begin{eqnarray}
\notag \textbf{S}_C(14)_1&&   =\langle\bar{q}g_s\sigma Gq\rangle \langle\bar{q}q\rangle^3 \int_{y_i}^{y_f}dye^{-\frac{\tilde{m}^2}{T^2}}\{(y-1) y\} \{\frac{7 \tilde{m}^2}{108 \pi^2}+\frac{7 \tilde{m}^4}{432 \pi^2  T^2}\}  \\
\notag
\notag && +m_c^2 \langle\bar{q}g_s\sigma Gq\rangle \langle\bar{q}q\rangle^3 \int_{y_i}^{y_f}dye^{-\frac{\tilde{m}^2}{T^2}}\{-\frac{7}{432 \pi^2}-\frac{7 \tilde{m}^2}{432 \pi^2  T^2}\}    \\
\notag
\notag && +m_c^2 \langle\bar{q}q\rangle^3 \langle\bar{q}g_s\sigma Gq\rangle \int_{y_i}^{y_f}dye^{-\frac{\tilde{m}^2}{T^2}}\{-\frac{5}{23328 \pi^4}-\frac{5 \tilde{m}^2}{23328 \pi^4  T^2}\}   \\
\notag
\notag && +\langle\bar{q}q\rangle^3 \langle\bar{q}g_s\sigma Gq\rangle \int_{y_i}^{y_f}dye^{-\frac{\tilde{m}^2}{T^2}}\{(y-1) y\} \{\frac{5 \tilde{m}^2}{5832 \pi^4}+\frac{5 \tilde{m}^4}{23328 \pi^4 T^2}\}  \, , \\
\notag
\end{eqnarray}

\begin{eqnarray}
\notag \textbf{S}_C(14)_2&& =m_c^2\langle g_s^2GG\rangle  \langle\bar{q}g_s\sigma Gq\rangle^2 \int_{y_i}^{y_f}dye^{-\frac{\tilde{m}^2}{T^2}}\{\frac{19}{1179648 \pi^6}+\frac{19 \tilde{m}^2}{1179648 \pi^6  T^2}\}   \\
\notag
\notag && +\langle g_s^2GG\rangle \langle\bar{q}g_s\sigma Gq\rangle^2 \int_{y_i}^{y_f}dye^{-\frac{\tilde{m}^2}{T^2}}\{(y-1) y\} \{-\frac{19 \tilde{m}^2}{294912 \pi^6}-\frac{19 \tilde{m}^4}{1179648 \pi^6  T^2}\}  \, , \\
\notag
\end{eqnarray}

\begin{eqnarray}
\notag \textbf{S}_C(14)_3&& =m_c^2 \langle\bar{q}q\rangle^2 \langle g_s^2GG\rangle^2 \int_{y_i}^{y_f}dye^{-\frac{\tilde{m}^2}{T^2}}\{\frac{5}{7962624 \pi^6}+\frac{5 \tilde{m}^2}{7962624 \pi^6  T^2}\}   \\
\notag
\notag && +\langle\bar{q}q\rangle^2 \langle g_s^2GG\rangle^2 \int_{y_i}^{y_f}dye^{-\frac{\tilde{m}^2}{T^2}}\{(y-1) y\} \{-\frac{5 \tilde{m}^2}{1990656 \pi^6}-\frac{5 \tilde{m}^4}{7962624 \pi^6  T^2}\}  \, , \\
\notag
\end{eqnarray}

\begin{eqnarray}
\notag \textbf{S}_C(16)_1&& =\frac{1}{6} \langle\bar{q}g_s\sigma Gq\rangle^2 \langle\bar{q}q\rangle^2 \int_{y_i}^{y_f}dye^{-\frac{\tilde{m}^2}{T^2}}\{(y-1) y\} \{-\frac{7}{32 \pi^2}+\frac{17 \tilde{m}^6}{192 \pi^2 T^6}-\frac{7 \tilde{m}^4}{64 \pi^2  T^4}-\frac{7 \tilde{m}^2}{32 \pi^2  T^2}\}    \\
\notag
\notag && +\frac{1}{6} \langle\bar{q}g_s\sigma Gq\rangle^2 \langle\bar{q}q\rangle^2 m_c^2 \int_{y_i}^{y_f}dye^{-\frac{\tilde{m}^2}{T^2}}\{\frac{7 \tilde{m}^4}{192 \pi^2  T^6}\}   \\
\notag
\notag && +\langle\bar{q}q\rangle^2 \langle\bar{q}g_s\sigma Gq\rangle^2 m_c^2 \int_{y_i}^{y_f}dye^{-\frac{\tilde{m}^2}{T^2}}\{\frac{1}{(y-1) y}\} \{-\frac{83}{331776 \pi^2  T^2}\}   \\
\notag
\notag && +\langle\bar{q}q\rangle^2 \langle\bar{q}g_s\sigma Gq\rangle^2 \int_{y_i}^{y_f}dye^{-\frac{\tilde{m}^2}{T^2}}\{\frac{83}{663552 \pi^2}+\frac{83 \tilde{m}^2}{663552 \pi^2  T^2}\}   \\
\notag
\notag && +  m_c^2 \langle\bar{q}g_s\sigma Gq\rangle^2 \langle\bar{q}q\rangle^2 \int_{y_i}^{y_f}dye^{-\frac{\tilde{m}^2}{T^2}}\{\frac{1}{y}\} \{-\frac{25 \tilde{m}^2}{497664 \pi^4  T^4}\}   \\
\notag
\notag && +  m_c^2  \langle\bar{q}g_s\sigma Gq\rangle^2 \langle\bar{q}q\rangle^2 \int_{y_i}^{y_f}dye^{-\frac{\tilde{m}^2}{T^2}}\{\frac{y}{(y-1)^2}\} \{-\frac{25 \tilde{m}^2}{497664 \pi^4  T^4}\}   \\
\notag
\notag &&+  \langle\bar{q}g_s\sigma Gq\rangle^2 \langle\bar{q}q\rangle^2 \int_{y_i}^{y_f}dye^{-\frac{\tilde{m}^2}{T^2}}\{\frac{y-1}{y}\} \{\frac{25}{373248 \pi^4}+\frac{25 \tilde{m}^2}{373248 \pi^4 T^2}-\frac{25 \tilde{m}^4 y}{746496 \pi^4  T^4}\}\\
\notag && +  \langle\bar{q}g_s\sigma Gq\rangle^2 \langle\bar{q}q\rangle^2 \int_{y_i}^{y_f}dye^{-\frac{\tilde{m}^2}{T^2}}\{\frac{y-1}{y}\} \{-\frac{25 \tilde{m}^2 y}{373248 \pi^4  T^2}-\frac{25 y}{373248 \pi^4}\}    \\
\notag
\notag && +  \langle\bar{q}g_s\sigma Gq\rangle^2 \langle\bar{q}q\rangle^2 \int_{y_i}^{y_f}dye^{-\frac{\tilde{m}^2}{T^2}}\{\frac{y-1}{y}\} \{-\frac{25}{1492992 \pi^4}-\frac{25 \tilde{m}^2}{1492992 \pi^4 T^2}+\frac{25 \tilde{m}^4 y}{746496 \pi^4  T^4}\}\\
\notag && +  \langle\bar{q}g_s\sigma Gq\rangle^2 \langle\bar{q}q\rangle^2 \int_{y_i}^{y_f}dye^{-\frac{\tilde{m}^2}{T^2}}\{\frac{y-1}{y}\}\{\frac{25 \tilde{m}^2 y}{373248 \pi^4  T^2}+\frac{25 y}{373248 \pi^4}\}   \\
\notag
\notag && +m_c^2 \langle\bar{q}q\rangle^2 \langle\bar{q}g_s\sigma Gq\rangle^2 \int_{y_i}^{y_f}dye^{-\frac{\tilde{m}^2}{T^2}}\{\frac{5 \tilde{m}^4}{186624 \pi^4  T^6}\}   \\
\notag
\notag && +\langle\bar{q}q\rangle^2 \langle\bar{q}g_s\sigma Gq\rangle^2 \int_{y_i}^{y_f}dye^{-\frac{\tilde{m}^2}{T^2}}\{(y-1) y\} \{-\frac{5}{31104 \pi^4}-\frac{5 \tilde{m}^6}{186624 \pi^4 T^6}\}\\
\notag && +\langle\bar{q}q\rangle^2 \langle\bar{q}g_s\sigma Gq\rangle^2 \int_{y_i}^{y_f}dye^{-\frac{\tilde{m}^2}{T^2}}\{(y-1) y\}\{-\frac{5 \tilde{m}^4}{62208 \pi^4  T^4}-\frac{5 \tilde{m}^2}{31104 \pi^4  T^2}\}  \, , \\
\notag
\end{eqnarray}

\begin{eqnarray}
\notag \textbf{S}_C(16)_2&& =2m_c^2 \langle\bar{q}q\rangle^4 \langle g_s^2GG\rangle   \int_{y_i}^{y_f}dye^{-\frac{\tilde{m}^2}{T^2}}\{\frac{1}{y^2}\} \{\frac{7}{10368 \pi^2  T^2}\}   \\
\notag
\notag && + m_c^4 \langle\bar{q}q\rangle^4 \langle g_s^2GG\rangle  \int_{y_i}^{y_f}dye^{-\frac{\tilde{m}^2}{T^2}}\{\frac{1}{y^3}\}\{-\frac{7}{15552 \pi^2  T^4}\}   \\
\notag
\notag && + m_c^2 \langle\bar{q}q\rangle^4 \langle g_s^2GG\rangle  \int_{y_i}^{y_f}dye^{-\frac{\tilde{m}^2}{T^2}}\{\frac{y-1}{y^2}\} \{\frac{7 \tilde{m}^2}{15552 \pi^2  T^4}\}   \\
\notag
\notag && +\langle g_s^2GG\rangle \langle\bar{q}q\rangle^4 \int_{y_i}^{y_f}dye^{-\frac{\tilde{m}^2}{T^2}}\{-\frac{7}{41472 \pi^2}-\frac{7 \tilde{m}^2}{41472 \pi^2  T^2}\}   \\
\notag
\notag && +m_c^2 \langle g_s^2GG\rangle \langle\bar{q}q\rangle^4 \int_{y_i}^{y_f}dye^{-\frac{\tilde{m}^2}{T^2}}\{\frac{1}{(y-1) y}\} \{\frac{7}{20736 \pi^2  T^2}\}   \\
\notag
\notag && +\frac{1}{6}m_c^2 \langle\bar{q}q\rangle^4 \langle g_s^2GG\rangle  \int_{y_i}^{y_f}dye^{-\frac{\tilde{m}^2}{T^2}}\{\frac{7 \tilde{m}^4}{2592 \pi^2  T^6}\}   \\
\notag
\notag &&+\frac{1}{6} \langle\bar{q}q\rangle^4 \langle g_s^2GG\rangle \int_{y_i}^{y_f}dye^{-\frac{\tilde{m}^2}{T^2}}\{(y-1) y\} \{-\frac{7}{432 \pi^2}-\frac{7 \tilde{m}^6}{2592 \pi^2  T^6}-\frac{7 \tilde{m}^4}{864 \pi^2  T^4}-\frac{7 \tilde{m}^2}{432 \pi^2  T^2}\}    \\
\notag
\notag && +2m_c^2  \langle g_s^2GG\rangle \langle\bar{q}q\rangle^4 \int_{y_i}^{y_f}dye^{-\frac{\tilde{m}^2}{T^2}}\{\frac{1}{y^2}\}\{\frac{1}{34992 \pi^4 T^2}\}   \\
\notag
\notag &&+m_c^4  \langle g_s^2GG\rangle \langle\bar{q}q\rangle^4  \int_{y_i}^{y_f}dye^{-\frac{\tilde{m}^2}{T^2}}\{\frac{1}{y^3}\} \{-\frac{1}{52488 \pi^4  T^4}\}    \\
\notag
\notag && +m_c^2  \langle g_s^2GG\rangle \langle\bar{q}q\rangle^4  \int_{y_i}^{y_f}dye^{-\frac{\tilde{m}^2}{T^2}}\{\frac{y-1}{y^2}\} \{\frac{\tilde{m}^2}{52488 \pi^4  T^4}\}   \\
\notag
\notag &&+2m_c^2 \langle g_s^2GG\rangle \langle\bar{q}q\rangle^4  \int_{y_i}^{y_f}dye^{-\frac{\tilde{m}^2}{T^2}}\{\frac{1}{y^2}\} \{-\frac{1}{93312 \pi^4 T^2}\}   \\
\notag
\notag && +m_c^4  \langle g_s^2GG\rangle\langle\bar{q}q\rangle^4  \int_{y_i}^{y_f}dye^{-\frac{\tilde{m}^2}{T^2}}\{\frac{1}{y^3}\} \{\frac{1}{139968 \pi^4  T^4}\}   \\
\notag
\notag && +m_c^2  \langle g_s^2GG\rangle \langle\bar{q}q\rangle^4  \int_{y_i}^{y_f}dye^{-\frac{\tilde{m}^2}{T^2}}\{\frac{y-1}{y^2}\} \{-\frac{\tilde{m}^2}{139968 \pi^4  T^4}\}   \\
\notag
\notag &&+\frac{1}{6} m_c^2 \langle\bar{q}q\rangle^4 \langle g_s^2GG\rangle \int_{y_i}^{y_f}dye^{-\frac{\tilde{m}^2}{T^2}}\{\frac{7 \tilde{m}^4}{186624 \pi^4  T^6}\}    \\
\notag
\notag && +\frac{1}{6} \langle\bar{q}q\rangle^4 \langle g_s^2GG\rangle \int_{y_i}^{y_f}dye^{-\frac{\tilde{m}^2}{T^2}}\{(y-1) y\} \{-\frac{7}{31104 \pi^4}-\frac{7 \tilde{m}^6}{186624 \pi^4  T^6}-\frac{7 \tilde{m}^4}{62208 \pi^4  T^4}-\frac{7 \tilde{m}^2}{31104 \pi^4  T^2}\}   \\
\notag
\notag && +2 m_c^2  \langle\bar{q}q\rangle^4 \langle g_s^2GG\rangle \int_{y_i}^{y_f}dye^{-\frac{\tilde{m}^2}{T^2}}\{\frac{1}{y^2}\} \{\frac{7}{20155392 \pi^6  T^2}\}   \\
\notag
\notag && +  m_c^4  \langle\bar{q}q\rangle^4 \langle g_s^2GG\rangle \int_{y_i}^{y_f}dye^{-\frac{\tilde{m}^2}{T^2}}\{\frac{1}{y^3}\} \{-\frac{7}{30233088 \pi^6  T^4}\}   \\
\notag
\notag && +  m_c^2  \langle\bar{q}q\rangle^4 \langle g_s^2GG\rangle \int_{y_i}^{y_f}dye^{-\frac{\tilde{m}^2}{T^2}}\{\frac{y-1}{y^2}\} \{\frac{7 \tilde{m}^2}{30233088 \pi^6  T^4}\}  \, , \\
\notag
\end{eqnarray}

\begin{eqnarray}
\notag \textbf{S}_C(16)_3&& =2 m_c^2 \langle\bar{q}g_s\sigma Gq\rangle \langle g_s^2GG\rangle^2 \langle\bar{q}q\rangle \int_{y_i}^{y_f}dye^{-\frac{\tilde{m}^2}{T^2}}\{\frac{1}{y^2}\} \{-\frac{1}{331776 \pi^6  T^2}\}   \\
\notag
\notag && + m_c^4  \langle\bar{q}g_s\sigma Gq\rangle \langle g_s^2GG\rangle^2 \langle\bar{q}q\rangle \int_{y_i}^{y_f}dye^{-\frac{\tilde{m}^2}{T^2}}\{\frac{1}{y^3}\} \{\frac{1}{497664 \pi^6  T^4}\}   \\
\notag
\notag && +  m_c^2 \langle\bar{q}g_s\sigma Gq\rangle \langle g_s^2GG\rangle^2 \langle\bar{q}q\rangle \int_{y_i}^{y_f}dye^{-\frac{\tilde{m}^2}{T^2}}\{\frac{y-1}{y^2}\}\{-\frac{\tilde{m}^2}{497664 \pi^6  T^4}\}   \\
\notag
\notag && +2m_c^2  \langle\bar{q}g_s\sigma Gq\rangle \langle g_s^2GG\rangle^2 \langle\bar{q}q\rangle \int_{y_i}^{y_f}dye^{-\frac{\tilde{m}^2}{T^2}}\{\frac{1}{y^2}\} \{\frac{1}{884736 \pi^6  T^2}\}   \\
\notag
\notag && + m_c^4  \langle\bar{q}g_s\sigma Gq\rangle \langle g_s^2GG\rangle^2 \langle\bar{q}q\rangle \int_{y_i}^{y_f}dye^{-\frac{\tilde{m}^2}{T^2}}\{\frac{1}{y^3}\}\{-\frac{1}{1327104 \pi^6  T^4}\}   \\
\notag
\notag && + m_c^2 \langle\bar{q}g_s\sigma Gq\rangle \langle g_s^2GG\rangle^2 \langle\bar{q}q\rangle  \int_{y_i}^{y_f}dye^{-\frac{\tilde{m}^2}{T^2}}\{\frac{y-1}{y^2}\} \{\frac{\tilde{m}^2}{1327104 \pi^6  T^4}\}   \\
\notag
\notag &&  +m_c^2 \langle\bar{q}q\rangle \langle\bar{q}g_s\sigma Gq\rangle \langle g_s^2GG\rangle^2 \int_{y_i}^{y_f}dye^{-\frac{\tilde{m}^2}{T^2}}\{\frac{1}{(y-1) y}\} \{-\frac{145}{84934656 \pi^6  T^2}\}  \\
\notag
\notag && +\langle\bar{q}q\rangle \langle\bar{q}g_s\sigma Gq\rangle \langle g_s^2GG\rangle^2 \int_{y_i}^{y_f}dye^{-\frac{\tilde{m}^2}{T^2}}\{\frac{259}{254803968 \pi^6}+\frac{259 \tilde{m}^2}{254803968 \pi^6  T^2}\} \, .  \\
\notag
\end{eqnarray}

Type D for $J_1$

\begin{eqnarray}
\notag \textbf{S}_D(12)_1&& =2 m_c^2 \langle\bar{q}g_s\sigma Gq\rangle \langle\bar{q}q\rangle \langle g_s^2GG\rangle \int_{y_i}^{y_f}dy\int_{z_i}^{1-y}dze^{-\frac{\bar{m}^2}{T^2}}\{\frac{y+z-1}{y^2}\} \{-\frac{5 \bar{m}^2 y}{55296 \pi^6}\}   \\
\notag
\notag && +2 m_c^2 \langle\bar{q}g_s\sigma Gq\rangle \langle\bar{q}q\rangle \langle g_s^2GG\rangle \int_{y_i}^{y_f}dy\int_{z_i}^{1-y}dze^{-\frac{\bar{m}^2}{T^2}}\{-\frac{z (y+z-1)}{y^2}\} \{\frac{5 \bar{m}^2}{55296 \pi^6}\} \, ,  \\
\notag
\end{eqnarray}

\begin{eqnarray}
\notag \textbf{S}_D(12)_2&& =2m_c^2 \langle\bar{q}q\rangle^4  \int_{y_i}^{y_f}dy\int_{z_i}^{1-y}dze^{-\frac{\bar{m}^2}{T^2}}\{\frac{y+z-1}{y}\} \{\frac{25 \bar{m}^2}{62208 \pi^4}\}  \\
\notag
\notag &&+2m_c^2  \langle\bar{q}q\rangle^4 \int_{y_i}^{y_f}dy\int_{z_i}^{1-y}dze^{-\frac{\bar{m}^2}{T^2}}\{-\frac{z (y+z-1)}{y^2}\} \{-\frac{25 \bar{m}^2}{62208 \pi^4}\}   \\
\notag
\notag &&+2 \langle\bar{q}q\rangle^4 \int_{y_i}^{y_f}dy\int_{z_i}^{1-y}dze^{-\frac{\bar{m}^2}{T^2}}\{-\frac{z (y+z-1)}{y}\} \{\frac{25 \bar{m}^4 y}{93312 \pi^4}\}   \\
\notag
\notag &&+\langle\bar{q}q\rangle^4 \int_{y_i}^{y_f}dy\int_{z_i}^{1-y}dze^{-\frac{\bar{m}^2}{T^2}}\{-\frac{z (y+z-1)}{y}\} \{-\frac{25 \bar{m}^4 y}{93312 \pi^4}\} \, ,  \\
\notag
\end{eqnarray}

\begin{eqnarray}
\notag \textbf{S}_D(12)_3&& =\langle \bar{\psi}\psi\rangle^2 \langle\bar{q}q\rangle^2 \int_{y_i}^{y_f}dy\int_{z_i}^{1-y}dze^{-\frac{\bar{m}^2}{T^2}}\{-(y+z-1)^2\} \{\frac{35 \bar{m}^4}{1119744 \pi^6 }\}  \\
\notag
\notag && +\langle \bar{\psi}\psi\rangle^2 \langle\bar{q}q\rangle^2 m_c^4 \int_{y_i}^{y_f}dy\int_{z_i}^{1-y}dze^{-\frac{\bar{m}^2}{T^2}}\{-\frac{(y+z-1)^2}{y^2 z^2}\} \{\frac{5}{373248 \pi^6}\}  \\
\notag
\notag && +\langle \bar{\psi}\psi\rangle^2 \langle\bar{q}q\rangle^2 m_c^2 \int_{y_i}^{y_f}dy\int_{z_i}^{1-y}dze^{-\frac{\bar{m}^2}{T^2}}\{-\frac{(y+z-1)^2}{y z^2}\} \{-\frac{5 \bar{m}^2 y}{373248 \pi^6}\}   \\
\notag
\notag &&+\langle \bar{\psi}\psi\rangle^2 \langle\bar{q}q\rangle^2 \int_{y_i}^{y_f}dy\int_{z_i}^{1-y}dze^{-\frac{\bar{m}^2}{T^2}}\{-(y+z-1)^2\} \{-\frac{5 \bar{m}^4}{1119744 \pi^6}\}   \\
\notag
\notag &&+\langle \bar{\psi}\psi\rangle^2 \langle\bar{q}q\rangle^2 m_c^2 \int_{y_i}^{y_f}dy\int_{z_i}^{1-y}dze^{-\frac{\bar{m}^2}{T^2}}\{\frac{(y+z-1)^2}{y z}\} \{-\frac{5 \bar{m}^2}{93312 \pi^6}\}  \, , \\
\notag
\end{eqnarray}

\begin{eqnarray}
\notag \textbf{S}_D(12)_4&& =\langle g_s^2GG\rangle^3 \int_{y_i}^{y_f}dy\int_{z_i}^{1-y}dze^{-\frac{\bar{m}^2}{T^2}}\{-(y+z-1)^2\} \{-\frac{103 \bar{m}^4}{195689447424 \pi^{10}}\}  \, ,\\
\notag
\end{eqnarray}

\begin{eqnarray}
\notag \textbf{S}_D(12)_5 &&=\langle g_s^3fGGG\rangle^2 \int_{y_i}^{y_f}dy\int_{z_i}^{1-y}dze^{-\frac{\bar{m}^2}{T^2}}\{-(y+z-1)^2\} \{\frac{\bar{m}^4}{150994944 \pi^{10}}\}  \, , \\
\notag
\end{eqnarray}

\begin{eqnarray}
\notag \textbf{S}_D(14)_1&&  =2 m_c^2 \langle\bar{q}g_s\sigma Gq\rangle^2 \langle g_s^2GG\rangle \int_{y_i}^{y_f}dy\int_{z_i}^{1-y}dze^{-\frac{\bar{m}^2}{T^2}}\{-\frac{1}{y^2}\} \{-\frac{5}{221184 \pi^6}+\frac{5 \bar{m}^2 y}{442368 \pi^6  T^2}+\frac{5 y}{442368 \pi^6}\}  \\
\notag
\notag &&  +2 m_c^2 \langle\bar{q}g_s\sigma Gq\rangle^2 \langle g_s^2GG\rangle \int_{y_i}^{y_f}dy\int_{z_i}^{1-y}dze^{-\frac{\bar{m}^2}{T^2}}\{\frac{z}{y^2}\} \{-\frac{5}{442368 \pi^6}-\frac{5 \bar{m}^2}{442368 \pi^6  T^2}\}  \\
\notag
\notag && +\langle g_s^2GG\rangle \langle\bar{q}g_s\sigma Gq\rangle^2 \int_{y_i}^{y_f}dy\int_{z_i}^{1-y}dze^{-\frac{\bar{m}^2}{T^2}}\{-1\} \{\frac{5 \bar{m}^2}{589824 \pi^6}\}  \\
\notag
\notag && +\langle g_s^2GG\rangle \langle\bar{q}g_s\sigma Gq\rangle^2 m_c^2 \int_{y_i}^{y_f}dy\int_{z_i}^{1-y}dze^{-\frac{\bar{m}^2}{T^2}}\{\frac{1}{y z}\} \{-\frac{5}{294912 \pi^6}\}  \, ,\\
\notag
\end{eqnarray}

\begin{eqnarray}
\notag \textbf{S}_D(14)_2&& =\frac{1}{2}m_c^4 \langle g_s^2GG\rangle^2 \langle\bar{q}q\rangle^2  \int_{y_i}^{y_f}dy\int_{z_i}^{1-y}dze^{-\frac{\bar{m}^2}{T^2}}\{-\frac{(y+z-1)^2}{y^2 z^2}\} \{-\frac{5 \bar{m}^2}{1990656 \pi^6 T^4}\}   \\
\notag
\notag &&+m_c^2\langle g_s^2GG\rangle^2 \langle\bar{q}q\rangle^2  \int_{y_i}^{y_f}dy\int_{z_i}^{1-y}dze^{-\frac{\bar{m}^2}{T^2}}\{\frac{(y+z-1)^3}{(y-1) y^2 z^2}\} \{\frac{5}{995328 \pi^6}-\frac{5 \bar{m}^2 y}{1990656 \pi^6  T^2}\}\\
\notag && +m_c^2\langle g_s^2GG\rangle^2 \langle\bar{q}q\rangle^2  \int_{y_i}^{y_f}dy\int_{z_i}^{1-y}dze^{-\frac{\bar{m}^2}{T^2}}\{\frac{(y+z-1)^3}{(y-1) y^2 z^2}\}\{-\frac{5 y}{1990656 \pi^6}\}    \\
\notag
\notag && +m_c^2 \langle g_s^2GG\rangle^2 \langle\bar{q}q\rangle^2 \int_{y_i}^{y_f}dy\int_{z_i}^{1-y}dze^{-\frac{\bar{m}^2}{T^2}}\{\frac{(y+z-1)^2}{(y-1) y^2 z}\} \{\frac{5}{1990656 \pi^6} (\frac{\bar{m}^2}{T^2}-1)\}   \\
\notag
\notag && +\frac{1}{2}m_c^2 \langle g_s^2GG\rangle^2 \langle\bar{q}q\rangle^2 \int_{y_i}^{y_f}dy\int_{z_i}^{1-y}dze^{-\frac{\bar{m}^2}{T^2}}\{\frac{(y+z-1)^2}{y z}\} \{\frac{5}{995328 \pi^6}+\frac{5 \bar{m}^4}{1990656 \pi^6  T^4}\}\\
\notag && + \frac{1}{2}m_c^2 \langle g_s^2GG\rangle^2 \langle\bar{q}q\rangle^2 \int_{y_i}^{y_f}dy\int_{z_i}^{1-y}dze^{-\frac{\bar{m}^2}{T^2}}\{\frac{(y+z-1)^2}{y z}\}\{\frac{5 \bar{m}^2}{995328 \pi^6  T^2}\}  \\
\notag
\notag && +2 m_c^2 \langle g_s^2GG\rangle^2 \langle\bar{q}q\rangle^2 \int_{y_i}^{y_f}dy\int_{z_i}^{1-y}dze^{-\frac{\bar{m}^2}{T^2}}\{\frac{z}{y^2}\} \{\frac{1}{248832 \pi^6 }-\frac{\bar{m}^2}{82944 \pi^6  T^2}\}  \\
\notag
\notag && +2 m_c^2 \langle g_s^2GG\rangle^2 \langle\bar{q}q\rangle^2 \int_{y_i}^{y_f}dy\int_{z_i}^{1-y}dze^{-\frac{\bar{m}^2}{T^2}}\{-\frac{1}{y^2}\} \{-\frac{1}{124416 \pi^6}+\frac{\bar{m}^2 y}{248832 \pi^6  T^2}+\frac{y}{248832 \pi^6}\}  \\
\notag
\notag && +2 m_c^2 \langle g_s^2GG\rangle^2 \langle\bar{q}q\rangle^2 \int_{y_i}^{y_f}dy\int_{z_i}^{1-y}dze^{-\frac{\bar{m}^2}{T^2}}\{\frac{z}{y^2}\} \{\frac{1}{663552 \pi^6}+\frac{\bar{m}^2}{663552 \pi^6  T^2}\}  \\
\notag
\notag &&+2 m_c^2 \langle g_s^2GG\rangle^2 \langle\bar{q}q\rangle^2 \int_{y_i}^{y_f}dy\int_{z_i}^{1-y}dze^{-\frac{\bar{m}^2}{T^2}}\{-\frac{1}{y^2}\} \{\frac{1}{331776 \pi^6 }-\frac{\bar{m}^2 y}{663552 \pi^6  T^2}-\frac{y}{663552 \pi^6}\}   \\
\notag
\notag && +m_c^2\langle g_s^2GG\rangle^2  \langle\bar{q}q\rangle^2 \int_{y_i}^{y_f}dy\int_{z_i}^{1-y}dze^{-\frac{\bar{m}^2}{T^2}}\{\frac{1}{y z}\} \{-\frac{145}{42467328 \pi^6}\}  \\
\notag
\notag && +\langle g_s^2GG\rangle^2 \langle\bar{q}q\rangle^2 \int_{y_i}^{y_f}dy\int_{z_i}^{1-y}dze^{-\frac{\bar{m}^2}{T^2}}\{-1\} \{-\frac{259 \bar{m}^2}{127401984 \pi^6}\}  \\
\notag
\notag && +\langle g_s^2GG\rangle^2 \langle\bar{q}q\rangle^2 \int_{y_i}^{y_f}dy\int_{z_i}^{1-y}dze^{-\frac{\bar{m}^2}{T^2}}\{y+z-1\} \{-\frac{83 \bar{m}^2}{15925248 \pi^6}-\frac{83 \bar{m}^4}{63700992 \pi^6  T^2}\}  \\
\notag
\notag && +m_c^2 \langle g_s^2GG\rangle^2 \langle\bar{q}q\rangle^2 \int_{y_i}^{y_f}dy\int_{z_i}^{1-y}dze^{-\frac{\bar{m}^2}{T^2}}\{\frac{(y+z-1)^3}{(y-1) y^2 z^2}\} \{\frac{7}{53747712 \pi^8}-\frac{7 \bar{m}^2 y}{107495424 \pi^8  T^2}\}\\
\notag && + m_c^2\langle g_s^2GG\rangle^2 \langle\bar{q}q\rangle^2 \int_{y_i}^{y_f}dy\int_{z_i}^{1-y}dze^{-\frac{\bar{m}^2}{T^2}}\{\frac{(y+z-1)^3}{(y-1) y^2 z^2}\}\{-\frac{7 y}{107495424 \pi^8}\} \\
\notag
\notag &&+m_c^2\langle g_s^2GG\rangle^2 \langle\bar{q}q\rangle^2  \int_{y_i}^{y_f}dy\int_{z_i}^{1-y}dze^{-\frac{\bar{m}^2}{T^2}}\{\frac{(y+z-1)^2}{(y-1) y^2 z}\} \{\frac{7 \bar{m}^2}{107495424 \pi^8  T^2}-\frac{7}{107495424 \pi^8}\}   \\
\notag
\notag &&+\frac{1}{2} m_c^2\langle g_s^2GG\rangle^2 \langle\bar{q}q\rangle^2  \int_{y_i}^{y_f}dy\int_{z_i}^{1-y}dze^{-\frac{\bar{m}^2}{T^2}}\{\frac{(y+z-1)^2}{y z}\} \{\frac{7}{53747712 \pi^8}+\frac{7 \bar{m}^4}{107495424 \pi^8  T^4}\}\\
\notag && +\frac{1}{2}m_c^2  \langle g_s^2GG\rangle^2 \langle\bar{q}q\rangle^2 \int_{y_i}^{y_f}dy\int_{z_i}^{1-y}dze^{-\frac{\bar{m}^2}{T^2}}\{\frac{(y+z-1)^2}{y z}\}\{\frac{7 \bar{m}^2}{53747712 \pi^8  T^2}\}   \\
\notag
\notag && +\frac{1}{2}m_c^4  \langle g_s^2GG\rangle^2 \langle\bar{q}q\rangle^2 \int_{y_i}^{y_f}dy\int_{z_i}^{1-y}dze^{-\frac{\bar{m}^2}{T^2}}\{-\frac{(y+z-1)^2}{y^2 z^2}\} \{-\frac{7 \bar{m}^2}{107495424 \pi^8  T^4}\} \, , \\
\notag
\end{eqnarray}

\begin{eqnarray}
\notag \textbf{S}_D(14)_3&& =2 m_c^2 \langle\bar{q}q\rangle^3 \langle\bar{q}g_s\sigma Gq\rangle \int_{y_i}^{y_f}dy\int_{z_i}^{1-y}dze^{-\frac{\bar{m}^2}{T^2}}\{-\frac{1}{y}\} \{-\frac{25}{124416 \pi^4}-\frac{25 \bar{m}^2}{124416 \pi^4  T^2}\}  \\
\notag
\notag &&+2 m_c^2 \langle\bar{q}q\rangle^3 \langle\bar{q}g_s\sigma Gq\rangle \int_{y_i}^{y_f}dy\int_{z_i}^{1-y}dze^{-\frac{\bar{m}^2}{T^2}}\{\frac{z}{y^2}\} \{\frac{25}{124416 \pi^4}+\frac{25 \bar{m}^2}{124416 \pi^4  T^2}\}   \\
\notag
\notag && +2 \langle\bar{q}g_s\sigma Gq\rangle \langle\bar{q}q\rangle^3 \int_{y_i}^{y_f}dy\int_{z_i}^{1-y}dze^{-\frac{\bar{m}^2}{T^2}}\{\frac{z}{y}\} \{\frac{25 \bar{m}^2}{93312 \pi^4}-\frac{25 \bar{m}^4 y}{186624 \pi^4  T^2}-\frac{25 \bar{m}^2 y}{46656 \pi^4}\}  \\
\notag
\notag && +2 \langle\bar{q}g_s\sigma Gq\rangle \langle\bar{q}q\rangle^3 \int_{y_i}^{y_f}dy\int_{z_i}^{1-y}dze^{-\frac{\bar{m}^2}{T^2}}\{\frac{z}{y}\} \{-\frac{25 \bar{m}^2}{373248 \pi^4}+\frac{25 \bar{m}^4 y}{186624 \pi^4  T^2}+\frac{25 \bar{m}^2 y}{46656 \pi^4}\}  \, ,\\
\notag
\end{eqnarray}

\begin{eqnarray}
\notag \textbf{S}_D(14)_4&& =m_c^4\langle\bar{q}g_s\sigma Gq\rangle \langle\bar{q}q\rangle \langle \bar{\psi}\psi\rangle^2  \int_{y_i}^{y_f}dy\int_{z_i}^{1-y}dze^{-\frac{\bar{m}^2}{T^2}}\{\frac{y+z-1}{y^2 z^2}\} \{-\frac{5}{373248 \pi^6  T^2}\}  \\
\notag
\notag &&+2m_c^2  \langle\bar{q}g_s\sigma Gq\rangle \langle\bar{q}q\rangle \langle \bar{\psi}\psi\rangle^2 \int_{y_i}^{y_f}dy\int_{z_i}^{1-y}dze^{-\frac{\bar{m}^2}{T^2}}\{\frac{y+z-1}{y z^2}\} \{-\frac{5}{373248 \pi^6}+\frac{5 \bar{m}^2 y}{746496 \pi^6  T^2}\}\\
\notag && +2m_c^2  \langle\bar{q}g_s\sigma Gq\rangle \langle\bar{q}q\rangle \langle \bar{\psi}\psi\rangle^2 \int_{y_i}^{y_f}dy\int_{z_i}^{1-y}dze^{-\frac{\bar{m}^2}{T^2}}\{\frac{y+z-1}{y z^2}\} \{\frac{5 y}{746496 \pi^6}\}   \\
\notag
\notag && +\langle\bar{q}g_s\sigma Gq\rangle \langle\bar{q}q\rangle \langle \bar{\psi}\psi\rangle^2 m_c^2 \int_{y_i}^{y_f}dy\int_{z_i}^{1-y}dze^{-\frac{\bar{m}^2}{T^2}}\{-\frac{y+z-1}{y z}\} \{\frac{5}{93312 \pi^6 }+\frac{5 \bar{m}^2}{93312 \pi^6  T^2}\}  \\
\notag
\notag && +\langle\bar{q}g_s\sigma Gq\rangle \langle\bar{q}q\rangle \langle \bar{\psi}\psi\rangle^2 \int_{y_i}^{y_f}dy\int_{z_i}^{1-y}dze^{-\frac{\bar{m}^2}{T^2}}\{\frac{(y+z-1)^2}{(y-1) y z}\}\{-\frac{5 \bar{m}^2 y}{559872 \pi^6}\}  \\
\notag
\notag &&+\langle\bar{q}g_s\sigma Gq\rangle \langle\bar{q}q\rangle \langle \bar{\psi}\psi\rangle^2 \int_{y_i}^{y_f}dy\int_{z_i}^{1-y}dze^{-\frac{\bar{m}^2}{T^2}}\{\frac{y+z-1}{(y-1) y}\} \{\frac{5 \bar{m}^2}{559872 \pi^6}\}   \\
\notag
\notag &&+\langle\bar{q}g_s\sigma Gq\rangle \langle\bar{q}q\rangle \langle \bar{\psi}\psi\rangle^2 \int_{y_i}^{y_f}dy\int_{z_i}^{1-y}dze^{-\frac{\bar{m}^2}{T^2}}\{y+z-1\} \{\frac{5 \bar{m}^2}{279936 \pi^6}+\frac{5 \bar{m}^4}{1119744 \pi^6  T^2}\}   \\
\notag
\notag && +\langle\bar{q}g_s\sigma Gq\rangle \langle\bar{q}q\rangle \langle \bar{\psi}\psi\rangle^2 \int_{y_i}^{y_f}dy\int_{z_i}^{1-y}dze^{-\frac{\bar{m}^2}{T^2}}\{\frac{(y+z-1)^2}{(y-1) y z}\} \{\frac{35 \bar{m}^2 y}{2239488 \pi^6}\}  \\
\notag
\notag && +\langle\bar{q}g_s\sigma Gq\rangle \langle\bar{q}q\rangle \langle \bar{\psi}\psi\rangle^2 \int_{y_i}^{y_f}dy\int_{z_i}^{1-y}dze^{-\frac{\bar{m}^2}{T^2}}\{\frac{y+z-1}{(y-1) y}\} \{-\frac{35 \bar{m}^2}{2239488 \pi^6}\}  \\
\notag
\notag &&  +\langle\bar{q}g_s\sigma Gq\rangle \langle\bar{q}q\rangle \langle \bar{\psi}\psi\rangle^2 \int_{y_i}^{y_f}dy\int_{z_i}^{1-y}dze^{-\frac{\bar{m}^2}{T^2}}\{y+z-1\} \{-\frac{35 \bar{m}^2}{279936 \pi^6}-\frac{35 \bar{m}^4}{1119744 \pi^6  T^2}\} \, , \\
\notag
\end{eqnarray}

\begin{eqnarray}
\notag \textbf{S}_D(16)_1&& =\frac{1}{2}m_c^2 \langle\bar{q}q\rangle \langle\bar{q}g_s\sigma Gq\rangle \langle g_s^2GG\rangle^2  \int_{y_i}^{y_f}dy\int_{z_i}^{1-y}dze^{-\frac{\bar{m}^2}{T^2}}\{-\frac{(y+z-1)^2}{(y-1) y^2 z^2}\} \{\frac{5 \bar{m}^2 y}{995328 \pi^6  T^4}\}\\
\notag && +\frac{1}{2}m_c^2 \langle\bar{q}q\rangle \langle\bar{q}g_s\sigma Gq\rangle \langle g_s^2GG\rangle^2  \int_{y_i}^{y_f}dy\int_{z_i}^{1-y}dze^{-\frac{\bar{m}^2}{T^2}}\{-\frac{(y+z-1)^2}{(y-1) y^2 z^2}\}\{-\frac{5}{497664 \pi^6  T^2}\}  \\
\notag
\notag && +\frac{1}{2}m_c^2 \langle\bar{q}q\rangle \langle\bar{q}g_s\sigma Gq\rangle \langle g_s^2GG\rangle^2  \int_{y_i}^{y_f}dy\int_{z_i}^{1-y}dze^{-\frac{\bar{m}^2}{T^2}}\{\frac{y+z-1}{y^2 z(1-y)}\} \{\frac{5}{497664 \pi^6 T^2}-\frac{5 \bar{m}^2}{995328 \pi^6  T^4}\}  \\
\notag
\notag &&+\frac{1}{6}m_c^2 \langle\bar{q}q\rangle \langle\bar{q}g_s\sigma Gq\rangle \langle g_s^2GG\rangle^2  \int_{y_i}^{y_f}dy\int_{z_i}^{1-y}dze^{-\frac{\bar{m}^2}{T^2}}\{-\frac{y+z-1}{y z}\} \{-\frac{5 \bar{m}^4}{663552 \pi^6 T^6}\}   \\
\notag
\notag && +\frac{1}{6} m_c^4 \langle\bar{q}q\rangle \langle\bar{q}g_s\sigma Gq\rangle \langle g_s^2GG\rangle^2 \int_{y_i}^{y_f}dy\int_{z_i}^{1-y}dze^{-\frac{\bar{m}^2}{T^2}}\{\frac{y+z-1}{y^2 z^2}\}\{\frac{5 \bar{m}^2}{663552 \pi^6 T^6}-\frac{5}{663552 \pi^6  T^4}\}  \\
\notag
\notag && +\frac{1}{2} \langle\bar{q}q\rangle \langle\bar{q}g_s\sigma Gq\rangle \langle g_s^2GG\rangle^2 \int_{y_i}^{y_f}dy\int_{z_i}^{1-y}dze^{-\frac{\bar{m}^2}{T^2}}\{-1\} \{\frac{83}{31850496 \pi^6}+\frac{83 \bar{m}^4}{63700992 \pi^6  T^4}\}\\
\notag && +\frac{1}{2} \langle\bar{q}q\rangle \langle\bar{q}g_s\sigma Gq\rangle \langle g_s^2GG\rangle^2 \int_{y_i}^{y_f}dy\int_{z_i}^{1-y}dze^{-\frac{\bar{m}^2}{T^2}}\{-1\} \{\frac{83 \bar{m}^2}{31850496 \pi^6  T^2}\} \, , \\
\notag
\end{eqnarray}

\begin{eqnarray}
\notag \textbf{S}_D(16)_2 &&=\frac{1}{2}  m_c^4\langle\bar{q}g_s\sigma Gq\rangle^2 \langle \bar{\psi}\psi\rangle^2 \int_{y_i}^{y_f}dy\int_{z_i}^{1-y}dze^{-\frac{\bar{m}^2}{T^2}}\{-\frac{1}{y^2 z^2}\} \{\frac{5}{1492992 \pi^6  T^4}\}   \\
\notag
\notag && +\frac{1}{2} m_c^2 \langle\bar{q}g_s\sigma Gq\rangle^2 \langle \bar{\psi}\psi\rangle^2 \int_{y_i}^{y_f}dy\int_{z_i}^{1-y}dze^{-\frac{\bar{m}^2}{T^2}}\{-\frac{1}{y z^2}\} \{\frac{5}{1492992 \pi^6 T^2}-\frac{5 \bar{m}^2 y}{2985984 \pi^6  T^4}\}  \\
\notag
\notag && +\frac{1}{2} m_c^2 \langle\bar{q}g_s\sigma Gq\rangle^2 \langle \bar{\psi}\psi\rangle^2 \int_{y_i}^{y_f}dy\int_{z_i}^{1-y}dze^{-\frac{\bar{m}^2}{T^2}}\{-\frac{1}{y z^2}\} \{\frac{5}{1492992 \pi^6 T^2}-\frac{5 \bar{m}^2 y}{2985984 \pi^6  T^4}\}  \\
\notag
\notag &&+\frac{1}{2}m_c^2 \langle\bar{q}g_s\sigma Gq\rangle^2 \langle \bar{\psi}\psi\rangle^2  \int_{y_i}^{y_f}dy\int_{z_i}^{1-y}dze^{-\frac{\bar{m}^2}{T^2}}\{\frac{1}{y z}\} \{\frac{5 \bar{m}^2}{373248 \pi^6  T^4}\}   \\
\notag
\notag && +\langle\bar{q}g_s\sigma Gq\rangle^2 \langle \bar{\psi}\psi\rangle^2 \int_{y_i}^{y_f}dy\int_{z_i}^{1-y}dze^{-\frac{\bar{m}^2}{T^2}}\{\frac{y+z-1}{y z(1-y)}\} \{-\frac{5}{2239488 \pi^6}+\frac{5 \bar{m}^2 y}{4478976 \pi^6  T^2}+\frac{5 y}{4478976 \pi^6}\}  \\
\notag
\notag && +\langle\bar{q}g_s\sigma Gq\rangle^2 \langle \bar{\psi}\psi\rangle^2 \int_{y_i}^{y_f}dy\int_{z_i}^{1-y}dze^{-\frac{\bar{m}^2}{T^2}}\{\frac{1}{y-y^2}\} \{\frac{5}{4478976 \pi^6}-\frac{5 \bar{m}^2}{4478976 \pi^6  T^2}\}  \\
\notag
\notag && +\frac{1}{2} \langle\bar{q}g_s\sigma Gq\rangle^2 \langle \bar{\psi}\psi\rangle^2 \int_{y_i}^{y_f}dy\int_{z_i}^{1-y}dze^{-\frac{\bar{m}^2}{T^2}}\{-1\} \{-\frac{5}{2239488 \pi^6}-\frac{5 \bar{m}^4}{4478976 \pi^6  T^4}-\frac{5 \bar{m}^2}{2239488 \pi^6  T^2}\}  \\
\notag
\notag && +\langle\bar{q}g_s\sigma Gq\rangle^2 \langle \bar{\psi}\psi\rangle^2 \int_{y_i}^{y_f}dy\int_{z_i}^{1-y}dze^{-\frac{\bar{m}^2}{T^2}}\{\frac{y+z-1}{y z(1-y)}\} \{\frac{35}{8957952 \pi^6}-\frac{35 \bar{m}^2 y}{17915904 \pi^6  T^2}-\frac{35 y}{17915904 \pi^6}\}  \\
\notag
\notag && +\langle\bar{q}g_s\sigma Gq\rangle^2 \langle \bar{\psi}\psi\rangle^2 \int_{y_i}^{y_f}dy\int_{z_i}^{1-y}dze^{-\frac{\bar{m}^2}{T^2}}\{\frac{1}{y-y^2}\} \{\frac{35 \bar{m}^2}{17915904 \pi^6 T^2}-\frac{35}{17915904 \pi^6}\}  \\
\notag
\notag && +\frac{1}{2} \langle\bar{q}g_s\sigma Gq\rangle^2 \langle \bar{\psi}\psi\rangle^2 \int_{y_i}^{y_f}dy\int_{z_i}^{1-y}dze^{-\frac{\bar{m}^2}{T^2}}\{-1\} \{\frac{35}{2239488 \pi^6}+\frac{35 \bar{m}^4}{4478976 \pi^6  T^4}+\frac{35 \bar{m}^2}{2239488 \pi^6  T^2}\} \, . \\
\notag
\end{eqnarray}

\clearpage
======================================================

Type A for $J_2$

\begin{eqnarray}
\notag \textbf{S}_A(12)_1&&        =m_c^2 \langle \bar{q}q\rangle^4 \int_{4m_c^2}^{s_0}ds\int_{y_i}^{y_f}dye^{-\frac{s}{T^2}} \{\frac{1}{144 \pi^2}\} \\
\notag &&        +\langle \bar{q}q\rangle^4  \int_{4m_c^2}^{s_0}ds\int_{y_i}^{y_f}dye^{-\frac{s}{T^2}} \{(y-1) y\}\{\frac{\tilde{m}^2-s}{72 \pi^2}-\frac{s}{144 \pi^2}\} \\
\notag &&        +  \langle \bar{q}q\rangle^4 \int_{4m_c^2}^{s_0}ds\int_{y_i}^{y_f}dye^{-\frac{s}{T^2}} \{(y-1) y\} \{\frac{\tilde{m}^2-s}{1944 \pi^4}-\frac{s}{3888 \pi^4}\} \\
\notag &&        +m_c^2 \langle \bar{q}q\rangle^4   \int_{4m_c^2}^{s_0}ds\int_{y_i}^{y_f}dye^{-\frac{s}{T^2}} \{\frac{1}{3888 \pi^4}\} \, ,\\
\notag
\end{eqnarray}

\begin{eqnarray}
\notag \textbf{S}_A(12)_1&&        =  m_c^2 \langle \bar{q}q\rangle \langle \bar{q}g_s\sigma Gq\rangle  \langle g_s^2 GG\rangle \int_{4m_c^2}^{s_0}ds\int_{y_i}^{y_f}dye^{-\frac{s}{T^2}} \{-\frac{1}{12288 \pi^6}\} \\
\notag &&        +  \langle \bar{q}q\rangle \langle \bar{q}g_s\sigma Gq\rangle  \langle g_s^2 GG\rangle  \int_{4m_c^2}^{s_0}ds\int_{y_i}^{y_f}dye^{-\frac{s}{T^2}}\{(y-1) y\} \{\frac{s}{12288 \pi^6}-\frac{\tilde{m}^2-s}{6144 \pi^6}\} \\
\notag &&        +m_c^2 \langle \bar{q}g_s\sigma Gq\rangle  \langle \bar{q}q\rangle \langle g_s^2 GG\rangle  \int_{4m_c^2}^{s_0}ds\int_{y_i}^{y_f}dye^{-\frac{s}{T^2}} \{-\frac{1}{36864 \pi^6}\} \\
\notag &&        +\langle \bar{q}g_s\sigma Gq\rangle  \langle \bar{q}q\rangle \langle g_s^2 GG\rangle \int_{4m_c^2}^{s_0}ds\int_{y_i}^{y_f}dye^{-\frac{s}{T^2}} \{(y-1) y\} \{\frac{s}{36864 \pi^6}-\frac{\tilde{m}^2-s}{18432 \pi^6}\} \\
\, ,\notag
\end{eqnarray}

\begin{eqnarray}
\notag \textbf{S}_A(14)_1&&        =\langle \bar{q}q\rangle^3 \langle \bar{q}g_s\sigma Gq\rangle  \int_{4m_c^2}^{s_0}ds\int_{y_i}^{y_f}dye^{-\frac{s}{T^2}} \{(y-1) y\} \{\frac{1}{24 \pi^2}\} \\
\notag &&        +\langle \bar{q}q\rangle^3 \langle \bar{q}g_s\sigma Gq\rangle  \int_{4m_c^2}^{s_0}ds\int_{y_i}^{y_f}dye^{-\frac{s}{T^2}} \{(y-1) y\} \{\frac{1}{1296 \pi^4}\}\, . \\
\notag
\end{eqnarray}

Type B for $J_2$

\begin{eqnarray}
\notag \textbf{S}_B(0)&&        =m_c^2 \int_{4m_c^2}^{s_0}ds\int_{y_i}^{y_f}dy\int_{z_i}^{1-y}dze^{-\frac{s}{T^2}} \{-(y+z-1)^5\} \{\frac{(s-\bar{m}^2)^6}{39321600 \pi^{10}}\} \\
\notag &&        +\int_{4m_c^2}^{s_0}ds\int_{y_i}^{y_f}dy\int_{z_i}^{1-y}dze^{-\frac{s}{T^2}} \{-y z (y+z-1)^5\} \{\frac{(s-\bar{m}^2)^7}{137625600 \pi^{10}}+\frac{s (s-\bar{m}^2)^6}{39321600 \pi^{10}}\}\, , \\
\notag
\end{eqnarray}

\begin{eqnarray}
\notag \textbf{S}_B(4)&&        =m_c^2 \langle g_s^2 GG\rangle \int_{4m_c^2}^{s_0}ds\int_{y_i}^{y_f}dy\int_{z_i}^{1-y}dze^{-\frac{s}{T^2}} \{\frac{(y+z-1)^5}{y^2}\} \{\frac{(y-1) (s-\bar{m}^2)^4}{23592960 \pi^{10}}+\frac{s y (s-\bar{m}^2)^3}{11796480 \pi^{10}}\} \\
\notag &&        +m_c^2 \langle g_s^2 GG\rangle \int_{4m_c^2}^{s_0}ds\int_{y_i}^{y_f}dy\int_{z_i}^{1-y}dze^{-\frac{s}{T^2}} \{-\frac{z (y+z-1)^5}{y^2}\} \{-\frac{(s-\bar{m}^2)^4}{23592960 \pi^{10}}-\frac{s (s-\bar{m}^2)^3}{11796480 \pi^{10}}\} \\
\notag &&        +  m_c^2 \langle g_s^2 GG\rangle \int_{4m_c^2}^{s_0}ds\int_{y_i}^{y_f}dy\int_{z_i}^{1-y}dze^{-\frac{s}{T^2}} \{-(y+z-1)^3\} \{\frac{(s-\bar{m}^2)^4}{1572864 \pi^{10}}\} \\
\notag &&        +  \langle g_s^2 GG\rangle \int_{4m_c^2}^{s_0}ds\int_{y_i}^{y_f}dy\int_{z_i}^{1-y}dze^{-\frac{s}{T^2}} \{-y z (y+z-1)^3\} \{\frac{(s-\bar{m}^2)^5}{3932160 \pi^{10}}+\frac{s (s-\bar{m}^2)^4}{1572864 \pi^{10}}\} \, ,\\
\notag
\end{eqnarray}

\begin{eqnarray}
\notag \textbf{S}_B(6)&&       =m_c^2 \langle \bar{q}q\rangle^2 \int_{4m_c^2}^{s_0}ds\int_{y_i}^{y_f}dy\int_{z_i}^{1-y}dze^{-\frac{s}{T^2}} \{(y+z-1)^2\} \{\frac{(s-\bar{m}^2)^3}{1536 \pi^6}\} \\
\notag &&        +\langle \bar{q}q\rangle^2 \int_{4m_c^2}^{s_0}ds\int_{y_i}^{y_f}dy\int_{z_i}^{1-y}dze^{-\frac{s}{T^2}} \{y z (y+z-1)^2\} \{\frac{(s-\bar{m}^2)^4}{3072 \pi^6}+\frac{s (s-\bar{m}^2)^3}{1536 \pi^6}\} \\
\notag &&        +m_c^2   \langle \bar{q}q\rangle^2 \int_{4m_c^2}^{s_0}ds\int_{y_i}^{y_f}dy\int_{z_i}^{1-y}dze^{-\frac{s}{T^2}} \{(y+z-1)^2\} \{\frac{(s-\bar{m}^2)^3}{27648 \pi^8}\} \\
\notag &&        +  \langle \bar{q}q\rangle^2 \int_{4m_c^2}^{s_0}ds\int_{y_i}^{y_f}dy\int_{z_i}^{1-y}dze^{-\frac{s}{T^2}} \{y z (y+z-1)^2\} \{\frac{(s-\bar{m}^2)^4}{165888 \pi^8}+\frac{s (s-\bar{m}^2)^3}{82944 \pi^8}\} \\
\notag &&        +m_c^2   \langle \bar{q}q\rangle^2 \int_{4m_c^2}^{s_0}ds\int_{y_i}^{y_f}dy\int_{z_i}^{1-y}dze^{-\frac{s}{T^2}} \{\frac{(y+z-1)^4}{y}\} \{-\frac{(s-\bar{m}^2)^3}{331776 \pi^8}-\frac{s (s-\bar{m}^2)^2}{221184 \pi^8}\} \\
\notag &&        +m_c^2   \langle \bar{q}q\rangle^2 \int_{4m_c^2}^{s_0}ds\int_{y_i}^{y_f}dy\int_{z_i}^{1-y}dze^{-\frac{s}{T^2}} \{\frac{z (y+z-1)^4}{y^2}\} \{-\frac{(s-\bar{m}^2)^3}{331776 \pi^8}-\frac{s (s-\bar{m}^2)^2}{221184 \pi^8}\} \\
\notag &&        +   \langle \bar{q}q\rangle^2 \int_{4m_c^2}^{s_0}ds\int_{y_i}^{y_f}dy\int_{z_i}^{1-y}dze^{-\frac{s}{T^2}} \{\frac{z (y+z-1)^4}{y}\} \{\frac{s^2 y (s-\bar{m}^2)^2}{331776 \pi^8}+\frac{y (s-\bar{m}^2)^4}{663552 \pi^8}\}\\
\notag && +\langle \bar{q}q\rangle^2 \int_{4m_c^2}^{s_0}ds\int_{y_i}^{y_f}dy\int_{z_i}^{1-y}dze^{-\frac{s}{T^2}} \{\frac{z (y+z-1)^4}{y}\}\{\frac{s y (s-\bar{m}^2)^3}{165888 \pi^8}-\frac{(s-\bar{m}^2)^4}{995328 \pi^8}-\frac{s (s-\bar{m}^2)^3}{497664 \pi^8}\} \\
\notag &&        +  \langle \bar{q}q\rangle^2 \int_{4m_c^2}^{s_0}ds\int_{y_i}^{y_f}dy\int_{z_i}^{1-y}dze^{-\frac{s}{T^2}} \{\frac{z (y+z-1)^4}{y}\} \{-\frac{s^2 y (s-\bar{m}^2)^2}{331776 \pi^8}-\frac{y (s-\bar{m}^2)^4}{663552 \pi^8}\}\\
\notag &&       +\langle \bar{q}q\rangle^2 \int_{4m_c^2}^{s_0}ds\int_{y_i}^{y_f}dy\int_{z_i}^{1-y}dze^{-\frac{s}{T^2}} \{\frac{z (y+z-1)^4}{y}\}\{-\frac{s y (s-\bar{m}^2)^3}{165888 \pi^8}+\frac{(s-\bar{m}^2)^4}{3981312 \pi^8}+\frac{s (s-\bar{m}^2)^3}{1990656 \pi^8}\} \, ,\\
\notag
\end{eqnarray}

\begin{eqnarray}
\notag \textbf{S}_B(8) &&        =m_c^2 \langle \bar{q}q\rangle \langle \bar{q}g_s\sigma Gq\rangle  \int_{4m_c^2}^{s_0}ds\int_{y_i}^{y_f}dy\int_{z_i}^{1-y}dze^{-\frac{s}{T^2}} \{-y-z+1\} \{-\frac{(s-\bar{m}^2)^2}{512 \pi^6}\} \\
\notag &&        +\langle \bar{q}q\rangle \langle \bar{q}g_s\sigma Gq\rangle  \int_{4m_c^2}^{s_0}ds\int_{y_i}^{y_f}dy\int_{z_i}^{1-y}dze^{-\frac{s}{T^2}} \{-y z (y+z-1)\} \{-\frac{(s-\bar{m}^2)^3}{768 \pi^6}-\frac{s (s-\bar{m}^2)^2}{512 \pi^6}\} \, ,\\
\notag
\end{eqnarray}

\begin{eqnarray}
\notag \textbf{S}_B(10)_1&&        =m_c^2 \langle \bar{q}g_s\sigma Gq\rangle ^2 \int_{4m_c^2}^{s_0}ds\int_{y_i}^{y_f}dy\int_{z_i}^{1-y}dze^{-\frac{s}{T^2}} \{-\frac{\bar{m}^2-s}{2048 \pi^6}\} \\
\notag &&        +\langle \bar{q}g_s\sigma Gq\rangle ^2 \int_{4m_c^2}^{s_0}ds\int_{y_i}^{y_f}dy\int_{z_i}^{1-y}dze^{-\frac{s}{T^2}} \{y z\} \{\frac{(s-\bar{m}^2)^2}{2048 \pi^6}+\frac{s (s-\bar{m}^2}{2048 \pi^6}\} \\
\notag &&        +\langle \bar{q}g_s\sigma Gq\rangle ^2 \int_{4m_c^2}^{s_0}ds\int_{y_i}^{y_f}dy\int_{z_i}^{1-y}dze^{-\frac{s}{T^2}} \{y z\} \{\frac{(s-\bar{m}^2)^2}{98304 \pi^6}+\frac{s (s-\bar{m}^2)}{98304 \pi^6}\} \\
\notag &&        +m_c^2 \langle \bar{q}g_s\sigma Gq\rangle ^2 \int_{4m_c^2}^{s_0}ds\int_{y_i}^{y_f}dy\int_{z_i}^{1-y}dze^{-\frac{s}{T^2}} \{-\frac{\bar{m}^2-s}{98304 \pi^6}\} \, ,\\
\notag
\end{eqnarray}

\begin{eqnarray}
\notag \textbf{S}_B(10)_2&&        =m_c^2 \langle \bar{q}q\rangle^2 \langle g_s^2 GG\rangle \int_{4m_c^2}^{s_0}ds\int_{y_i}^{y_f}dy\int_{z_i}^{1-y}dze^{-\frac{s}{T^2}} \{-\frac{(y+z-1)^2}{y^2}\} \{\frac{s y}{9216 \pi^6}-\frac{(y-1) (\bar{m}^2-s)}{4608 \pi^6}\} \\
\notag &&        +m_c^2 \langle \bar{q}q\rangle^2 \langle g_s^2 GG\rangle \int_{4m_c^2}^{s_0}ds\int_{y_i}^{y_f}dy\int_{z_i}^{1-y}dze^{-\frac{s}{T^2}} \{\frac{z (y+z-1)^2}{y^2}\} \{\frac{\bar{m}^2-s}{4608 \pi^6}-\frac{s}{9216 \pi^6}\} \\
\notag &&       +  m_c^2 \langle \bar{q}q\rangle^2 \langle g_s^2 GG\rangle \int_{4m_c^2}^{s_0}ds\int_{y_i}^{y_f}dy\int_{z_i}^{1-y}dze^{-\frac{s}{T^2}} \{-\frac{\bar{m}^2-s}{6144 \pi^6}\} \\
\notag &&        +  \langle \bar{q}q\rangle^2 \langle g_s^2 GG\rangle \int_{4m_c^2}^{s_0}ds\int_{y_i}^{y_f}dy\int_{z_i}^{1-y}dze^{-\frac{s}{T^2}} \{y z\} \{\frac{(s-\bar{m}^2)^2}{6144 \pi^6}+\frac{s (s-\bar{m}^2}{6144 \pi^6}\} \\
\notag &&        +m_c^2 \langle \bar{q}q\rangle^2 \langle g_s^2 GG\rangle \int_{4m_c^2}^{s_0}ds\int_{y_i}^{y_f}dy\int_{z_i}^{1-y}dze^{-\frac{s}{T^2}} \{-\frac{\bar{m}^2-s}{9216 \pi^6}\} \\
\notag &&        +\langle \bar{q}q\rangle^2 \langle g_s^2 GG\rangle \int_{4m_c^2}^{s_0}ds\int_{y_i}^{y_f}dy\int_{z_i}^{1-y}dze^{-\frac{s}{T^2}} \{y z\} \{\frac{(s-\bar{m}^2)^2}{9216 \pi^6}+\frac{s (s-\bar{m}^2)}{9216 \pi^6}\} \\
\notag &&        +m_c^2   \langle g_s^2 GG\rangle \langle \bar{q}q\rangle^2 \int_{4m_c^2}^{s_0}ds\int_{y_i}^{y_f}dy\int_{z_i}^{1-y}dze^{-\frac{s}{T^2}} \{-\frac{(y+z-1)^2}{y^2}\} \{\frac{(1-y) (\bar{m}^2-s)}{248832 \pi^8}+\frac{s y}{497664 \pi^8}\} \\
\notag &&        +m_c^2   \langle g_s^2 GG\rangle \langle \bar{q}q\rangle^2 \int_{4m_c^2}^{s_0}ds\int_{y_i}^{y_f}dy\int_{z_i}^{1-y}dze^{-\frac{s}{T^2}} \{\frac{z (y+z-1)^2}{y^2}\} \{\frac{\bar{m}^2-s}{248832 \pi^8}-\frac{s}{497664 \pi^8}\} \, ,\\
\notag
\end{eqnarray}

\begin{eqnarray}
\notag \textbf{S}_B(12)_1&&        =m_c^2 \langle \bar{q}q\rangle^4   \int_{4m_c^2}^{s_0}ds\int_{y_i}^{y_f}dy\int_{z_i}^{1-y}dze^{-\frac{s}{T^2}} \{\frac{y+z-1}{y}\} \{\frac{1}{1296 \pi^4}\} \\
\notag &&        +m_c^2   \langle \bar{q}q\rangle^4 \int_{4m_c^2}^{s_0}ds\int_{y_i}^{y_f}dy\int_{z_i}^{1-y}dze^{-\frac{s}{T^2}} \{-\frac{z (y+z-1)}{y^2}\} \{-\frac{1}{1296 \pi^4}\} \\
\notag &&        +  \langle \bar{q}q\rangle^4 \int_{4m_c^2}^{s_0}ds\int_{y_i}^{y_f}dy\int_{z_i}^{1-y}dze^{-\frac{s}{T^2}} \{-\frac{z (y+z-1)}{y}\} \{-\frac{\bar{m}^2 y}{648 \pi^4}+\frac{\bar{m}^2}{972 \pi^4}+\frac{s y}{324 \pi^4}-\frac{s}{648 \pi^4}\}\\
\notag &&        +  \langle \bar{q}q\rangle^4 \int_{4m_c^2}^{s_0}ds\int_{y_i}^{y_f}dy\int_{z_i}^{1-y}dze^{-\frac{s}{T^2}} \{-\frac{z (y+z-1)}{y}\} \{\frac{\bar{m}^2 y}{648 \pi^4}-\frac{\bar{m}^2}{3888 \pi^4}-\frac{s y}{324 \pi^4}+\frac{s}{2592 \pi^4}\} \, ,\\
\notag
\end{eqnarray}

\begin{eqnarray}
\notag \textbf{S}_B(12)_2&&        =m_c^2 \langle \bar{q}g_s\sigma Gq\rangle  \langle \bar{q}q\rangle \langle g_s^2 GG\rangle \int_{4m_c^2}^{s_0}ds\int_{y_i}^{y_f}dy\int_{z_i}^{1-y}dze^{-\frac{s}{T^2}} \{\frac{y+z-1}{y^2}\} \{\frac{1-y}{4608 \pi^6}\} \\
\notag &&        +m_c^2 \langle \bar{q}g_s\sigma Gq\rangle  \langle \bar{q}q\rangle \langle g_s^2 GG\rangle \int_{4m_c^2}^{s_0}ds\int_{y_i}^{y_f}dy\int_{z_i}^{1-y}dze^{-\frac{s}{T^2}} \{-\frac{z (y+z-1)}{y^2}\} \{\frac{1}{4608 \pi^6}\} \, .\\
\notag
\end{eqnarray}

 Type C for $J_2$

\begin{eqnarray}
\notag \textbf{S}_C(14)&&         =m_c^2 \langle \bar{q}q\rangle^3 \langle \bar{q}g_s\sigma Gq\rangle  \int_{y_i}^{y_f}dye^{-\frac{\tilde{m}^2}{T^2}} \{-\frac{1}{144 \pi^2}-\frac{\tilde{m}^2}{144 \pi^2 T^2}\} \\
\notag &&         +\langle \bar{q}q\rangle^3 \langle \bar{q}g_s\sigma Gq\rangle  \int_{y_i}^{y_f}dye^{-\frac{\tilde{m}^2}{T^2}} \{(y-1) y\} \{\frac{\tilde{m}^2}{36 \pi^2}+\frac{\tilde{m}^4}{144 \pi^2 T^2}\} \\
\notag &&         +m_c^2 \langle \bar{q}q\rangle^3 \langle \bar{q}g_s\sigma Gq\rangle  \int_{y_i}^{y_f}dye^{-\frac{\tilde{m}^2}{T^2}}  \{-\frac{1}{7776 \pi^4}-\frac{\tilde{m}^2}{7776 \pi^4 T^2}\} \\
\notag &&         +\langle \bar{q}q\rangle^3 \langle \bar{q}g_s\sigma Gq\rangle  \int_{y_i}^{y_f}dye^{-\frac{\tilde{m}^2}{T^2}} \{(y-1) y\} \{\frac{\tilde{m}^2}{1944 \pi^4}+\frac{\tilde{m}^4}{7776 \pi^4 T^2}\}\, , \\
\notag
\end{eqnarray}

\begin{eqnarray}
\notag \textbf{S}_C(16)_1&&         =\langle \bar{q}q\rangle^2 \langle \bar{q}g_s\sigma Gq\rangle ^2 \int_{y_i}^{y_f}dye^{-\frac{\tilde{m}^2}{T^2}} \{(y-1) y\} \{-\frac{1}{64 \pi^2}-\frac{\tilde{m}^6}{384 \pi^2 T^6}-\frac{\tilde{m}^4}{128 \pi^2 T^4}-\frac{\tilde{m}^2}{64 \pi^2 T^2}\} \\
\notag &&         +m_c^2 \langle \bar{q}q\rangle^2 \langle \bar{q}g_s\sigma Gq\rangle ^2 \int_{y_i}^{y_f}dye^{-\frac{\tilde{m}^2}{T^2}} \{\frac{\tilde{m}^4}{384 \pi^2 T^6}\} \\
\notag &&         +m_c^2 \langle \bar{q}q\rangle^2 \langle \bar{q}g_s\sigma Gq\rangle ^2 \int_{y_i}^{y_f}dye^{-\frac{\tilde{m}^2}{T^2}}  \{\frac{\tilde{m}^4}{62208 \pi^4 T^6}\} \\
\notag &&         +\langle \bar{q}q\rangle^2 \langle \bar{q}g_s\sigma Gq\rangle ^2 \int_{y_i}^{y_f}dye^{-\frac{\tilde{m}^2}{T^2}} \{(y-1) y\} \{-\frac{1}{10368 \pi^4}-\frac{\tilde{m}^6}{62208 \pi^4 T^6}-\frac{\tilde{m}^4}{20736 \pi^4 T^4}-\frac{\tilde{m}^2}{10368 \pi^4 T^2}\} \, ,\\
\notag
\end{eqnarray}

\begin{eqnarray}
\notag \textbf{S}_C(16)_2&&         =m_c^2 \langle g_s^2 GG\rangle \langle \bar{q}q\rangle^4 \int_{y_i}^{y_f}dye^{-\frac{\tilde{m}^2}{T^2}} \{\frac{1}{y^2}\} \{\frac{1}{1728 \pi^2 T^2}\} \\
\notag &&         +m_c^4 \langle g_s^2 GG\rangle \langle \bar{q}q\rangle^4 \int_{y_i}^{y_f}dye^{-\frac{\tilde{m}^2}{T^2}} \{\frac{1}{y^3}\} \{-\frac{1}{5184 \pi^2 T^4}\} \\
\notag &&         +m_c^2 \langle g_s^2 GG\rangle \langle \bar{q}q\rangle^4 \int_{y_i}^{y_f}dye^{-\frac{\tilde{m}^2}{T^2}} \{\frac{y-1}{y^2}\} \{\frac{\tilde{m}^2}{5184 \pi^2 T^4}\} \\
\notag &&         +m_c^2 \langle \bar{q}q\rangle^4 \langle g_s^2 GG\rangle \int_{y_i}^{y_f}dye^{-\frac{\tilde{m}^2}{T^2}} \{\frac{\tilde{m}^4}{5184 \pi^2 T^6}\} \\
\notag &&         +\langle \bar{q}q\rangle^4 \langle g_s^2 GG\rangle \int_{y_i}^{y_f}dye^{-\frac{\tilde{m}^2}{T^2}} \{(y-1) y\} \{-\frac{1}{864 \pi^2}-\frac{\tilde{m}^6}{5184 \pi^2 T^6}-\frac{\tilde{m}^4}{1728 \pi^2 T^4}-\frac{\tilde{m}^2}{864 \pi^2 T^2}\}\, .\\
\notag
\end{eqnarray}

Type D for $J_2$

\begin{eqnarray}
\notag \textbf{S}_D(12)_1&&           =m_c^2 \langle \bar{q}g_s\sigma Gq\rangle  \langle \bar{q}q\rangle \langle g_s^2 GG\rangle \int_{y_i}^{y_f}dy\int_{z_i}^{1-y}dze^{-\frac{\bar{m}^2}{T^2}} \{\frac{y+z-1}{y^2}\} \{-\frac{\bar{m}^2 y}{9216 \pi^6}\} \\
\notag &&            +m_c^2 \langle \bar{q}g_s\sigma Gq\rangle  \langle \bar{q}q\rangle \langle g_s^2 GG\rangle \int_{y_i}^{y_f}dy\int_{z_i}^{1-y}dze^{-\frac{\bar{m}^2}{T^2}} \{-\frac{z (y+z-1)}{y^2}\} \{\frac{\bar{m}^2}{9216 \pi^6}\} \, ,\\
\notag
\end{eqnarray}

\begin{eqnarray}
\notag \textbf{S}_D(12)_2&&            =m_c^2 \langle \bar{q}q\rangle^4   \int_{y_i}^{y_f}dy\int_{z_i}^{1-y}dze^{-\frac{\bar{m}^2}{T^2}} \{\frac{y+z-1}{y}\} \{\frac{\bar{m}^2}{2592 \pi^4}\} \\
\notag &&            +m_c^2   \langle \bar{q}q\rangle^4 \int_{y_i}^{y_f}dy\int_{z_i}^{1-y}dze^{-\frac{\bar{m}^2}{T^2}} \{-\frac{z (y+z-1)}{y^2}\} \{-\frac{\bar{m}^2}{2592 \pi^4}\} \\
\notag &&            +  \langle \bar{q}q\rangle^4 \int_{y_i}^{y_f}dy\int_{z_i}^{1-y}dze^{-\frac{\bar{m}^2}{T^2}} \{-\frac{z (y+z-1)}{y}\} \{\frac{\bar{m}^4 y}{3888 \pi^4}\} \\
\notag &&            +  \langle \bar{q}q\rangle^4 \int_{y_i}^{y_f}dy\int_{z_i}^{1-y}dze^{-\frac{\bar{m}^2}{T^2}} \{-\frac{z (y+z-1)}{y}\} \{-\frac{\bar{m}^4 y}{3888 \pi^4}\} \, .\\
\notag
\end{eqnarray}

\clearpage
======================================================

Type A for $J_3$

\begin{eqnarray}
\notag  \textbf{S}_A(12)&&        =\langle \bar{q}q\rangle^4 \int_{4m_c^2}^{s_0}ds\int_{y_i}^{y_f}dye^{-\frac{s}{T^2}} \{(y-1) y\} \{\frac{\tilde{m}^2-s}{216 \pi^2}-\frac{s}{432 \pi^2}\} \\
\notag  &&        +\langle \bar{q}q\rangle^4 \int_{4m_c^2}^{s_0}ds\int_{y_i}^{y_f}dye^{-\frac{s}{T^2}} \{(y-1) y\} \{\frac{\tilde{m}^2-s}{432 \pi^2}-\frac{s}{216 \pi^2}\} \\
\notag  &&        -m_c^2 \langle \bar{q}q\rangle^4 \int_{4m_c^2}^{s_0}ds\int_{y_i}^{y_f}dye^{-\frac{s}{T^2}} \{\frac{1}{144 \pi^2}\} \\
\notag  &&        +\langle \bar{q}q\rangle^4 \int_{4m_c^2}^{s_0}ds\int_{y_i}^{y_f}dye^{-\frac{s}{T^2}} \{(y-1) y\} \{\frac{\tilde{m}^2-s}{11664 \pi^4}-\frac{s}{5832 \pi^4}\} \\
\notag  &&        -m_c^2 \langle \bar{q}q\rangle^4 \int_{4m_c^2}^{s_0}ds\int_{y_i}^{y_f}dye^{-\frac{s}{T^2}}  \{\frac{1}{3888 \pi^4}\} \\
\notag  &&        +\langle \bar{q}q\rangle^4 \int_{4m_c^2}^{s_0}ds\int_{y_i}^{y_f}dye^{-\frac{s}{T^2}} \{(y-1) y\} \{\frac{\tilde{m}^2-s}{5832 \pi^4}-\frac{s}{11664 \pi^4}\} \, ,\\
\notag
\end{eqnarray}

\begin{eqnarray}
\notag  \textbf{S}_A(14)&&        =\langle \bar{q}g_s\sigma Gq\rangle \langle \bar{q}q\rangle^3 \int_{4m_c^2}^{s_0}ds\int_{y_i}^{y_f}dye^{-\frac{s}{T^2}} \{(y-1) y\} \{\frac{1}{72 \pi^2}\} \\
\notag  &&        +\langle \bar{q}g_s\sigma Gq\rangle \langle \bar{q}q\rangle^3 \int_{4m_c^2}^{s_0}ds\int_{y_i}^{y_f}dye^{-\frac{s}{T^2}} \{(y-1) y\} \{\frac{1}{144 \pi^2}\} \\
\notag  &&        +\langle \bar{q}q\rangle^3 \langle \bar{q}g_s\sigma Gq\rangle \int_{4m_c^2}^{s_0}ds\int_{y_i}^{y_f}dye^{-\frac{s}{T^2}} \{(y-1) y\} \{\frac{1}{7776 \pi^4}\} \\
\notag  &&        +\langle \bar{q}q\rangle^3 \langle \bar{q}g_s\sigma Gq\rangle \int_{4m_c^2}^{s_0}ds\int_{y_i}^{y_f}dye^{-\frac{s}{T^2}} \{(y-1) y\} \{\frac{1}{3888 \pi^4}\} \, .\\
\notag
\end{eqnarray}

Type B for $J_3$

\begin{eqnarray}
\notag \textbf{S}_B(0) &&        =\int_{4m_c^2}^{s_0}ds\int_{y_i}^{y_f}dy\int_{z_i}^{1-y}dze^{-\frac{s}{T^2}} \{-y z (y+z-1)^5\} \{\frac{(s-\bar{m}^2)^7}{412876800 \pi^{10}}+\frac{s (s-\bar{m}^2)^6}{117964800 \pi^{10}}\} \\
\notag  &&        +\int_{4m_c^2}^{s_0}ds\int_{y_i}^{y_f}dy\int_{z_i}^{1-y}dze^{-\frac{s}{T^2}} \{-y z (y+z-1)^5\} \{\frac{(s-\bar{m}^2)^7}{825753600 \pi^{10}}+\frac{s (s-\bar{m}^2)^6}{58982400 \pi^{10}}\} \\
\notag  &&        -m_c^2 \int_{4m_c^2}^{s_0}ds\int_{y_i}^{y_f}dy\int_{z_i}^{1-y}dze^{-\frac{s}{T^2}} \{-(y+z-1)^5\} \{\frac{(s-\bar{m}^2)^6}{39321600 \pi^{10}}\} \, ,\\
\notag
\end{eqnarray}

\begin{eqnarray}
\notag \textbf{S}_B(4) &&         =m_c^2 \langle g_s^2 GG\rangle \int_{4m_c^2}^{s_0}ds\int_{y_i}^{y_f}dy\int_{z_i}^{1-y}dze^{-\frac{s}{T^2}} \{-\frac{z (y+z-1)^5}{y^2}\} \{-\frac{(s-\bar{m}^2)^4}{141557760 \pi^{10}}-\frac{s (s-\bar{m}^2)^3}{17694720 \pi^{10}}\} \\
\notag  &&         -m_c^2 \langle g_s^2 GG\rangle \int_{4m_c^2}^{s_0}ds\int_{y_i}^{y_f}dy\int_{z_i}^{1-y}dze^{-\frac{s}{T^2}} \{\frac{(y+z-1)^5}{y^2}\} \{\frac{s y (s-\bar{m}^2)^3}{11796480 \pi^{10}}-\frac{(1-y) (s-\bar{m}^2)^4}{23592960 \pi^{10}}\} \\
\notag  &&         +m_c^2 \langle g_s^2 GG\rangle \int_{4m_c^2}^{s_0}ds\int_{y_i}^{y_f}dy\int_{z_i}^{1-y}dze^{-\frac{s}{T^2}} \{-\frac{z (y+z-1)^5}{y^2}\} \{-\frac{(s-\bar{m}^2)^4}{70778880 \pi^{10}}-\frac{s (s-\bar{m}^2)^3}{35389440 \pi^{10}}\} \\
\notag  &&         +\langle g_s^2 GG\rangle \int_{4m_c^2}^{s_0}ds\int_{y_i}^{y_f}dy\int_{z_i}^{1-y}dze^{-\frac{s}{T^2}} \{-y z (y+z-1)^3\} \{\frac{(s-\bar{m}^2)^5}{23592960 \pi^{10}}+\frac{s (s-\bar{m}^2)^4}{2359296 \pi^{10}}\} \\
\notag  &&         -m_c^2 \langle g_s^2 GG\rangle \int_{4m_c^2}^{s_0}ds\int_{y_i}^{y_f}dy\int_{z_i}^{1-y}dze^{-\frac{s}{T^2}} \{-(y+z-1)^3\} \{\frac{(s-\bar{m}^2)^4}{1572864 \pi^{10}}\} \\
\notag  &&         +\langle g_s^2 GG\rangle \int_{4m_c^2}^{s_0}ds\int_{y_i}^{y_f}dy\int_{z_i}^{1-y}dze^{-\frac{s}{T^2}} \{-y z (y+z-1)^3\} \{\frac{(s-\bar{m}^2)^5}{11796480 \pi^{10}}+\frac{s (s-\bar{m}^2)^4}{4718592 \pi^{10}}\} \, ,\\
\notag
\end{eqnarray}

\begin{eqnarray}
\notag \textbf{S}_B(6) &&            =\langle \bar{q}q\rangle^2 \int_{4m_c^2}^{s_0}ds\int_{y_i}^{y_f}dy\int_{z_i}^{1-y}dze^{-\frac{s}{T^2}} \{y z (y+z-1)^2\} \{\frac{(s-\bar{m}^2)^4}{9216 \pi^6}+\frac{s (s-\bar{m}^2)^3}{4608 \pi^6}\} \\
\notag  &&            +\langle \bar{q}q\rangle^2 \int_{4m_c^2}^{s_0}ds\int_{y_i}^{y_f}dy\int_{z_i}^{1-y}dze^{-\frac{s}{T^2}} \{y z (y+z-1)^2\} \{\frac{(s-\bar{m}^2)^4}{18432 \pi^6}+\frac{s (s-\bar{m}^2)^3}{2304 \pi^6}\} \\
\notag  &&            -m_c^2 \langle \bar{q}q\rangle^2 \int_{4m_c^2}^{s_0}ds\int_{y_i}^{y_f}dy\int_{z_i}^{1-y}dze^{-\frac{s}{T^2}} \{(y+z-1)^2\} \{\frac{(s-\bar{m}^2)^3}{1536 \pi^6}\} \\
\notag  &&            +\langle \bar{q}q\rangle^2 \int_{4m_c^2}^{s_0}ds\int_{y_i}^{y_f}dy\int_{z_i}^{1-y}dze^{-\frac{s}{T^2}} \{y z (y+z-1)^2\} \{\frac{(s-\bar{m}^2)^4}{995328 \pi^8}+\frac{s (s-\bar{m}^2)^3}{124416 \pi^8}\} \\
\notag  &&            -m_c^2 \langle \bar{q}q\rangle^2 \int_{4m_c^2}^{s_0}ds\int_{y_i}^{y_f}dy\int_{z_i}^{1-y}dze^{-\frac{s}{T^2}} \{(y+z-1)^2\} \{\frac{(s-\bar{m}^2)^3}{82944 \pi^8}\} \\
\notag  &&            +\langle \bar{q}q\rangle^2 \int_{4m_c^2}^{s_0}ds\int_{y_i}^{y_f}dy\int_{z_i}^{1-y}dze^{-\frac{s}{T^2}} \{y z (y+z-1)^2\} \{\frac{(s-\bar{m}^2)^4}{497664 \pi^8}+\frac{s (s-\bar{m}^2)^3}{248832 \pi^8}\}\, , \\
\notag
\end{eqnarray}

\begin{eqnarray}
\notag \textbf{S}_B(8) &&               =\langle \bar{q}g_s\sigma Gq\rangle \langle \bar{q}q\rangle \int_{4m_c^2}^{s_0}ds\int_{y_i}^{y_f}dy\int_{z_i}^{1-y}dze^{-\frac{s}{T^2}} \{-y z (y+z-1)\} \{-\frac{(s-\bar{m}^2)^3}{2304 \pi^6}-\frac{s (s-\bar{m}^2)^2}{1536 \pi^6}\} \\
\notag  &&               +\langle \bar{q}g_s\sigma Gq\rangle \langle \bar{q}q\rangle \int_{4m_c^2}^{s_0}ds\int_{y_i}^{y_f}dy\int_{z_i}^{1-y}dze^{-\frac{s}{T^2}} \{-y z (y+z-1)\} \{-\frac{(s-\bar{m}^2)^3}{4608 \pi^6}-\frac{s (s-\bar{m}^2)^2}{768 \pi^6}\} \\
\notag  &&               -m_c^2 \langle \bar{q}g_s\sigma Gq\rangle \langle \bar{q}q\rangle \int_{4m_c^2}^{s_0}ds\int_{y_i}^{y_f}dy\int_{z_i}^{1-y}dze^{-\frac{s}{T^2}} \{-y-z+1\} \{-\frac{(s-\bar{m}^2)^2}{512 \pi^6}\}\, , \\
\notag
\end{eqnarray}

\begin{eqnarray}
\notag \textbf{S}_B(10)_1 &&               =m_c^2 \langle \bar{q}q\rangle^2 \langle g_s^2 GG\rangle \int_{4m_c^2}^{s_0}ds\int_{y_i}^{y_f}dy\int_{z_i}^{1-y}dze^{-\frac{s}{T^2}} \{\frac{z (y+z-1)^2}{y^2}\} \{\frac{\bar{m}^2-s}{27648 \pi^6}-\frac{s}{13824 \pi^6}\} \\
\notag  &&               -m_c^2 \langle g_s^2 GG\rangle \langle \bar{q}q\rangle^2 \int_{4m_c^2}^{s_0}ds\int_{y_i}^{y_f}dy\int_{z_i}^{1-y}dze^{-\frac{s}{T^2}} \{-\frac{(y+z-1)^2}{y^2}\} \{\frac{(1-y) (\bar{m}^2-s)}{4608 \pi^6}+\frac{s y}{9216 \pi^6}\} \\
\notag  &&               +m_c^2 \langle g_s^2 GG\rangle \langle \bar{q}q\rangle^2 \int_{4m_c^2}^{s_0}ds\int_{y_i}^{y_f}dy\int_{z_i}^{1-y}dze^{-\frac{s}{T^2}} \{\frac{z (y+z-1)^2}{y^2}\} \{\frac{\bar{m}^2-s}{13824 \pi^6}-\frac{s}{27648 \pi^6}\} \\
\notag  &&               +\langle \bar{q}q\rangle^2 \langle g_s^2 GG\rangle \int_{4m_c^2}^{s_0}ds\int_{y_i}^{y_f}dy\int_{z_i}^{1-y}dze^{-\frac{s}{T^2}} \{y z\} \{\frac{(s-\bar{m}^2)^2}{36864 \pi^6}+\frac{s (s-\bar{m}^2)}{9216 \pi^6}\} \\
\notag  &&                -m_c^2 \langle g_s^2 GG\rangle \langle \bar{q}q\rangle^2 \int_{4m_c^2}^{s_0}ds\int_{y_i}^{y_f}dy\int_{z_i}^{1-y}dze^{-\frac{s}{T^2}} \{-\frac{\bar{m}^2-s}{6144 \pi^6}\} \\
\notag  &&               +\langle g_s^2 GG\rangle \langle \bar{q}q\rangle^2 \int_{4m_c^2}^{s_0}ds\int_{y_i}^{y_f}dy\int_{z_i}^{1-y}dze^{-\frac{s}{T^2}} \{y z\} \{\frac{(s-\bar{m}^2)^2}{18432 \pi^6}+\frac{s (s-\bar{m}^2)}{18432 \pi^6}\} \\
\notag  &&               +\langle \bar{q}q\rangle^2 \langle g_s^2 GG\rangle \int_{4m_c^2}^{s_0}ds\int_{y_i}^{y_f}dy\int_{z_i}^{1-y}dze^{-\frac{s}{T^2}} \{y z\} \{\frac{(s-\bar{m}^2)^2}{55296 \pi^6}+\frac{s (s-\bar{m}^2)}{13824 \pi^6}\} \\
\notag  &&                -m_c^2 \langle \bar{q}q\rangle^2 \langle g_s^2 GG\rangle \int_{4m_c^2}^{s_0}ds\int_{y_i}^{y_f}dy\int_{z_i}^{1-y}dze^{-\frac{s}{T^2}} \{-\frac{\bar{m}^2-s}{9216 \pi^6}\} \\
\notag  &&               +\langle \bar{q}q\rangle^2 \langle g_s^2 GG\rangle \int_{4m_c^2}^{s_0}ds\int_{y_i}^{y_f}dy\int_{z_i}^{1-y}dze^{-\frac{s}{T^2}} \{y z\} \{\frac{(s-\bar{m}^2)^2}{27648 \pi^6}+\frac{s (s-\bar{m}^2)}{27648 \pi^6}\}\, , \\
\notag
\end{eqnarray}

\begin{eqnarray}
\notag \textbf{S}_B(10)_2 &&                       +\langle \bar{q}g_s\sigma Gq\rangle^2 \int_{4m_c^2}^{s_0}ds\int_{y_i}^{y_f}dy\int_{z_i}^{1-y}dze^{-\frac{s}{T^2}} \{(y z\} \{\frac{(s-\bar{m}^2)^2}{6144 \pi^6}+\frac{s (s-\bar{m}^2)}{6144 \pi^6}\} \\
\notag  &&                       +\langle \bar{q}g_s\sigma Gq\rangle^2 \int_{4m_c^2}^{s_0}ds\int_{y_i}^{y_f}dy\int_{z_i}^{1-y}dze^{-\frac{s}{T^2}} \{y z\} \{\frac{(s-\bar{m}^2)^2}{12288 \pi^6}+\frac{s (s-\bar{m}^2)}{3072 \pi^6}\} \\
\notag  &&                       -m_c^2 \langle \bar{q}g_s\sigma Gq\rangle^2 \int_{4m_c^2}^{s_0}ds\int_{y_i}^{y_f}dy\int_{z_i}^{1-y}dze^{-\frac{s}{T^2}} \{-\frac{\bar{m}^2-s}{2048 \pi^6}\} \\
\notag  &&                       +\langle \bar{q}g_s\sigma Gq\rangle^2 \int_{4m_c^2}^{s_0}ds\int_{y_i}^{y_f}dy\int_{z_i}^{1-y}dze^{-\frac{s}{T^2}} \{y z\} \{\frac{(s-\bar{m}^2)^2}{589824 \pi^6}+\frac{s (s-\bar{m}^2)}{147456 \pi^6}\} \\
\notag  &&                       -m_c^2 \langle \bar{q}g_s\sigma Gq\rangle^2 \int_{4m_c^2}^{s_0}ds\int_{y_i}^{y_f}dy\int_{z_i}^{1-y}dze^{-\frac{s}{T^2}} \{-\frac{\bar{m}^2-s}{98304 \pi^6}\} \\
\notag  &&                       +\langle \bar{q}g_s\sigma Gq\rangle^2 \int_{4m_c^2}^{s_0}ds\int_{y_i}^{y_f}dy\int_{z_i}^{1-y}dze^{-\frac{s}{T^2}} \{y z\} \{\frac{(s-\bar{m}^2)^2}{294912 \pi^6}+\frac{s (s-\bar{m}^2)}{294912 \pi^6}\}\, . \\
\notag
\end{eqnarray}

Type C for $J_3$

\begin{eqnarray}
\notag \textbf{S}_C(14) &&                  =\langle \bar{q}g_s\sigma Gq\rangle \langle \bar{q}q\rangle^3 \int_{y_i}^{y_f}dye^{-\frac{\tilde{m}^2}{T^2}} \{(y-1) y\} \{\frac{\tilde{m}^2}{108 \pi^2}+\frac{\tilde{m}^4}{432 \pi^2 T^2}\} \\
\notag  &&                  +\langle \bar{q}g_s\sigma Gq\rangle \langle \bar{q}q\rangle^3 \int_{y_i}^{y_f}dye^{-\frac{\tilde{m}^2}{T^2}} \{(y-1) y\} \{\frac{5 \tilde{m}^2}{432 \pi^2}+\frac{\tilde{m}^4}{216 \pi^2 T^2}\} \\
\notag  &&                  -m_c^2 \langle \bar{q}g_s\sigma Gq\rangle \langle \bar{q}q\rangle^3 \int_{y_i}^{y_f}dye^{-\frac{\tilde{m}^2}{T^2}} \{-\frac{1}{144 \pi^2}-\frac{\tilde{m}^2}{144 \pi^2 T^2}\} \\
\notag  &&                  +\langle \bar{q}q\rangle^3 \langle \bar{q}g_s\sigma Gq\rangle \int_{y_i}^{y_f}dye^{-\frac{\tilde{m}^2}{T^2}} \{(y-1) y\} \{\frac{5 \tilde{m}^2}{23328 \pi^4}+\frac{\tilde{m}^4}{11664 \pi^4 T^2}\} \\
\notag  &&                  -m_c^2 \langle \bar{q}q\rangle^3 \langle \bar{q}g_s\sigma Gq\rangle \int_{y_i}^{y_f}dye^{-\frac{\tilde{m}^2}{T^2}} \{-\frac{1}{7776 \pi^4}-\frac{\tilde{m}^2}{7776 \pi^4 T^2}\} \\
\notag  &&                  +\langle \bar{q}q\rangle^3 \langle \bar{q}g_s\sigma Gq\rangle \int_{y_i}^{y_f}dye^{-\frac{\tilde{m}^2}{T^2}} \{(y-1) y\} \{\frac{\tilde{m}^2}{5832 \pi^4}+\frac{\tilde{m}^4}{23328 \pi^4 T^2}\} \, ,\\
\notag
\end{eqnarray}

\begin{eqnarray}
\notag \textbf{S}_C(16)_1 &&                  =\langle \bar{q}g_s\sigma Gq\rangle^2 \langle \bar{q}q\rangle^2 \int_{y_i}^{y_f}dye^{-\frac{\tilde{m}^2}{T^2}} \{(y-1) y\} \{-\frac{1}{192 \pi^2}-\frac{\tilde{m}^6}{1152 \pi^2 T^6}-\frac{\tilde{m}^4}{384 \pi^2 T^4}-\frac{\tilde{m}^2}{192 \pi^2 T^2}\} \\
\notag  &&                  +\langle \bar{q}g_s\sigma Gq\rangle^2 \langle \bar{q}q\rangle^2 \int_{y_i}^{y_f}dye^{-\frac{\tilde{m}^2}{T^2}} \{(y-1) y\} \{\frac{1}{192 \pi^2}-\frac{\tilde{m}^6}{576 \pi^2 T^6}-\frac{\tilde{m}^4}{384 \pi^2 T^4}\} \\
\notag  &&                  -m_c^2 \langle \bar{q}g_s\sigma Gq\rangle^2 \langle \bar{q}q\rangle^2 \int_{y_i}^{y_f}dye^{-\frac{\tilde{m}^2}{T^2}} \{\frac{\tilde{m}^4}{64 \pi^2 T^6}\} \\
\notag  &&                  +\langle \bar{q}q\rangle^2 \langle \bar{q}g_s\sigma Gq\rangle^2 \int_{y_i}^{y_f}dye^{-\frac{\tilde{m}^2}{T^2}} \{(y-1) y\} \{\frac{1}{31104 \pi^4}-\frac{\tilde{m}^6}{93312 \pi^4 T^6}-\frac{\tilde{m}^4}{62208 \pi^4 T^4}\} \\
\notag  &&                  -m_c^2 \langle \bar{q}q\rangle^2 \langle \bar{q}g_s\sigma Gq\rangle^2 \int_{y_i}^{y_f}dye^{-\frac{\tilde{m}^2}{T^2}} \{\frac{\tilde{m}^4}{62208 \pi^4 T^6}\} \\
\notag  &&                  +\langle \bar{q}q\rangle^2 \langle \bar{q}g_s\sigma Gq\rangle^2 \int_{y_i}^{y_f}dye^{-\frac{\tilde{m}^2}{T^2}} \{(y-1) y\} \{-\frac{1}{31104 \pi^4}-\frac{\tilde{m}^6}{186624 \pi^4 T^6}-\frac{\tilde{m}^4}{62208 \pi^4 T^4}-\frac{\tilde{m}^2}{31104 \pi^4 T^2}\} \, ,\\
\notag
\end{eqnarray}

\begin{eqnarray}
\notag \textbf{S}_C(16)_2 &&                   =m_c^2 \langle \bar{q}q\rangle^4 \langle g_s^2 GG\rangle \int_{y_i}^{y_f}dye^{-\frac{\tilde{m}^2}{T^2}} \{\frac{y-1}{y^2}\} \{\frac{\tilde{m}^2}{7776 \pi^2 T^4}-\frac{1}{5184 \pi^2 T^2}\} \\
\notag  &&                     -m_c^2 \langle \bar{q}q\rangle^4 \langle g_s^2 GG\rangle \int_{y_i}^{y_f}dye^{-\frac{\tilde{m}^2}{T^2}} \{\frac{1}{y^2}\} \{\frac{1}{1728 \pi^2 T^2}\} \\
\notag  &&                   -m_c^4 \langle \bar{q}q\rangle^4 \langle g_s^2 GG\rangle \int_{y_i}^{y_f}dye^{-\frac{\tilde{m}^2}{T^2}} \{\frac{1}{y^3}\} \{-\frac{1}{5184 \pi^2 T^4}\} \\
\notag  &&                   +m_c^2 \langle g_s^2 GG\rangle \langle \bar{q}q\rangle^4 \int_{y_i}^{y_f}dye^{-\frac{\tilde{m}^2}{T^2}} \{\frac{y-1}{y^2}\} \{\frac{\tilde{m}^2}{15552 \pi^2 T^4}\} \\
\notag  &&                   +\langle \bar{q}q\rangle^4 \langle g_s^2 GG\rangle \int_{y_i}^{y_f}dye^{-\frac{\tilde{m}^2}{T^2}} \{(y-1) y\} \{\frac{1}{2592 \pi^2}-\frac{\tilde{m}^6}{7776 \pi^2 T^6}-\frac{\tilde{m}^4}{5184 \pi^2 T^4}\} \\
\notag  &&                   -m_c^2 \langle \bar{q}q\rangle^4 \langle g_s^2 GG\rangle \int_{y_i}^{y_f}dye^{-\frac{\tilde{m}^2}{T^2}} \{\frac{\tilde{m}^4}{5184 \pi^2 T^6}\} \\
\notag  &&                   +\langle \bar{q}q\rangle^4 \langle g_s^2 GG\rangle \int_{y_i}^{y_f}dye^{-\frac{\tilde{m}^2}{T^2}} \{(y-1) y\} \{-\frac{1}{2592 \pi^2}-\frac{\tilde{m}^6}{15552 \pi^2 T^6}-\frac{\tilde{m}^4}{5184 \pi^2 T^4}-\frac{\tilde{m}^2}{2592 \pi^2 T^2}\} \, .\\
\notag
\end{eqnarray}

\clearpage
======================================================

Type A for $J_4$

\begin{eqnarray}
\notag  \textbf{S}_A(12)&&        =\langle \bar{q}q\rangle^4 \int_{4m_c^2}^{s_0}ds\int_{y_i}^{y_f}dye^{-\frac{s}{T^2}} \{(y-1) y\} \{\frac{\tilde{m}^2-s}{216 \pi^2}-\frac{s}{432 \pi^2}\} \\
\notag  &&        +\langle \bar{q}q\rangle^4 \int_{4m_c^2}^{s_0}ds\int_{y_i}^{y_f}dye^{-\frac{s}{T^2}} \{(y-1) y\} \{\frac{\tilde{m}^2-s}{432 \pi^2}-\frac{s}{216 \pi^2}\} \\
\notag  &&        +m_c^2 \langle \bar{q}q\rangle^4 \int_{4m_c^2}^{s_0}ds\int_{y_i}^{y_f}dye^{-\frac{s}{T^2}} \{\frac{1}{144 \pi^2}\} \\
\notag  &&        +\langle \bar{q}q\rangle^4 \int_{4m_c^2}^{s_0}ds\int_{y_i}^{y_f}dye^{-\frac{s}{T^2}} \{(y-1) y\} \{\frac{\tilde{m}^2-s}{11664 \pi^4}-\frac{s}{5832 \pi^4}\} \\
\notag  &&        +m_c^2 \langle \bar{q}q\rangle^4 \int_{4m_c^2}^{s_0}ds\int_{y_i}^{y_f}dye^{-\frac{s}{T^2}}  \{\frac{1}{3888 \pi^4}\} \\
\notag  &&        +\langle \bar{q}q\rangle^4 \int_{4m_c^2}^{s_0}ds\int_{y_i}^{y_f}dye^{-\frac{s}{T^2}} \{(y-1) y\} \{\frac{\tilde{m}^2-s}{5832 \pi^4}-\frac{s}{11664 \pi^4}\} \, ,\\
\notag
\end{eqnarray}

\begin{eqnarray}
\notag  \textbf{S}_A(14)&&        =\langle \bar{q}g_s\sigma Gq\rangle \langle \bar{q}q\rangle^3 \int_{4m_c^2}^{s_0}ds\int_{y_i}^{y_f}dye^{-\frac{s}{T^2}} \{(y-1) y\} \{\frac{1}{72 \pi^2}\} \\
\notag  &&        +\langle \bar{q}g_s\sigma Gq\rangle \langle \bar{q}q\rangle^3 \int_{4m_c^2}^{s_0}ds\int_{y_i}^{y_f}dye^{-\frac{s}{T^2}} \{(y-1) y\} \{\frac{1}{144 \pi^2}\} \\
\notag  &&        +\langle \bar{q}q\rangle^3 \langle \bar{q}g_s\sigma Gq\rangle \int_{4m_c^2}^{s_0}ds\int_{y_i}^{y_f}dye^{-\frac{s}{T^2}} \{(y-1) y\} \{\frac{1}{7776 \pi^4}\} \\
\notag  &&        +\langle \bar{q}q\rangle^3 \langle \bar{q}g_s\sigma Gq\rangle \int_{4m_c^2}^{s_0}ds\int_{y_i}^{y_f}dye^{-\frac{s}{T^2}} \{(y-1) y\} \{\frac{1}{3888 \pi^4}\} \, .\\
\notag
\end{eqnarray}

Type B for $J_4$

\begin{eqnarray}
\notag \textbf{S}_B(0) &&        =\int_{4m_c^2}^{s_0}ds\int_{y_i}^{y_f}dy\int_{z_i}^{1-y}dze^{-\frac{s}{T^2}} \{-y z (y+z-1)^5\} \{\frac{(s-\bar{m}^2)^7}{412876800 \pi^{10}}+\frac{s (s-\bar{m}^2)^6}{117964800 \pi^{10}}\} \\
\notag  &&        +\int_{4m_c^2}^{s_0}ds\int_{y_i}^{y_f}dy\int_{z_i}^{1-y}dze^{-\frac{s}{T^2}} \{-y z (y+z-1)^5\} \{\frac{(s-\bar{m}^2)^7}{825753600 \pi^{10}}+\frac{s (s-\bar{m}^2)^6}{58982400 \pi^{10}}\} \\
\notag  &&        +m_c^2 \int_{4m_c^2}^{s_0}ds\int_{y_i}^{y_f}dy\int_{z_i}^{1-y}dze^{-\frac{s}{T^2}} \{-(y+z-1)^5\} \{\frac{(s-\bar{m}^2)^6}{39321600 \pi^{10}}\}\, , \\
\notag
\end{eqnarray}

\begin{eqnarray}
\notag \textbf{S}_B(4) &&         =m_c^2 \langle g_s^2 GG\rangle \int_{4m_c^2}^{s_0}ds\int_{y_i}^{y_f}dy\int_{z_i}^{1-y}dze^{-\frac{s}{T^2}} \{-\frac{z (y+z-1)^5}{y^2}\} \{-\frac{(s-\bar{m}^2)^4}{141557760 \pi^{10}}-\frac{s (s-\bar{m}^2)^3}{17694720 \pi^{10}}\} \\
\notag  &&         +m_c^2 \langle g_s^2 GG\rangle \int_{4m_c^2}^{s_0}ds\int_{y_i}^{y_f}dy\int_{z_i}^{1-y}dze^{-\frac{s}{T^2}} \{\frac{(y+z-1)^5}{y^2}\} \{\frac{s y (s-\bar{m}^2)^3}{11796480 \pi^{10}}-\frac{(1-y) (s-\bar{m}^2)^4}{23592960 \pi^{10}}\} \\
\notag  &&         +m_c^2 \langle g_s^2 GG\rangle \int_{4m_c^2}^{s_0}ds\int_{y_i}^{y_f}dy\int_{z_i}^{1-y}dze^{-\frac{s}{T^2}} \{-\frac{z (y+z-1)^5}{y^2}\} \{-\frac{(s-\bar{m}^2)^4}{70778880 \pi^{10}}-\frac{s (s-\bar{m}^2)^3}{35389440 \pi^{10}}\} \\
\notag  &&         +\langle g_s^2 GG\rangle \int_{4m_c^2}^{s_0}ds\int_{y_i}^{y_f}dy\int_{z_i}^{1-y}dze^{-\frac{s}{T^2}} \{-y z (y+z-1)^3\} \{\frac{(s-\bar{m}^2)^5}{23592960 \pi^{10}}+\frac{s (s-\bar{m}^2)^4}{2359296 \pi^{10}}\} \\
\notag  &&         +m_c^2 \langle g_s^2 GG\rangle \int_{4m_c^2}^{s_0}ds\int_{y_i}^{y_f}dy\int_{z_i}^{1-y}dze^{-\frac{s}{T^2}} \{-(y+z-1)^3\} \{\frac{(s-\bar{m}^2)^4}{1572864 \pi^{10}}\} \\
\notag  &&         +\langle g_s^2 GG\rangle \int_{4m_c^2}^{s_0}ds\int_{y_i}^{y_f}dy\int_{z_i}^{1-y}dze^{-\frac{s}{T^2}} \{-y z (y+z-1)^3\} \{\frac{(s-\bar{m}^2)^5}{11796480 \pi^{10}}+\frac{s (s-\bar{m}^2)^4}{4718592 \pi^{10}}\} \, ,\\
\notag
\end{eqnarray}

\begin{eqnarray}
\notag \textbf{S}_B(6)_1 &&            =\langle \bar{q}q\rangle^2 \int_{4m_c^2}^{s_0}ds\int_{y_i}^{y_f}dy\int_{z_i}^{1-y}dze^{-\frac{s}{T^2}} \{y z (y+z-1)^2\} \{\frac{(s-\bar{m}^2)^4}{9216 \pi^6}+\frac{s (s-\bar{m}^2)^3}{4608 \pi^6}\} \\
\notag  &&            +\langle \bar{q}q\rangle^2 \int_{4m_c^2}^{s_0}ds\int_{y_i}^{y_f}dy\int_{z_i}^{1-y}dze^{-\frac{s}{T^2}} \{y z (y+z-1)^2\} \{\frac{(s-\bar{m}^2)^4}{18432 \pi^6}+\frac{s (s-\bar{m}^2)^3}{2304 \pi^6}\} \\
\notag  &&            +m_c^2 \langle \bar{q}q\rangle^2 \int_{4m_c^2}^{s_0}ds\int_{y_i}^{y_f}dy\int_{z_i}^{1-y}dze^{-\frac{s}{T^2}} \{(y+z-1)^2\} \{\frac{(s-\bar{m}^2)^3}{1536 \pi^6}\} \\
\notag  &&            +\langle \bar{q}q\rangle^2 \int_{4m_c^2}^{s_0}ds\int_{y_i}^{y_f}dy\int_{z_i}^{1-y}dze^{-\frac{s}{T^2}} \{y z (y+z-1)^2\} \{\frac{(s-\bar{m}^2)^4}{995328 \pi^8}+\frac{s (s-\bar{m}^2)^3}{124416 \pi^8}\} \\
\notag  &&            +m_c^2 \langle \bar{q}q\rangle^2 \int_{4m_c^2}^{s_0}ds\int_{y_i}^{y_f}dy\int_{z_i}^{1-y}dze^{-\frac{s}{T^2}} \{(y+z-1)^2\} \{\frac{(s-\bar{m}^2)^3}{82944 \pi^8}\} \\
\notag  &&            +\langle \bar{q}q\rangle^2 \int_{4m_c^2}^{s_0}ds\int_{y_i}^{y_f}dy\int_{z_i}^{1-y}dze^{-\frac{s}{T^2}} \{y z (y+z-1)^2\} \{\frac{(s-\bar{m}^2)^4}{497664 \pi^8}+\frac{s (s-\bar{m}^2)^3}{248832 \pi^8}\}\, , \\
\notag
\end{eqnarray}

\begin{eqnarray}
\notag \textbf{S}_B(6)_2 &&               =m_c^2 \langle \bar{q}q\rangle^2 \int_{4m_c^2}^{s_0}ds\int_{y_i}^{y_f}dy\int_{z_i}^{1-y}dze^{-\frac{s}{T^2}} \{\frac{(y+z-1)^4}{y}\} \{-\frac{(s-\bar{m}^2)^3}{331776 \pi^8}-\frac{s (s-\bar{m}^2)^2}{221184 \pi^8}\} \\
\notag  &&               +m_c^2 \langle \bar{q}q\rangle^2 \int_{4m_c^2}^{s_0}ds\int_{y_i}^{y_f}dy\int_{z_i}^{1-y}dze^{-\frac{s}{T^2}} \{\frac{z (y+z-1)^4}{y^2}\} \{-\frac{(s-\bar{m}^2)^3}{995328 \pi^8}-\frac{s (s-\bar{m}^2)^2}{663552 \pi^8}\} \\
\notag  &&               +m_c^2 \langle \bar{q}q\rangle^2 \int_{4m_c^2}^{s_0}ds\int_{y_i}^{y_f}dy\int_{z_i}^{1-y}dze^{-\frac{s}{T^2}} \{\frac{z (y+z-1)^4}{y^2}\} \{-\frac{(s-\bar{m}^2)^3}{1990656 \pi^8}-\frac{s (s-\bar{m}^2)^2}{331776 \pi^8}\} \\
\notag  &&               +\langle \bar{q}q\rangle^2 \int_{4m_c^2}^{s_0}ds\int_{y_i}^{y_f}dy\int_{z_i}^{1-y}dze^{-\frac{s}{T^2}} \{\frac{z (y+z-1)^4}{y}\} \{\frac{s^2 y (s-\bar{m}^2)^2}{995328 \pi^8}+\frac{y (s-\bar{m}^2)^4}{1990656 \pi^8}\}\\
\notag  &&  +\langle \bar{q}q\rangle^2 \int_{4m_c^2}^{s_0}ds\int_{y_i}^{y_f}dy\int_{z_i}^{1-y}dze^{-\frac{s}{T^2}} \{\frac{z (y+z-1)^4}{y}\}\{\frac{s y (s-\bar{m}^2)^3}{497664 \pi^8}-\frac{(s-\bar{m}^2)^4}{2985984 \pi^8}-\frac{s (s-\bar{m}^2)^3}{1492992 \pi^8}\} \\
\notag  &&               +\langle \bar{q}q\rangle^2 \int_{4m_c^2}^{s_0}ds\int_{y_i}^{y_f}dy\int_{z_i}^{1-y}dze^{-\frac{s}{T^2}} \{\frac{z (y+z-1)^4}{y}\} \{\frac{s^2 y (s-\bar{m}^2)^2}{497664 \pi^8}+\frac{y (s-\bar{m}^2)^4}{3981312 \pi^8}\}\\
\notag && +\langle \bar{q}q\rangle^2 \int_{4m_c^2}^{s_0}ds\int_{y_i}^{y_f}dy\int_{z_i}^{1-y}dze^{-\frac{s}{T^2}} \{\frac{z (y+z-1)^4}{y}\}\{\frac{s y (s-\bar{m}^2)^3}{331776 \pi^8}-\frac{(s-\bar{m}^2)^4}{5971968 \pi^8}-\frac{s (s-\bar{m}^2)^3}{746496 \pi^8}\} \\
\notag  &&               +\langle \bar{q}q\rangle^2 \int_{4m_c^2}^{s_0}ds\int_{y_i}^{y_f}dy\int_{z_i}^{1-y}dze^{-\frac{s}{T^2}} \{\frac{z (y+z-1)^4}{y}\} \{\frac{s^2 y (s-\bar{m}^2)^2}{995328 \pi^8}-\frac{y (s-\bar{m}^2)^4}{1990656 \pi^8}\}\\
\notag  && +\langle \bar{q}q\rangle^2 \int_{4m_c^2}^{s_0}ds\int_{y_i}^{y_f}dy\int_{z_i}^{1-y}dze^{-\frac{s}{T^2}} \{\frac{z (y+z-1)^4}{y}\}\{-\frac{s y (s-\bar{m}^2)^3}{497664 \pi^8}+\frac{(s-\bar{m}^2)^4}{11943936 \pi^8}+\frac{s (s-\bar{m}^2)^3}{5971968 \pi^8}\} \\
\notag  &&               +\langle \bar{q}q\rangle^2 \int_{4m_c^2}^{s_0}ds\int_{y_i}^{y_f}dy\int_{z_i}^{1-y}dze^{-\frac{s}{T^2}} \{\frac{z (y+z-1)^4}{y}\} \{\frac{s^2 y (s-\bar{m}^2)^2}{497664 \pi^8}-\frac{y (s-\bar{m}^2)^4}{3981312 \pi^8}\}\\
\notag  && +\langle \bar{q}q\rangle^2 \int_{4m_c^2}^{s_0}ds\int_{y_i}^{y_f}dy\int_{z_i}^{1-y}dze^{-\frac{s}{T^2}} \{\frac{z (y+z-1)^4}{y}\}\{-\frac{s y (s-\bar{m}^2)^3}{331776 \pi^8}+\frac{(s-\bar{m}^2)^4}{23887872 \pi^8}+\frac{s (s-\bar{m}^2)^3}{2985984 \pi^8}\} \, ,\\
\notag
\end{eqnarray}

\begin{eqnarray}
\notag \textbf{S}_B(8) &&               =\langle \bar{q}g_s\sigma Gq\rangle \langle \bar{q}q\rangle \int_{4m_c^2}^{s_0}ds\int_{y_i}^{y_f}dy\int_{z_i}^{1-y}dze^{-\frac{s}{T^2}} \{-y z (y+z-1)\} \{-\frac{(s-\bar{m}^2)^3}{2304 \pi^6}-\frac{s (s-\bar{m}^2)^2}{1536 \pi^6}\} \\
\notag  &&               +\langle \bar{q}g_s\sigma Gq\rangle \langle \bar{q}q\rangle \int_{4m_c^2}^{s_0}ds\int_{y_i}^{y_f}dy\int_{z_i}^{1-y}dze^{-\frac{s}{T^2}} \{-y z (y+z-1)\} \{-\frac{(s-\bar{m}^2)^3}{4608 \pi^6}-\frac{s (s-\bar{m}^2)^2}{768 \pi^6}\} \\
\notag  &&               +m_c^2 \langle \bar{q}g_s\sigma Gq\rangle \langle \bar{q}q\rangle \int_{4m_c^2}^{s_0}ds\int_{y_i}^{y_f}dy\int_{z_i}^{1-y}dze^{-\frac{s}{T^2}} \{-y-z+1\} \{-\frac{(s-\bar{m}^2)^2}{512 \pi^6}\} \, ,\\
\notag
\end{eqnarray}

\begin{eqnarray}
\notag \textbf{S}_B(10)_1 &&               =m_c^2 \langle \bar{q}q\rangle^2 \langle g_s^2 GG\rangle \int_{4m_c^2}^{s_0}ds\int_{y_i}^{y_f}dy\int_{z_i}^{1-y}dze^{-\frac{s}{T^2}} \{\frac{z (y+z-1)^2}{y^2}\} \{\frac{\bar{m}^2-s}{27648 \pi^6}-\frac{s}{13824 \pi^6}\} \\
\notag  &&               +m_c^2 \langle g_s^2 GG\rangle \langle \bar{q}q\rangle^2 \int_{4m_c^2}^{s_0}ds\int_{y_i}^{y_f}dy\int_{z_i}^{1-y}dze^{-\frac{s}{T^2}} \{-\frac{(y+z-1)^2}{y^2}\} \{\frac{(1-y) (\bar{m}^2-s)}{4608 \pi^6}+\frac{s y}{9216 \pi^6}\} \\
\notag  &&               +m_c^2 \langle g_s^2 GG\rangle \langle \bar{q}q\rangle^2 \int_{4m_c^2}^{s_0}ds\int_{y_i}^{y_f}dy\int_{z_i}^{1-y}dze^{-\frac{s}{T^2}} \{\frac{z (y+z-1)^2}{y^2}\} \{\frac{\bar{m}^2-s}{13824 \pi^6}-\frac{s}{27648 \pi^6}\} \\
\notag  &&               +\langle \bar{q}q\rangle^2 \langle g_s^2 GG\rangle \int_{4m_c^2}^{s_0}ds\int_{y_i}^{y_f}dy\int_{z_i}^{1-y}dze^{-\frac{s}{T^2}} \{y z\} \{\frac{(s-\bar{m}^2)^2}{36864 \pi^6}+\frac{s (s-\bar{m}^2)}{9216 \pi^6}\} \\
\notag  &&                +m_c^2 \langle g_s^2 GG\rangle \langle \bar{q}q\rangle^2 \int_{4m_c^2}^{s_0}ds\int_{y_i}^{y_f}dy\int_{z_i}^{1-y}dze^{-\frac{s}{T^2}} \{-\frac{\bar{m}^2-s}{6144 \pi^6}\} \\
\notag  &&               +\langle g_s^2 GG\rangle \langle \bar{q}q\rangle^2 \int_{4m_c^2}^{s_0}ds\int_{y_i}^{y_f}dy\int_{z_i}^{1-y}dze^{-\frac{s}{T^2}} \{y z\} \{\frac{(s-\bar{m}^2)^2}{18432 \pi^6}+\frac{s (s-\bar{m}^2)}{18432 \pi^6}\} \\
\notag  &&               +\langle \bar{q}q\rangle^2 \langle g_s^2 GG\rangle \int_{4m_c^2}^{s_0}ds\int_{y_i}^{y_f}dy\int_{z_i}^{1-y}dze^{-\frac{s}{T^2}} \{y z\} \{\frac{(s-\bar{m}^2)^2}{55296 \pi^6}+\frac{s (s-\bar{m}^2)}{13824 \pi^6}\} \\
\notag  &&                +m_c^2 \langle \bar{q}q\rangle^2 \langle g_s^2 GG\rangle \int_{4m_c^2}^{s_0}ds\int_{y_i}^{y_f}dy\int_{z_i}^{1-y}dze^{-\frac{s}{T^2}} \{-\frac{\bar{m}^2-s}{9216 \pi^6}\} \\
\notag  &&               +\langle \bar{q}q\rangle^2 \langle g_s^2 GG\rangle \int_{4m_c^2}^{s_0}ds\int_{y_i}^{y_f}dy\int_{z_i}^{1-y}dze^{-\frac{s}{T^2}} \{y z\} \{\frac{(s-\bar{m}^2)^2}{27648 \pi^6}+\frac{s (s-\bar{m}^2)}{27648 \pi^6}\} \, ,\\
\notag
\end{eqnarray}

\begin{eqnarray}
\notag \textbf{S}_B(10)_2 &&                       +\langle \bar{q}g_s\sigma Gq\rangle^2 \int_{4m_c^2}^{s_0}ds\int_{y_i}^{y_f}dy\int_{z_i}^{1-y}dze^{-\frac{s}{T^2}} \{(y z\} \{\frac{(s-\bar{m}^2)^2}{6144 \pi^6}+\frac{s (s-\bar{m}^2)}{6144 \pi^6}\} \\
\notag  &&                       +\langle \bar{q}g_s\sigma Gq\rangle^2 \int_{4m_c^2}^{s_0}ds\int_{y_i}^{y_f}dy\int_{z_i}^{1-y}dze^{-\frac{s}{T^2}} \{y z\} \{\frac{(s-\bar{m}^2)^2}{12288 \pi^6}+\frac{s (s-\bar{m}^2)}{3072 \pi^6}\} \\
\notag  &&                       +m_c^2 \langle \bar{q}g_s\sigma Gq\rangle^2 \int_{4m_c^2}^{s_0}ds\int_{y_i}^{y_f}dy\int_{z_i}^{1-y}dze^{-\frac{s}{T^2}} \{-\frac{\bar{m}^2-s}{2048 \pi^6}\} \\
\notag  &&                       +\langle \bar{q}g_s\sigma Gq\rangle^2 \int_{4m_c^2}^{s_0}ds\int_{y_i}^{y_f}dy\int_{z_i}^{1-y}dze^{-\frac{s}{T^2}} \{y z\} \{\frac{(s-\bar{m}^2)^2}{589824 \pi^6}+\frac{s (s-\bar{m}^2)}{147456 \pi^6}\} \\
\notag  &&                       +m_c^2 \langle \bar{q}g_s\sigma Gq\rangle^2 \int_{4m_c^2}^{s_0}ds\int_{y_i}^{y_f}dy\int_{z_i}^{1-y}dze^{-\frac{s}{T^2}} \{-\frac{\bar{m}^2-s}{98304 \pi^6}\} \\
\notag  &&                       +\langle \bar{q}g_s\sigma Gq\rangle^2 \int_{4m_c^2}^{s_0}ds\int_{y_i}^{y_f}dy\int_{z_i}^{1-y}dze^{-\frac{s}{T^2}} \{y z\} \{\frac{(s-\bar{m}^2)^2}{294912 \pi^6}+\frac{s (s-\bar{m}^2)}{294912 \pi^6}\} \, ,\\
\notag
\end{eqnarray}

\begin{eqnarray}
\notag \textbf{S}_B(12) &&               =m_c^2\langle \bar{q}q\rangle^4  \int_{4m_c^2}^{s_0}ds\int_{y_i}^{y_f}dy\int_{z_i}^{1-y}dze^{-\frac{s}{T^2}} \{\frac{y+z-1}{y}\} \{\frac{1}{1296 \pi^4}\} \\
\notag  &&               +m_c^2\langle \bar{q}q\rangle^4  \int_{4m_c^2}^{s_0}ds\int_{y_i}^{y_f}dy\int_{z_i}^{1-y}dze^{-\frac{s}{T^2}} \{-\frac{z (y+z-1)}{y^2}\} \{-\frac{1}{3888 \pi^4}\} \\
\notag  &&               +m_c^2\langle \bar{q}q\rangle^4  \int_{4m_c^2}^{s_0}ds\int_{y_i}^{y_f}dy\int_{z_i}^{1-y}dze^{-\frac{s}{T^2}} \{-\frac{z (y+z-1)}{y^2}\} \{-\frac{1}{7776 \pi^4}\} \\
\notag  &&               +\langle \bar{q}q\rangle^4 \int_{4m_c^2}^{s_0}ds\int_{y_i}^{y_f}dy\int_{z_i}^{1-y}dze^{-\frac{s}{T^2}} \{-\frac{z (y+z-1)}{y}\} \{\frac{\bar{m}^2-s}{2916 \pi^4}-\frac{s}{5832 \pi^4}\} \\
\notag  &&               +m_c^2 \langle \bar{q}q\rangle^4 \int_{4m_c^2}^{s_0}ds\int_{y_i}^{y_f}dy\int_{z_i}^{1-y}dze^{-\frac{s}{T^2}} \{-\frac{z (y+z-1)}{y^2})\} \{\frac{1}{5832 \pi^4}\} \\
\notag  &&               +\langle \bar{q}q\rangle^4 \int_{4m_c^2}^{s_0}ds\int_{y_i}^{y_f}dy\int_{z_i}^{1-y}dze^{-\frac{s}{T^2}} \{-\frac{z (y+z-1)}{y})\} \{\frac{\bar{m}^2-s}{5832 \pi^4}-\frac{s}{2916 \pi^4}\} \\
\notag  &&               +m_c^2 \langle \bar{q}q\rangle^4 \int_{4m_c^2}^{s_0}ds\int_{y_i}^{y_f}dy\int_{z_i}^{1-y}dze^{-\frac{s}{T^2}} \{-\frac{z (y+z-1)}{y^2}\} \{\frac{1}{11664 \pi^4}\} \\
\notag  &&               +\langle \bar{q}q\rangle^4 \int_{4m_c^2}^{s_0}ds\int_{y_i}^{y_f}dy\int_{z_i}^{1-y}dze^{-\frac{s}{T^2}} \{-\frac{z (y+z-1)}{y})\} \{\frac{\bar{m}^2 y}{1944 \pi^4}-\frac{\bar{m}^2}{11664 \pi^4}-\frac{s y}{972 \pi^4}+\frac{s}{7776 \pi^4}\} \\
\notag  &&               +\langle \bar{q}q\rangle^4 \int_{4m_c^2}^{s_0}ds\int_{y_i}^{y_f}dy\int_{z_i}^{1-y}dze^{-\frac{s}{T^2}} \{-\frac{z (y+z-1)}{y}\} \{\frac{\bar{m}^2 y}{3888 \pi^4}-\frac{\bar{m}^2}{23328 \pi^4}-\frac{s y}{972 \pi^4}+\frac{s}{7776 \pi^4}\}\, . \\
\notag
\end{eqnarray}

Type C for $J_4$

\begin{eqnarray}
\notag \textbf{S}_C(14) &&                  =\langle \bar{q}g_s\sigma Gq\rangle \langle \bar{q}q\rangle^3 \int_{y_i}^{y_f}dye^{-\frac{\tilde{m}^2}{T^2}} \{(y-1) y\} \{\frac{\tilde{m}^2}{108 \pi^2}+\frac{\tilde{m}^4}{432 \pi^2 T^2}\} \\
\notag  &&                  +\langle \bar{q}g_s\sigma Gq\rangle \langle \bar{q}q\rangle^3 \int_{y_i}^{y_f}dye^{-\frac{\tilde{m}^2}{T^2}} \{(y-1) y\} \{\frac{5 \tilde{m}^2}{432 \pi^2}+\frac{\tilde{m}^4}{216 \pi^2 T^2}\} \\
\notag  &&                  +m_c^2 \langle \bar{q}g_s\sigma Gq\rangle \langle \bar{q}q\rangle^3 \int_{y_i}^{y_f}dye^{-\frac{\tilde{m}^2}{T^2}} \{-\frac{1}{144 \pi^2}-\frac{\tilde{m}^2}{144 \pi^2 T^2}\} \\
\notag  &&                  +\langle \bar{q}q\rangle^3 \langle \bar{q}g_s\sigma Gq\rangle \int_{y_i}^{y_f}dye^{-\frac{\tilde{m}^2}{T^2}} \{(y-1) y\} \{\frac{5 \tilde{m}^2}{23328 \pi^4}+\frac{\tilde{m}^4}{11664 \pi^4 T^2}\} \\
\notag  &&                  +m_c^2 \langle \bar{q}q\rangle^3 \langle \bar{q}g_s\sigma Gq\rangle \int_{y_i}^{y_f}dye^{-\frac{\tilde{m}^2}{T^2}} \{-\frac{1}{7776 \pi^4}-\frac{\tilde{m}^2}{7776 \pi^4 T^2}\} \\
\notag  &&                  +\langle \bar{q}q\rangle^3 \langle \bar{q}g_s\sigma Gq\rangle \int_{y_i}^{y_f}dye^{-\frac{\tilde{m}^2}{T^2}} \{(y-1) y\} \{\frac{\tilde{m}^2}{5832 \pi^4}+\frac{\tilde{m}^4}{23328 \pi^4 T^2}\}\, , \\
\notag
\end{eqnarray}

\begin{eqnarray}
\notag \textbf{S}_C(16)_1 &&                  =\langle \bar{q}g_s\sigma Gq\rangle^2 \langle \bar{q}q\rangle^2 \int_{y_i}^{y_f}dye^{-\frac{\tilde{m}^2}{T^2}} \{(y-1) y\} \{-\frac{1}{192 \pi^2}-\frac{\tilde{m}^6}{1152 \pi^2 T^6}-\frac{\tilde{m}^4}{384 \pi^2 T^4}-\frac{\tilde{m}^2}{192 \pi^2 T^2}\} \\
\notag  &&                  +\langle \bar{q}g_s\sigma Gq\rangle^2 \langle \bar{q}q\rangle^2 \int_{y_i}^{y_f}dye^{-\frac{\tilde{m}^2}{T^2}} \{(y-1) y\} \{\frac{1}{192 \pi^2}-\frac{\tilde{m}^6}{576 \pi^2 T^6}-\frac{\tilde{m}^4}{384 \pi^2 T^4}\} \\
\notag  &&                  +m_c^2 \langle \bar{q}g_s\sigma Gq\rangle^2 \langle \bar{q}q\rangle^2 \int_{y_i}^{y_f}dye^{-\frac{\tilde{m}^2}{T^2}} \{\frac{\tilde{m}^4}{64 \pi^2 T^6}\} \\
\notag  &&                  +\langle \bar{q}q\rangle^2 \langle \bar{q}g_s\sigma Gq\rangle^2 \int_{y_i}^{y_f}dye^{-\frac{\tilde{m}^2}{T^2}} \{(y-1) y\} \{\frac{1}{31104 \pi^4}-\frac{\tilde{m}^6}{93312 \pi^4 T^6}-\frac{\tilde{m}^4}{62208 \pi^4 T^4}\} \\
\notag  &&                  +m_c^2 \langle \bar{q}q\rangle^2 \langle \bar{q}g_s\sigma Gq\rangle^2 \int_{y_i}^{y_f}dye^{-\frac{\tilde{m}^2}{T^2}} \{\frac{\tilde{m}^4}{62208 \pi^4 T^6}\} \\
\notag  &&                  +\langle \bar{q}q\rangle^2 \langle \bar{q}g_s\sigma Gq\rangle^2 \int_{y_i}^{y_f}dye^{-\frac{\tilde{m}^2}{T^2}} \{(y-1) y\} \{-\frac{1}{31104 \pi^4}-\frac{\tilde{m}^6}{186624 \pi^4 T^6}-\frac{\tilde{m}^4}{62208 \pi^4 T^4}-\frac{\tilde{m}^2}{31104 \pi^4 T^2}\}\, , \\
\notag
\end{eqnarray}

\begin{eqnarray}
\notag \textbf{S}_C(16)_2 &&                   =m_c^2 \langle \bar{q}q\rangle^4 \langle g_s^2 GG\rangle \int_{y_i}^{y_f}dye^{-\frac{\tilde{m}^2}{T^2}} \{\frac{y-1}{y^2}\} \{\frac{\tilde{m}^2}{7776 \pi^2 T^4}-\frac{1}{5184 \pi^2 T^2}\} \\
\notag  &&                     +m_c^2 \langle \bar{q}q\rangle^4 \langle g_s^2 GG\rangle \int_{y_i}^{y_f}dye^{-\frac{\tilde{m}^2}{T^2}} \{\frac{1}{y^2}\} \{\frac{1}{1728 \pi^2 T^2}\} \\
\notag  &&                   +m_c^4 \langle \bar{q}q\rangle^4 \langle g_s^2 GG\rangle \int_{y_i}^{y_f}dye^{-\frac{\tilde{m}^2}{T^2}} \{\frac{1}{y^3}\} \{-\frac{1}{5184 \pi^2 T^4}\} \\
\notag  &&                   +m_c^2 \langle g_s^2 GG\rangle \langle \bar{q}q\rangle^4 \int_{y_i}^{y_f}dye^{-\frac{\tilde{m}^2}{T^2}} \{\frac{y-1}{y^2}\} \{\frac{\tilde{m}^2}{15552 \pi^2 T^4}\} \\
\notag  &&                   +\langle \bar{q}q\rangle^4 \langle g_s^2 GG\rangle \int_{y_i}^{y_f}dye^{-\frac{\tilde{m}^2}{T^2}} \{(y-1) y\} \{\frac{1}{2592 \pi^2}-\frac{\tilde{m}^6}{7776 \pi^2 T^6}-\frac{\tilde{m}^4}{5184 \pi^2 T^4}\} \\
\notag  &&                   +m_c^2 \langle \bar{q}q\rangle^4 \langle g_s^2 GG\rangle \int_{y_i}^{y_f}dye^{-\frac{\tilde{m}^2}{T^2}} \{\frac{\tilde{m}^4}{5184 \pi^2 T^6}\} \\
\notag  &&                   +\langle \bar{q}q\rangle^4 \langle g_s^2 GG\rangle \int_{y_i}^{y_f}dye^{-\frac{\tilde{m}^2}{T^2}} \{(y-1) y\} \{-\frac{1}{2592 \pi^2}-\frac{\tilde{m}^6}{15552 \pi^2 T^6}-\frac{\tilde{m}^4}{5184 \pi^2 T^4}-\frac{\tilde{m}^2}{2592 \pi^2 T^2}\} \, .\\
\notag
\end{eqnarray}

Type D for $J_4$

\begin{eqnarray}
\notag \textbf{S}_D(12) &&                   = m_c^2\langle \bar{q}q\rangle^4 \int_{y_i}^{y_f}dy\int_{z_i}^{1-y}dze^{-\frac{\bar{m}^2}{T^2}} \{\frac{y+z-1}{y}\} \{\frac{\bar{m}^2}{2592 \pi^4}\} \\
\notag  &&                   +m_c^2\langle \bar{q}q\rangle^4  \int_{y_i}^{y_f}dy\int_{z_i}^{1-y}dze^{-\frac{\bar{m}^2}{T^2}} \{-\frac{z (y+z-1)}{y^2}\} \{-\frac{\bar{m}^2}{7776 \pi^4}\} \\
\notag  &&                   +m_c^2\langle \bar{q}q\rangle^4  \int_{y_i}^{y_f}dy\int_{z_i}^{1-y}dze^{-\frac{\bar{m}^2}{T^2}} \{-\frac{z (y+z-1)}{y^2}\} \{-\frac{\bar{m}^2}{3888 \pi^4}\} \\
\notag  &&                   +m_c^2\langle \bar{q}q\rangle^4  \int_{y_i}^{y_f}dy\int_{z_i}^{1-y}dze^{-\frac{\bar{m}^2}{T^2}} \{-\frac{z (y+z-1)}{y^2}\} \{\frac{\bar{m}^2}{11664 \pi^4}\} \\
\notag  &&                   +m_c^2\langle \bar{q}q\rangle^4  \int_{y_i}^{y_f}dy\int_{z_i}^{1-y}dze^{-\frac{\bar{m}^2}{T^2}} \{-\frac{z (y+z-1)}{y^2}\} \{\frac{\bar{m}^2}{5832 \pi^4}\} \\
\notag  &&                   +\langle \bar{q}q\rangle^4 \int_{y_i}^{y_f}dy\int_{z_i}^{1-y}dze^{-\frac{\bar{m}^2}{T^2}} \{-\frac{z (y+z-1)}{y}\} \{-\frac{\bar{m}^4 y}{11664 \pi^4}\} \\
\notag  &&                   +\langle \bar{q}q\rangle^4 \int_{y_i}^{y_f}dy\int_{z_i}^{1-y}dze^{-\frac{\bar{m}^2}{T^2}} \{-\frac{z (y+z-1)}{y}\} \{-\frac{\bar{m}^4 y}{5832 \pi^4}\} \, .\\
\notag
\end{eqnarray}

\end{document}